%% file: LukasBarth_aGeometricFrameworkToComparePDEsAndClassicalFieldTheories.tex
\documentclass[graybox, dvipsnames]{svmult}

\usepackage{type1cm}        
\usepackage{makeidx}         
\usepackage{graphicx}        
\usepackage{multicol}        
\usepackage[bottom]{footmisc}
\usepackage{geometry}
\usepackage{float}

\usepackage{newtxtext}       %
\usepackage{amssymb}

\makeindex             

\usepackage[titles]{tocloft}
\usepackage{enumitem}
\usepackage{cite} 
\usepackage[hyperfootnotes=false, colorlinks=true,
  urlcolor=black,
  linkcolor=Blue,
  citecolor=Violet,
  pagebackref=true,
  pdftitle={A geometric framework to compare classical field theories},
  pdfauthor={Lukas Barth},
  pdfkeywords={Classical field theory, PDE, partial differential equation, symmetry, symmetries, Bäcklund transformation, transfer of solutions, jet bundle, jet space, quotient equation, electrodynamics, hydrodynamics, comparison of theories}
]{hyperref}
\input{programmingCodeDisplay.tex}
\usepackage{mathtools}
\usepackage{tikz-cd}
\usepackage{setspace}

\usepackage[T1]{fontenc}
\usepackage{textcomp}
\usepackage[most]{tcolorbox}

\newtcolorbox[auto counter,number within=section,
                          ]{greendefinition}[1][]{enhanced jigsaw,
  colback=white!50!green,
  coltext={black},
  colframe={black},
  coltitle={black},
  boxrule=0pt,
  arc=1mm,
  outer arc=1mm,
  boxsep=5pt,
  left=2pt,
  right=2pt,
  bottom=2pt,
  top=2pt,
  breakable,
  before skip=3mm,
  after skip=3mm,
  title={Definition~\thetcbcounter.},
  attach title to upper=\quad,
  fonttitle=\bfseries,
  #1}
\newtcolorbox[auto counter,
  number within=section,
  number freestyle={\noexpand\thechapter.\noexpand\arabic{\tcbcounter}}]{myexample}[2][]{%
    enhanced,
    breakable,
    fonttitle=\bfseries,
    title=Example~\thetcbcounter: #2,
    #1
}
\newtcolorbox[auto counter, number within=section]{blueproposition}[1][]{enhanced jigsaw,
  colback=white, 
  coltext={black},
  coltitle={black},
  boxrule=0pt,
  arc=1mm,
  auto outer arc,
  boxsep=5pt,
  left=2pt,
  right=2pt,
  bottom=2pt,
  top=2pt,
  borderline west={1mm}{0mm}{white!60!blue},
  before skip=3mm,
  after skip=3mm,
  breakable,
  title={Proposition~\thetcbcounter.},
  attach title to upper=\quad,
  fonttitle=\bfseries,
  #1}
\newtcolorbox[auto counter, number within=section]{bluecorollary}[1][]{enhanced jigsaw,
  colback=white, 
  coltext={black},
  colframe={black},
  coltitle={black},
  boxrule=0pt,
  frame hidden,
  arc=1mm,
  auto outer arc,
  boxsep=5pt,
  left=2pt,
  right=2pt,
  bottom=2pt,
  top=2pt,
  borderline west={1mm}{0mm}{white!60!blue},
  before skip=3mm,
  after skip=3mm,
  breakable,
  title={Corollary~\thetcbcounter.},
  attach title to upper=\quad,
  fonttitle=\bfseries,
  #1}
\newtcolorbox[auto counter,number within=section
                          ]{debox}[1][]{enhanced jigsaw,
  colback=green!0!white,
  coltext={black},
  colframe={black},
  coltitle={black},
  boxrule=0pt,
  frame hidden,
  borderline west={1mm}{0mm}{white!50!green},
  arc=0mm,
  auto outer arc,
  boxsep=5pt,
  left=4pt,
  breakable,
  right=4pt,
  bottom=0pt,
  top=0pt,
  before skip=3mm,
  after skip=3mm,
  title={Definition~\thetcbcounter.},
  attach title to upper=\quad,
  fonttitle=\bfseries,
  #1}
\newtcolorbox[auto counter,number within=section
                          ]{yellowRemark}[1][breakable]{enhanced jigsaw,
  colback=green!0!white,
  coltext={black},
  colframe={black},
  coltitle={black},
  boxrule=0pt,
  frame hidden,
  breakable,
  borderline west={1mm}{0mm}{white!50!yellow},
  arc=0mm,
  auto outer arc,
  boxsep=5pt,
  left=4pt,
  right=4pt,
  bottom=0pt,
  top=0pt,
  before skip=3mm,
  after skip=3mm,
  title={Remark~\thetcbcounter.},
  attach title to upper=\quad,
  fonttitle=\bfseries,
  #1}
\newtcolorbox[auto counter,number within=section
                          ]{exampleCommand}[1][breakable]{enhanced jigsaw,
  colback=green!0!white,
  coltext={black},
  colframe={black},
  coltitle={black},
  boxrule=0pt,
  frame hidden,
  breakable,
  borderline west={1mm}{0mm}{white!50!yellow},
  arc=0mm,
  auto outer arc,
  boxsep=5pt,
  left=4pt,
  right=4pt,
  bottom=0pt,
  top=0pt,
  before skip=3mm,
  after skip=3mm,
  title={Example~\thetcbcounter.},
  attach title to upper=\quad,
  fonttitle=\bfseries,
  #1}
\newtcolorbox[auto counter,number within=section
                          ]{yellowCitation}[1][breakable]{enhanced jigsaw,
  colback=green!0!white,
  coltext={black},
  colframe={black},
  coltitle={black},
  boxrule=0pt,
  frame hidden,
  breakable,
  borderline west={1mm}{0mm}{white!50!yellow},
  arc=0mm,
  auto outer arc,
  boxsep=5pt,
  left=4pt,
  right=4pt,
  bottom=0pt,
  top=0pt,
  before skip=3mm,
  after skip=3mm,
  #1}
\newtcolorbox[auto counter,number within=section
                          ]{yellowQuestion}[1][breakable]{enhanced jigsaw,
  colback=green!0!white,
  coltext={black},
  colframe={black},
  coltitle={black},
  boxrule=0pt,
  frame hidden,
  breakable,
  borderline west={1mm}{0mm}{white!50!yellow},
  arc=0mm,
  auto outer arc,
  boxsep=5pt,
  left=4pt,
  right=4pt,
  bottom=0pt,
  top=0pt,
  before skip=3mm,
  after skip=3mm,
  title={Question~\thetcbcounter.},
  attach title to upper=\quad,
  fonttitle=\bfseries,
  #1}

\newtcolorbox{blueshadedcvbox}[1][]{enhanced jigsaw,
  colback=white!70!blue,
  coltext={black},
  boxrule=0pt,
  arc=4mm,
  auto outer arc,
  boxsep=7pt,
  left=4pt,
  right=2pt,
  bottom=2pt,
  top=2pt,
  #1}
\newtcolorbox[auto counter,number within=section
                          ]{greenshadedcvbox}[1][]{enhanced jigsaw,
  colback=green!0!white,
  coltext={black},
  colframe={black},
  coltitle={black},
  boxrule=0pt,
  frame hidden,
  borderline west={1mm}{0mm}{white!50!green},
  arc=0mm,
  auto outer arc,
  boxsep=5pt,
  left=4pt,
  breakable,
  right=4pt,
  bottom=0pt,
  top=0pt,
  before skip=3mm,
  after skip=3mm,
  title={Definition~\thetcbcounter.},
  attach title to upper=\quad,
  fonttitle=\bfseries,
  #1}
\newtcolorbox{yellowshadedcvbox}[1][]{enhanced jigsaw,
  colback=white!30!yellow,
  coltext={black},
  boxrule=0pt,
  arc=4mm,
  auto outer arc,
  boxsep=7pt,
  left=4pt,
  right=2pt,
  bottom=2pt,
  top=2pt,
  #1}

\newcommand{\bt}[1][normal]{\begin{tikzcd}[ampersand replacement = \&, column sep=#1,row sep=#1]}
\newcommand{\et}{\end{tikzcd}}
\newcommand{\cor}{\mathrel{\widehat{=}}}
\newcommand{\Ra}{\mathrel{ \Rightarrow }}
\newcommand{\ra}{\mathrel{ \rightarrow }}

\newcommand{\te}[1]{\text{#1}}
\newcommand{\tf}[1]{\textbf{#1}}
\newcommand{\ti}[1]{\textit{#1}}
\newcommand{\ov}[1]{\mathrel{ \overset{ \text{(#1)} }{=}  }}
\newcommand{\br}[1]{\mathrel{ \left(#1\right)  }}
\newcommand{\Br}[1]{\mathrel{ \left[#1\right]  }}
\newcommand{\bbr}[1]{\mathrel{ \left\{#1\right\}  }}

\newcommand{\de}[1]{\mathcal{#1}}
\newcommand{\bo}[1]{\mathbf{#1}}
\newcommand{\eq}[1]{\begin{rcases}\begin{dcases}#1\end{dcases}\end{rcases}}
\newcommand{\un}[2]{\mathrel{ \underbrace{#1}_{#2}  }}
\newcommand{\tv}{\mathrel{\text{\tpitchfork}}}
\makeatletter
\newcommand{\tpitchfork}{%
    \raise-0.1ex
    \vbox{
    \baselineskip\z@skip
    \lineskip-.52ex
    \lineskiplimit\maxdimen
    \m@th
    \ialign{##\crcr\hidewidth\smash{$-$}\hidewidth\crcr$\pitchfork$\crcr}
  }%
}
\makeatother
\newcommand{\thirdoffive}[5]{#3}
\newcommand{\assignnameref}[2]{%
  \gdef#2{}
  \ifcsname r@#1\endcsname 
    \xdef#2{\expandafter\expandafter\expandafter\thirdoffive\csname r@#1\endcsname}
  \fi
}
\newcommand{\mb}[1]{\mathbb{#1}}
\newcommand{\bc}[2]{\begin{bluecorollary}[label=#1]#2\end{bluecorollary}}
\newcommand{\pr}[1]{\begin{proof}#1\end{proof}}
\newcommand{\mytoc}[1]{\hyperref[#1]{\ref{#1} $ \qquad $ \nameref{#1} \dotfill\pageref{#1}}}
\newcommand{\mytoct}[1]{\hyperref[#1]{\ref{#1}$ \qquad $\nameref{#1} \dotfill\pageref{#1}}}
\newcommand{\mytoca}[1]{\hyperref[#1]{\ref{#1}$ \quad ~~$\nameref{#1} \dotfill\pageref{#1}}}

\newcommand{\bp}[2]{\begin{blueproposition}[label=#1]#2\end{blueproposition}}

\newcommand{\gd}[2]{\begin{debox}[label=#1]#2\end{debox}}
\newcommand{\yr}[2]{\begin{yellowRemark}[label=#1]#2\end{yellowRemark}}
\newcommand{\ye}[2]{\begin{exampleCommand}[label=#1]#2\end{exampleCommand}}

\makeatletter
\def\toclevel@title{-1}
\def\toclevel@author{0}
\makeatother

\begin{document}
\title*{A geometric framework\\to compare classical field theories\\and to transfer solutions between PDEs}
\author{Lukas Silvester Barth}

\institute{Lukas Silvester Barth \at Max Planck Institute for Mathematics in the Sciences, Inselstraße 22, 04103 Leipzig, Germany\\\email{Lukas.Barth@mis.mpg.de}}
%
%
\maketitle
\vspace{-13ex}
\abstract*{~In this contribution, 
a mathematical framework is constructed
to relate and compare non-linear partial differential equations (PDEs) in the category of smooth manifolds.
In particular, it can be used to
compare those aspects of field theories
(e.g. of classical (Newtonian) mechanics,
hydrodynamics, electrodynamics, relativity theory,
classical Yang-Mills theory and so on)
that are described by such equations.\\
Employing a geometric (jet space) approach, a suitable notion of shared structure of
two systems of PDEs is identified.
It is proven that this shared structure can serve to transfer solutions
from one theory to another and a generalization of so-called Bäcklund transformations is derived
that can be used to generate non-trivial solutions of some non-linear PDEs.\\
A procedure (based on formal integrability) is introduced with which one can
explicitly compute the minimal consistency conditions that two systems of PDEs
need to fulfill in order to share structure under a given correspondence.
Furthermore, it is shown how symmetry groups can be used to
identify useful correspondences and structure that is shared up to symmetries.
Thereby, the role that Bäcklund transformations play in the theory of quotient equations is clarified.\\
Explicit examples illustrate the general ideas throughout the text and in the last chapter, the framework is applied to systems related to
electrodynamics and hydrodynamics.
}

\abstract{~In this contribution, 
a mathematical framework is constructed
to relate and compare non-linear partial differential equations (PDEs) in the category of smooth manifolds.
In particular, it can be used to
compare those aspects of field theories
(e.g. of classical (Newtonian) mechanics,
hydrodynamics, electrodynamics, relativity theory,
classical Yang-Mills theory and so on)
that are described by such equations.\\
Employing a geometric (jet space) approach, a suitable notion of shared structure of
two systems of PDEs is identified.
It is proven that this shared structure can serve to transfer solutions
from one theory to another and a generalization of so-called Bäcklund transformations is derived
that can be used to generate non-trivial solutions of some non-linear PDEs.\\
A procedure (based on formal integrability) is introduced with which one can
explicitly compute the minimal consistency conditions that two systems of PDEs
need to fulfill in order to share structure under a given correspondence.
Furthermore, it is shown how symmetry groups can be used to
identify useful correspondences and structure that is shared up to symmetries.
Thereby, the role that Bäcklund transformations play in the theory of quotient equations is clarified.\\
Explicit examples illustrate the general ideas throughout the text and in the last chapter, the framework is applied to systems related to
electrodynamics and hydrodynamics.
}
~

\section*{Contents}

\mytoc{sec:introMethods}\\
\mytoc{sec:notationPreliminaries}\\
\mytoc{sec:corrInt}\\
\mytoc{sec:constistencyConditionsSection}\\
\mytoc{sec:formalIntegrability}\\
\mytoc{sec:sharedStructure}\\
\mytoc{sec:BaecklundCorrespondencesSection}\\
\mytoc{sec:equivUpToSym}\\
\mytoc{chap:applications}\\
\mytoct{sec:conlusion}\\
~\\
\mytoca{sec:introgeomethod}\\
\mytoca{app:formalIntSolver}\\
\mytoca{app:axForm}


~
\begin{acknowledgement}
  Since this work builds on the research of my Master's thesis, I want to express
  my gratitude towards my two former supervisors, James Owen Weatherall (University of California, Irvine) and Ion Stamatescu (University of Heidelberg).
  Furthermore, I obtained valuable comments and support from Luca Vitagliano (University of Salerno) and Igor Khavkine (Czech Academy of Sciences).
  Finally, I'd like to thank my dear friend Thomas Mikhail for his continuous feedback.
  \end{acknowledgement}

\newpage
\section{Introduction}
\label{sec:introMethods}

Studying relationships of different theories
can serve to identify their underlying central features.
Once shared structure of two theories is known,
methods for solving a problem in one domain can be transferred to another.
In the long run, a structured overview could set
free innovation for the development of these theories.

\subsection{Previous attempts to compare theories}
\label{sec:previousAttempts}
    In the physics literature, comparisons were usually restricted to analogies of two specific theories
    established by juxtaposition of the corresponding equations of motion. For instance,
    \cite{analogyNSMW1998} introduced new effective quantities to rewrite the Navier-Stokes equations
    in a form very similar to Maxwell's equations.
    \cite{Goulart2008} established an analogy between general relativity and electrodynamics
    by showing that a certain linear combination of derivatives of the Faraday tensor has an
    irreducible representation with 16 components, 10 of which can be associated with the 10 components
    of the Weyl tensor of general relativity.
    \cite{visser1997SonicBlackHole} explained the analogy of mathematical aspects
    of the description of black holes and supersonic flows which resulted in research of so-called analogue
    experiments (cf. \cite{sonicBlacHoleExp2017}).
    All those analogies are however rather specific and a general framework for comparisons is missing.\\
    In the philosophy of science literature, some more abstract, category theoretical
    approaches are outlined.
    Weatherall uses groupoids (categories
  in which all morphisms are isomorphisms)
  to compare theories that differ in their formulation but describe the same physics
  (cf. \cite{weather2014}, \cite{weather2015}).
  More specifically, the objects in those groupoids are the formal solutions of systems of PDEs
  and the morphisms are symmetries of the underlying spacetime that preserve those solutions.
  Weatherall then defines an equivalence of two such theories as a categorical equivalence between
  their corresponding groupoids that preserves the empirical content of the physical theories.
  This idea was subsequently used by others to compare formulations of other theories,
  e.g. \cite{rosen2015} compare the geometric and algebraic formulation of general relativity
  and \cite{barrett2017} compares the Lagrangian and Hamiltonian formulation of classical mechanics.\\
  The problem of this approach is however that categorical equivalence can only serve
  to render equivalent formulations that differ up to invertible (symmetry) transformations
  but is not capable of providing a framework to compare entirely different theories, to identify their intersection
  or subtheories. And it does not provide any means for understanding which solutions can be
  transferred from one theory to another.\\
  Comparisons between the Hamiltonian
  and the Lagrangian view of mechanics are also discussed in
  the mathematics literature, see e.g. \cite{abraham2008foundations} or \cite{Rom_n_Roy_2009}. But again, such
    discussions are not aimed at the formalization of a general framework
    for the comparison of theories. The most general discussion of relationships
    between systems of differential equations, known to me,
    involves the powerful concept of so-called coverings in the category of diffieties
    (cf. \cite{coverings1984}, \cite{nonlocalTrends1989}, \cite{Krasilshchik1999}).
    However, coverings were constructed to investigate generalized, nonlocal symmetries of PDEs
    and are not designed for the comparison of arbitrary systems of equations.
    Furthermore, since they are defined over infinitely prolonged differential equations,
    they can not serve to find integrability conditions
    (which requires the inclusion of methods of formal integrability at the level
    of finitely prolonged equations)
    that arise upon the comparison of different theories.\\
    Apart from these mathematical approaches, there is also literature that
    discusses the differences and transitions of physical theories heuristically.
    For instance, the ideas regarding
    the structure of scientific progress developed by \cite{Kuhn1996} are well-known.
    Kuhn describes progress in a recurring loop of eras with three stages which might roughly be described as follows:
    Confusion about how to describe a process in nature, determination of a unifying model
    and finally application of this model - until new ideas and experiments lead to
    another stage of confusion.\\
Another example of a heuristic discussion of the conceptual structure of physical theories is provided by
\cite{Stama2013}. He takes into account the role of the symbols that we use for the description of
physics and emphasises as a guideline the so-called Hertzian principle (cf. \cite{Hertz1894}).
  According this principle,
the concordance of reality and symbolic description must be such that
any consequences of an initial experimental setup due to the laws of nature
must correspond to thought consequences of the symbols that describe this initial setup
due to the laws of the mathematical formalism.
Stamatescu also discusses the transition of theories
and the development of their concepts.
The problem with more heuristic discussions is that they are very hard
to formalize.
Indeed, the geometric framework presented here can not account for
transitions of physical theories. One reason for that is that a suitable meta-theory for the description of such transitions must involve the experimental bounds / limits of a physical theory but those are highly non-trivial to determine in a complete and precise way as the discussion in appendix \ref{sec:phenomenaED} about the empirical limits of electrodynamics is supposed to demonstrate.
However, the present framework
might be extended in the future to at least formally incorporate the description of transitions of theories along the lines
suggested in the outlook \ref{sec:outlook}.\\
As a final remark on previous approaches, it should be mentioned that the present work builds on research of my Master's thesis
but contains several generalizations.
For example, the notion of a correspondence between theories was generalized
from a differential operator to a correspondence
on the natural product bundle, which now allows for more implicit comparisons of systems of equations.
Moreover, the present approach is conceptually cleaner because the two compared theories
determine the natural space in which the intersection takes place before the
correspondence is imposed. Furthermore, both,
the compared theories as well as the correspondence, are all treated as geometric spaces.
Most importantly, the approach in the thesis did not allow for a generalization of Bäcklund transformations whose inclusion allows for a much more powerful transfer of solutions.

\subsection{Requirements for the framework}
\label{sec:requirements}

A classical field theory is here understood
as a system of partial differential equations (PDEs) on some manifold
(possibly called spacetime), together with
a physical interpretation. This physical interpretation specifies
\begin{itemize}
\itemsep0em
  \item
how the mathematical quantities are related to experimental measurements,
\item
which initial / boundary conditions are physically plausible
\item
and strictly speaking should also include
validity bounds for the mathematical formalism.\footnote{For example,
  classical mechanics is only valid on certain scales,
  only produces predictions within acceptable errors up to certain velocities
etc.}
\end{itemize}
In this article however, only the PDEs themselves are compared
without considering their interpretation for two reasons:
\begin{enumerate}
  \item The aspiration of the present work is to identify common causes.
  It is desired to understand which models are structurally similar even if they
  can be associated with different experimental setups
  because exactly this abstraction facilitates
  to obtain a new intuition for the phenomena described by the equations and to
  transfer methods.
  If desired, it is always possible to impose an interpretation later to discriminate theories further. 
  \item To take into account the validity bounds that go along with an interpretation would require a lot of work,
  both because those bounds are not always clearly defined
  and because one might have to add inequalities that restrict the range of the variables. The discussion about the empirical bounds of electrodynamics in appendix \ref{sec:phenomenaED} is supposed to illustrate the associated difficulties.
\end{enumerate}
The previous subsection shows that there are many different aspects
of classical field theories that can be compared. Some approaches
focus on symmetries, orbit spaces and conservation laws, some on the dynamics,
others on structural similarities or on the solution spaces. However,
if the underlying systems of PDEs of the field theories are equivalent,
then all of those aspects are equivalent as well. At the same time,
each single aspect can also be studied at the level of the PDEs.
As a conclusion, a very wholesome approach to the comparison
of the mathematical structures of field theories consists in the formulation
of a framework that compares PDEs.\\
Such a framework then should be able to answer the following questions
in a mathematically precise way.
\begin{enumerate}[label=(Q.\arabic*)]
\itemsep0em
  \item Are two systems of PDEs equivalent?
    \label{req:1}
  \item Do two PDEs share any subsystem?
    \label{req:2}
  \item When are two systems equivalent up to a symmetry?
    \label{req:3}
  \item How to transfer solutions from one system to another?
    \label{req:4}
\end{enumerate}
It is important for the framework to provide an answer
to the last question because it requires a degree of formalization
that exceeds a purely heuristic comparison and because
the determination of the space of common solutions
is arguably one of the best measures for the similarity of two theories.

\subsection{Methods}

To summarise the above, the aim of this article is to compare field theories by comparing
their PDEs, preferably in a well-defined category. In order to do that,
one needs to define what two systems of PDEs have in common
but there is usually no canonical way to define this common part.
However, if one could comprehend a PDE as a geometric object,
then the common part could be naturally identified as the intersection
of those objects in a suitable space.
Fortunately, the language of \ti{jet spaces}, in which PDEs are understood
as submanifolds, allows for such an approach
which is the main reason that the present framework is formulated in this language.
Another important reason is that it also allows for the implementation
of methods from the area of \ti{formal integrability}
that serve to calculate the minimal consistency conditions that arise
when comparing two systems of PDEs.\\
Jet spaces arose with Cartan's concept of a prolongation and were defined by Ehresmann
in 1953. Their theory steadily evolved,
giving rise to the theories of formal integrability (cf. \cite{goldschmidt1967}, \cite{bryantGoldschmidt}),
involution (cf. \cite{seiler2009involution}), differential Galois theory
(cf. \cite{pommaret1994galoisgrouptheory}), to the invention of so-called
diffieties which generalise algebraic varieties (cf. \cite{vinogradov1984diffiety}),
and a whole new calculus called secondary calculus (cf. \cite{vinogradov2001}, \cite{luca2010}).
Furthermore, they were used for the study of
(variational) boundary value problems (cf. \cite{Vinogradov_2007}, \cite{moreno2012geometry}, \cite{Vitagliano_2014}),
control theory (here the algebraic reformulation is particularly useful, cf. \cite{Pommaret1991control}, \cite{sorokina2013poisson}),
the application of (co)homological methods and moving frames
to PDEs (cf. \cite{Krasilshchik1998homMeth}, \cite{Kogan2003}, \cite{Valiquette2011},
\cite{Valiquette2015}),
and especially to investigate local and nonlocal symmetries
(cf. \cite{olver1995}, \cite{Krasilshchik1999}, \cite{kruglikov2007symmetry}),
their invariants and quotients (cf. \cite{kruglikov2015global}, \cite{schneider2020solutions}).
\\
The present framework is restricted to jet spaces in the category of smooth manifolds,
i.e. PDEs are assumed to be smooth submanifolds. However,
this does not imply that their solutions are necessarily smooth
or that the framework can only compare the spaces of smooth solutions of two systems of PDEs.
Instead, this smooth category is a convenient setting to study
certain singular solutions as well, like e.g. shock waves, whose singularity
vanishes on higher order jet spaces (cf. \cite{SeilerKant2011}, \cite{Vitagliano_2014}).
However, distributional solutions are indeed excluded in the present framework.

\subsection{Outline}

As a first step, the necessary mathematical preliminaries about jet spaces are summarized in section \ref{sec:notationPreliminaries}, while a more detailed introduction to the geometric theory of PDEs is given in appendix \ref{sec:introgeomethod}.\\
In section \ref{sec:corrInt}, it is described how one can define a \ti{correspendence} between two given systems of PDEs, each
represented as a submanifold of a jet space,
by another submanifold (subject to some conditions) in the fibered product of those jet spaces. The correspondence connects the two systems and gives rise to a meaningful notion of an \ti{intersection}, that is also defined in the same section.\\
However, this intersection is only meaningful if certain topological consistency conditions are met, which are discussed in section \ref{sec:constistencyConditionsSection} and this intersection only has (formal) solutions
if certain \ti{integrability conditions} are
fulfilled. The latter can in turn be calculated with methods
of formal integrability, a detailed introduction to which is hence provided in section \ref{sec:formalIntegrability}. Moreover, in applying the theory, one often has to compute the rank of somewhat larger tensorial systems. Thus, a program was written that performs this computation, which is provided in appendix \ref{app:formalIntSolver}.\\
In section \ref{sec:sharedStructure}, all previous material is combined by defining \ti{shared structure} of two systems of PDEs as an intersection that satisfies all the consistency and integrability conditions discussed before. 
Solutions of this shared structure
are furthermore shown to be solutions of both intersected theories in subsection \ref{sec:solutionTransfer}, which demonstrates that the so-defined notion of shared structure is a meaningful one. In particular, \ref{req:2} is answered in this way
because the shared structure corresponds to a subsystem with shared solutions. Moreover, since formal integrability serves to calculate the \ti{minimal
integrability conditions}, it is the largest possible subsystem given a chosen correspondence.\\
Building on the theorems of subsection \ref{sec:solutionTransfer}, an especially useful way to \ti{transfer solutions} is elaborated in section \ref{sec:BaecklundCorrespondencesSection}, in which so-called \ti{Bäcklund transformations} are generalized to arbitrary order. A couple of examples demonstrate their power of generating solutions of non-linear PDEs. 
As a consequence, \ref{req:4} can be answered.
Also \ref{req:1} is answered by defining two subsystems to be equivalent if their shared structure possesses all solutions of both theories.\\
To answer \ref{req:3}, section \ref{sec:equivUpToSym} is devoted to the investigation of symmetries, quotient equations and differential syzygies of PDEs and in definition \ref{def:equivalenceUpToSymmetr} a precise notion of \ti{equivalence up to symmetry} of two systems of PDEs is provided. Furthermore, it is shown in the same section how \ti{symmetries of PDEs can be used to find useful Bäcklund correspondences}, which is important because such correspondences are otherwise usually hard to find. In this regard, it is also shown in proposition \ref{prop:quotientCorAndBaeckCor} that Bäcklund correspondences are in fact generalized symmetries in a precise sense, as one might expect.\\
Finally, section \ref{chap:applications} demonstrates various applications of the theory developed in the previously mentioned sections, in particular to various aspects of electrodynamics and the shared structure of magneto-statics and the Navier-Stokes equation. For completeness, appendix \ref{app:axForm} provides an axiomatic derivation of Maxwell's equations and discusses the difficulty of determining the experimental validity bounds of a theory like electrodynamics.\\
The final section \ref{sec:conlusion} summarizes the results and gives an outlook to possible future directions.

\newpage
\section{Notation and preliminaries}
\label{sec:notationPreliminaries}
The aim of this section is to fix the notation and to introduce subsequently necessary notions.
To make this document as self-contained as possible, a more gentle \textbf{introduction to the geometric theory of PDEs is given in appendix \ref{sec:introgeomethod}}.\\
Furthermore, the reader unfamiliar with manifolds is referred
to \cite{tu2010introduction} and more details about fibered manifolds and jet bundles are provided in \cite{saunders1989}. A shorter introduction to jet bundles, besides the one in the appendix, can be found in section 2 of chapter 3 of \cite{Krasilshchik1999}. An advanced
introduction that also includes the preliminaries for
the theory of formal integrability is given in the article \cite{goldschmidt1967}.
\begin{enumerate}
  \item
    $M$ denotes a smooth manifold with dimension $m$.
    A point of $M$ is denoted by $x$.
  \item
    $\pi:E\ra M$ denotes a smooth fibered manifold over $M$ with dimension $d:=m+e$,
    i.e. $e$ is the dimension of the fiber.\footnote{
    A fibered manifold $\pi:E\ra M$ is a differentiable manifold $E$
    together with a differentiable surjective submersion $\pi$ called projection.\\
    A surjective submersion is a differentiable surjective map
    such that its pushforward $\pi_*$ is also surjective at each point.\\
    A fiber bundle is a fibered manifold with a local trivialization.\\
    A vector bundle is a fiber bundle in which the fibers are vector spaces
  and whose transition maps are linear.}
    $p$ denotes a point of $E$. The local coordinates of $E$ may be expressed
    by $C_E=(x^i,u^j)$.
    The convention is used that tuples like $(x^i, u^j)$ stand for tuples like
    $(x^1,\cdots, x^m, u^1,\cdots,u^e)$. Often
    $\xi:F\ra M$ also denotes a fibered manifold with local coordinates $C_F=(x^i,w^h)$ and dimension $m+f$.
  \item
    Let $\alpha = \alpha_1 \cdots \alpha_n$ be a multi-index. It is a tuple of $n\in\mathbb{N}_0$
    numbers $\alpha_i\in \bbr{0,1,\cdots,m=\te{dim}(M)}$
    for which one defines the length $|\alpha|=n$. The tuple is commutative,
    i.e. $\alpha_i\alpha_j=\alpha_j\alpha_i$.
    One can multiply multi-indices as follows:
    \begin{equation}
        \alpha\sigma := \alpha_1 \cdots \alpha_n \sigma_1 \cdots \sigma_l
        \qquad\Ra\qquad |\alpha\sigma|=n+l.
      \label{eq:multiplicationOfMultiIndices}
    \end{equation}
    If $s:U\subset M\ra E$ is a section of our fibered manifold $\pi:E\ra M$,
    and $i\in\bbr{0,\cdots,m}$ an index and $\alpha=\alpha_1 \cdots \alpha_n$ a multi-index, then define
    \begin{equation}
      s^j_i := \frac{\partial s^j}{\partial x^i},\qquad
      s^j_\alpha := \frac{\partial^{n} s^j}{\partial x^{\alpha_1} \cdots \partial x^{\alpha_n}}
      \quad
      \Ra\quad
      s^j_{\alpha i} = \frac{\partial^{n+1} s^j}{\partial x^{\alpha_1} \cdots\partial x^{\alpha_n} \partial x^i}.
      \label{eq:multiIndexDifferentiation}
    \end{equation}
  \item
    $J^k(E)$ denotes the $k$-th order jet bundle of $E$. It can be endowed
    with the structure of a smooth manifold. Locally, $J^k(E)$
    may be described by the coordinates $(x^i,u^j,u^j_\sigma)$ where
    $\sigma$ is a multi-index and $1\le |\sigma| \le k$.
    Please note that in contrast to (\ref{eq:multiIndexDifferentiation}),
    $u^j_\sigma$ is not the derivative of $u^j$.
    Here the multi-index only serves as a label.
    $\pi^n_m:J^n(E)\ra J^m(E)$ denotes the projection for all $0\le m\le n$.
    $J^0(E):=E$ and $\pi^n:J^n(E)\ra M$.
    $J^n(E)_x:=(\pi^n)^{-1}(x)$ denotes the fiber of $J^n(E)$ over $x\in M$.
    Counting local coordinates, one obtains
    \begin{equation}
      \begin{split}
      \te{dim}(J^k(E))&=m+e
      \begin{pmatrix}
        m+k\\
        k
      \end{pmatrix},\qquad\quad
      \te{dim}(J^{k}(E))-\te{dim}(J^{k-1}(E))=e
      \begin{pmatrix}
        m-1+k\\
        k
      \end{pmatrix},
      \end{split}
      \label{eq:dimJets}
    \end{equation}
    where $\te{dim}J^{k}(E)-\te{dim}J^{k-1}(E)$ is the dimension of the fiber of $\pi^k_{k-1}:J^k(E)\ra J^{k-1}(E)$.
  \item
    Let $s_E:M\ra E$ and $s:=s_F:M\ra F$ be sections. $j^l(s)$ denotes the $l$-th prolongation of $s$.
    If $s(x)=(x^i,s^h(x))$ are the local coordinates of the section, then one can use
    the multi-index notation to give an explicit formulation of the prolongation\footnote{
    By Borel's lemma, given any point $\theta\in J^k(E)$, one can always find a section $s_E$
    such that $j^k(s_E)(x)=\theta$. However, given a submanifold $O$
  of $J^k(E)$, it is not always possible
to find a section $s:\pi(O)\ra E$ whose prolongation lies in $O$.}
    \begin{equation}
      j^l(s)(x) = (x^i,s^h(x),s^h_\alpha(x)),~1\le |\alpha|\le l
      \label{eq:prolongationExplicit}
    \end{equation}
  \item
    A \textit{differential equation} $\de{E}$ is defined to be a fibered submanifold of $J^k(E)$. One can show that this is a geometric generalization of the usual notion of a (possibly non-linear) partial differential equation.
  \item
    An essential notion in the algebro-geometric theory of PDEs, that is also heavily used in the present article, is the \textit{differential consequence} or \textit{prolongation} of a differential equation. To prolong a differential equation $\de{E} \subset J^k(E)$ to a submanifold in $J^{k+l}(E)$, one needs
    the concept of repeated Jets: Since $\de{E}$ is a fibered submanifold of $J^k(E)$, one can consider the fibered manifold $\pi^k|_{\de{E}}:\de{E}\to M$ and one can consider the space of jets of sections of $\pi^k|_{\de{E}}$, called  $J^l(\de{E})$. Since $\de{E}$ is a submanifold of $J^k(E)$, $J^l(\de{E})$ is naturally a submanifold of the jet bundle $J^l(J^k(E))$.\\  
    If $J^k(E)$ is locally described by the coordinates $(x^i,u^j_\sigma)$,
    then the coordinates
    of $J^l(J^k(E))$ are $(x^i, (u^j_\sigma)_\alpha)$ where $|\sigma|\le k$ and $|\alpha|\le l$.\footnote{
    Note that this is not the same as $u^j_{\sigma\alpha}$ because one ``double-counts''
    those coordinates that arise from jets of sections whose derivatives would usually commute.}
    The subset of \textit{repeated jets} in $J^l(J^k(E))$ consists of the image of the embedding
    \begin{equation}
      i_{k,l}:J^{k+l}(E)\ra J^l(J^k(E)),~j^{k+l}(s)(x)\mapsto j^l(j^k(s))(x)
      \label{eq:repeatedJet}
    \end{equation}
    In local coordinates, this embedding reads
    $(x^i,u^j_{\sigma\alpha}=s^j_{\sigma\alpha}(x))\mapsto (x^i,(u^j_\sigma)_\alpha=s^j_{\sigma\alpha}(x))$.
    One can show that it is well defined (see \cite{saunders1989}).\\
    Now one can prolong a fibered submanifold $\de{E} \subset J^k(E)$ to a submanifold in
    $J^{k+l}(E)$ as follows.
    First take the intersection $J^l(\de{E}) \cap i_{k,l}(J^{k+l}(E))$ within $J^l(J^k(E))$. In this intersection
    are only points of the form $j^l(j^k(s))(x)$ and therefore the projection
    $p:J^l(\de{E}) \cap i_{k,l}(J^{k+l}(E))\ra J^{k+l}(E),~j^l(j^k(s))(x) \mapsto
    j^{k+l}(s)(x)$ is well-defined.
    Thus, define the $l$-th \ti{prolongation of a PDE $\de{E}\subset J^k(E)$ (into $J^{k+l}(E)$)} by
    \begin{equation}
      P^l(\de{E}) := p(J^l(\de{E}) \cap i_{k,l}(J^{k+l}(E)))
      \label{eq:prolSubmfdJetSpace}
    \end{equation}
    An intersection must not necessarily be a smooth manifold and therefore, a prolongation
    does not always exist in the category of smooth manifolds. In particular, the intersection might be empty.
  \item
    Define the total differential operators $D_i^k,~i\in\bbr{1,\cdots,m=\te{dim}(M)}$
    as vector fields on $J^k(E)$ locally by
    \begin{equation}
      D_i^k := \frac{\partial}{\partial x^i} +
      \sum_{j=1}^e \sum_{|\sigma| < k} u^j_{\sigma i} \frac{\partial}{\partial u^j_\sigma},\qquad
      D_i := D_i^\infty.
      \label{eq:totalDiffOperator}
    \end{equation}
    If $\alpha=\alpha_1 \cdots \alpha_n$ is a multi-index, define
    $D_\alpha := D_{\alpha_1} \circ \cdots \circ D_{\alpha_n}$.
  \item
    If $\pi:E\ra M$ and $\pi':E'\ra M'$ are fibered smooth manifolds,
    then a smooth map $\Phi:E\ra E'$ is called a \textit{morphism of fibered (smooth) manifolds}
    if there exists a map $\phi:M\ra M'$ such that $\pi'\circ\Phi=\phi\circ\pi$.
    A special case is $M=M'$, $\phi=\te{id}$.
    Then the map $\Phi$ is a morphism of fibered manifolds if $\pi'\circ\Phi=\pi$.
    In the following, a morphism of fibered manifolds shall always refer to
    this special case if nothing else is mentioned.
  \item
    A differential operator $\varphi:J\subset J^k(E)\ra F$
    is defined as a morphism of fibered manifolds. Its $l$-th prolongation is defined by
    \begin{equation}
      p^l(\varphi):P^l(J) \ra J^l(F),\qquad
      j^{k+l}(x)\mapsto j^l(\varphi(j^k(s)(x)))
      \label{eq:prolDiffOp}
    \end{equation}
    In local coordinates, it is given by
    \begin{equation}
      p^l(\varphi)(x^i,u^j_{\sigma \alpha}) = (x^i,D_\alpha \varphi^h(x^i,u^j_{\sigma})),
      ~
      0 \le |\sigma| \le k,~
      0 \le |\alpha| \le l.
      \label{eq:prolongvarphiExplicit}
    \end{equation}
    (Most often, one considers $J=J^k(E)$ and then $P^l(J)=J^{k+l}(E)$.)
  \item 
    Let $s:M\to F$ be a section. Define the kernel of a differential operator by
    \begin{equation}
      \ker_s(\varphi):=\bbr{\theta\in J~|~\varphi(\theta)=s(\pi^k(\theta)) }.
      \label{eq:defKer}
    \end{equation}
  \item
    Proposition 2.1 of \cite{goldschmidt1967} includes the statement that
    for any morphism $\varphi: A \to B$ of fibered manifolds (over the same base space)
    and any section $s$ of $B$,
    $\ker_s(\varphi)$ is a fibered submanifold of $A$ if
    \begin{equation}
      s(M)\subset \varphi(J) \te{ and } \te{rank}(\varphi) \te{ is locally constant.}
      \label{eq:PDEcriteria}
    \end{equation}
    This holds in particular for a differential operator $\varphi:J\subset J^k(E)\to F$
    (which, by definition, is a morphism of fibered manifolds)
    and therefore $\ker_s(\varphi)$ is a fibered submanifold
    of $J\subset J^k(E)$ and hence
    a differential equation whenever (\ref{eq:PDEcriteria}) holds for any differential operator $\varphi$.
  \item
    If $\de{E}=\ker_s(\varphi)$ is a differential equation, then
    the following equality holds,
    \begin{equation}
      \begin{split}
        P^l(\de{E}) &= \ker_{j^l(s)}(p^l(\varphi))= \bbr{ \theta \in P^l(J)~\big|~D_\alpha\varphi^h(\theta)=D_\alpha s^h(\pi^{k+l}(\theta)),
        |\alpha|\le l}
      \end{split}
      \label{eq:prolongRk}
    \end{equation}
  \item
    For a section $s:U\subset M\to E$, denote by $\Gamma_s^k$ the image of $j^k(s):U\subset
    M\to J^k(E)$
    and, for any section $s$ and for any point $\theta\in \Gamma_s^k$,
    call $T_\theta\Gamma_s^k\subset T_\theta J^k(E)$ an $R$-plane.
    The span of all $R$-planes at a point $\theta \in J^k(E)$ is denoted by $\de{C}_\theta$
    and is called Cartan-plane. The map
      $\de{C}: J^k(E)\to TJ^k(E),~\theta\mapsto \de{C}_\theta$
      is called Cartan distribution (sometimes also Vessiot distribution).
  \item An integral submanifold of the Cartan distribution
    is defined to be a submanifold $W\subset J^k(E)$ such that $T_\theta W\subset \de{C}_\theta$
    for all $\theta \in W$. An integral submanifold $W$ is called locally maximal
    if no open subset of $W$ can be embedded into an integral submanifold of greater dimension.
  \item\label{solution}
    A solution $S$ of a differential equation $\de{E}$ is a
    locally maximal, $\text{dim}(M)$-dimensional integral submanifold
      of $\de{C}$ with $S\subset \de{E}$. As emphasized before,
      this definition includes certain singular solutions
      (cf.
      \cite{Vitagliano_2014}).
  \end{enumerate}

\newpage
\section{Correspondence and intersection}
\label{sec:corrInt}
This section develops the framework for the comparison of systems of differential equations.
To this end, the most important concepts are those of a correspondence and an intersection which are described below.
\subsection{Motivating example}
\label{sec:correspondenceMotivatingExample}
Consider the equations of magneto-statics and of the viscous Navier-Stokes equation
(in a dimensionless form):
\begin{enumerate}
  \item Magneto-statics:
    \begin{equation}
      \begin{split}
        \nabla\times\bo{B}&=\bo{j},\qquad
        \nabla\cdot\bo{B}= 0.
      \end{split}
      \label{eq:m-static}
    \end{equation}
    Here $\bo{B}=(B^1,B^2,B^3)^T$ denotes the magnetic field vector and
    $\bo{j}=(j^1,j^2,j^3)^T$ the
    charge current density.
  \item Viscous, incompressible Navier-Stokes equations (without external forcing):
    \begin{equation}
      \begin{split}
        \br{\frac{\partial }{\partial t} + \bo{u}\cdot\nabla}\bo{u}&= -\nabla \br{\frac{p}{\rho}}+\nu\Delta\bo{u},
        \qquad
        \nabla\cdot \bo{u}=0.
      \end{split}
      \label{eq:navier-stokes}
    \end{equation}
    Here $\bo{u}$ is the velocity vector, $p$ is the pressure, $\rho$ is the density and $\nu$
    is the viscosity coefficient.
\end{enumerate}
Now let us make the following additional assumptions that might occur in some physical settings:
\begin{equation}
  \begin{split}
    &\text{1) The current density $\bo{j}$ is the gradient of a function $\psi$,
  i.e. $\bo{j}=-\nabla\psi$},\\
  &\text{2) The velocity flow is static, i.e. $0=d\bo{u}/dt=\partial u/\partial t + (\bo{u}\cdot\nabla)\bo{u}$.}
  \end{split}
  \label{eq:assumptions}
\end{equation}
If we apply those assumptions to the equations above and use the vector identity
$\Delta\bo{u}=\nabla(\nabla\cdot\bo{u})-\nabla\times(\nabla\times\bo{u})$ as well as $\nabla\cdot\bo{u}=0$
and $\nabla\cdot(\nabla\times\bo{u})=0$ (because of grad $\circ$ rot $=0$),
the systems of equations
above become:
\begin{equation}
  \begin{split}
    \nabla\times\bo{B}&=-\nabla\psi,\qquad\nabla\cdot\bo{B}=0\\
    \te{and }\qquad \nabla\times(\nabla\times\bo{u})&=-\nabla\phi,\qquad\nabla\cdot(\nabla\times\bo{u})=0,
    \qquad \nabla\cdot\bo{u}=0.
  \end{split}
  \label{eq:comparison}
\end{equation}
where $\phi:=p/(\rho\nu)$.
It is apparent that those equations aquire a similar form under the ``correspondence''
\begin{equation}
  \bo{B}=\nabla\times\bo{u}.
  \label{eq:correspondence}
\end{equation}
Or, put differently, if one replaced $\bo{B}$ by $\nabla\times\bo{u}$, then the
system of all equations together would be consistent.\\
And in fact, because $\nabla\cdot\bo{B}=0$, we can use the Poincaré Lemma (in any star-shaped region) to
conclude that there exists a vector potential $\bo{A}$ such that $\nabla\times\bo{A}=\bo{B}$
and because gauge transformations do not change the physics of classical electro-dynamics
(and in particular of magneto-statics),
we can use them to gauge $\bo{A}$ in such a way that $\nabla\cdot\bo{A}=0$. Therefore, under
the above assumptions, there is
a direct correspondence between $\bo{A}$ (in some gauge) and $\bo{u}$.
The physical interpretation is that a static fluid velocity field behaves like the vector potential
of magneto-statics with certain charge current densities.\footnote{
  Of course the initial and boundary conditions additionally influence the solutions.
}
This can give a new intuition about
the corresponding physical phenomena.\\
As this example illustrates in an intuitive way,
(\ref{eq:comparison}) and
(\ref{eq:correspondence}) describe ``shared structure'' of
the equations (\ref{eq:m-static}) and (\ref{eq:navier-stokes})
under the conditions
(\ref{eq:assumptions}).
But what is the appropriate space in which the correspondence (\ref{eq:correspondence}) holds
and in which the shared structure can be obtained? Is there a way to compute the
assumptions (\ref{eq:assumptions}) instead of guessing them,
given the correspondence (\ref{eq:correspondence})?
And how to generalize the procedure?
To answer those and other questions already motivated in the introduction,
a general framework is constructed in the next subsections.

\subsection{Formal definitions}

Suppose that $\pi:E\to M$ and $\xi:F\to M$ are fibered manifolds with the same base space
(a generalization to different base spaces is work in progress).
Suppose further that we are given two PDEs $\mathcal{E}\subset J^k(E)$ and
$\mathcal{F}\subset J^l(F)$.
We want to relate the PDEs in a space in which we can embed both of them.
A natural choice is their pullback in the category of smooth manifolds,
i.e. their so-called fibered product
\begin{equation}
  J := J^k(E)\times_M J^l(F):=\bbr{J^k(E)_x \times J^l(F)_x~|~x\in M}.
\end{equation}
Now the canonical projections $\pi_E:J\to J^k(E)$ and $\pi_F:J\to J^l(F)$ allow to pull
$\mathcal{E}$ and $\mathcal{F}$ back to $J$:
\begin{equation}
  \mathcal{E}_J := \pi_E^{-1}(\mathcal{E})\qquad\text{ and }\qquad\mathcal{F}_J:=\pi_F^{-1}(\mathcal{F}).
\end{equation}
$\pi_E$ and $\pi_F$ can also be used to pull back the Cartan distribution defined on
$J^k(E)$ and $J^l(F)$: If $\Sigma_E$ is the module of Cartan forms
(the differential forms that annihilate the Cartan distribution)
on $J^k(E)$ and $\Sigma_F$ is the module of Cartan forms on $J^l(F)$,
then the module $\Sigma$ on $J^k(E)\times_M J^l(F)$ is generated by
$\pi_E^{*}\Sigma_E$ and $\pi_F^*\Sigma_F$.\\
Though the two equations are now pulled back into a natural common space,
they are not yet related. Directly intersecting $\de{E}_J$ and $\de{F}_J$
would result in a space
\begin{equation}
  \de{EF}:=\de{E}_J\cap \de{F}_J
  \label{eq:EF}
\end{equation}
that is big enough to
accomodate all solutions of both $\de{E}$ and $\de{F}$, even if
$\de{E}$ and $\de{F}$ are completely unrelated.
Therefore, one additionally
needs to intersect $\de{E}_J$ and $\de{F}_J$
with a third submanifold $\Phi\subset J$ in order to relate them.\\
But what kind of submanifold is $\Phi$ supposed to be?
One would not like $\Phi$ to be of the form $\pi_E^{-1}(\phi)$ or $\pi_F^{-1}(\psi)$
for $\phi\subset J^k(E)$ and $\psi\subset J^l(F)$
because this would only impose additional relations on one
of the pulled back equations. Instead $\Phi$ is supposed
to relate the fibers of $J^k(E)$ with those of $J^l(F)$
without imposing such additional conditions.
To ensure that, one might require that $\Phi$ is large enough
to fulfill $\de{E}\subset\pi_E(\Phi)$ and $\de{F}\subset\pi_F(\Phi)$.
This would in particular imply that $M\subset\Phi$, i.e. $\Phi$
would not impose any relations on $M$.
However, the condition $\de{E}\subset \pi_E(\Phi)$ might be considered
too weak because it does not necessarily
ensure that $\Phi$ does not impose any relations on $\de{E}_J$ locally.
At the same time, the condition
$\de{E}\subset \pi_E(\Phi)$ in a different sense might also be considered
too strong because it does not allow to restrict the comparison of
the PDEs to a particular open neighbourhood (for example, by
  adding some inequalities to the local definition of
$\Phi$).
Both issues can be resolved, however, by requiring instead
that for all open $W\subset \de{E}_J\cap \Phi$, one has $\pi_E(W)$ open in $\de{E}$.
\\
The previous condition ensures that the dimension of $\Phi$ is locally sufficiently large.
At the same time, it should not be arbitrarily large because this
would again not impose any relations and thus render the intersection meaningless.
Since every submanifold can locally be described by a set of equations
where the number of independent equations is equal to the codimension of the submanifold
(see also subsection \ref{sec:existenceSmoothInter}),
the codimension of $\Phi$ quantifies the number of (global) relations it imposes.
To ensure that the dependent variables of at least either $\de{E}_J$ or $\de{F}_J$ are
determined in terms of the other, this codimension should at least
equal $n(x):=\text{min}(e(x),f(x))$ where $e(x):=\text{dim}(E_x)$ and $f(x):=\text{dim}(F_x)$ are
the dimensions of the fibers of $E$ and $F$ over $x$.
(Often they are constant and do not depend on $x$. They are always locally
constant because we work in the category of smooth manifolds.)
The above thoughts can be summarized in the following definitions.
\gd{diagonality}{
  Let $p:Y\to X$ and $q:Z\to X$ be fibered manifolds,
  let $S_Y$ be a submanifold of $Y$ and $S_Z$ be a submanifold of $Z$,
  let $Y\times_X Z$ be the fibered product of $Y$ and $Z$ over $X$
  and let $\pi_Y:Y\times_X Z\to Y$ and $\pi_Z:Y\times_X Z\to Z$
  be the canonical projections.\\
  A submanifold $S\subset Y\times_X Z$ is called \textit{almost diagonal}
  iff for all open subsets $U\subset S$, the set
  $\pi_Y(U)$ is an open subset of $Y$ and the set $\pi_Z(U)$
  is an open subset of $Z$.\\
  A submanifold $S\subset Y\times_X Z$ is called \textit{almost diagonal to $S_Y$ and $S_Z$}
  iff for all open subsets $U\subset S$,
  the set $\pi_Y(U)\cap S_Y$ is an open subset of $S_Y$
  and the set $\pi_Z(U)\cap S_Z$
  is an open subset of $S_Z$.
}
As said above, intuitively, the definition is supposed to ensure that the submanifold $S$ is defined by equations,
that either only relate fiber coordinates of $Y$ with fiber coordinates of $Z$
within the fibered product $Y\times_XZ$, or,
if it imposes additional relations on the coordinates of $Y$ or $Z$ alone within the
fibered product,
then those relations must already be imposed by $\pi_Y^{-1}(S_Y)$ and $\pi_Z^{-1}(S_Z)$.
The previous definition now allows to define a correspondence.
\gd{implicitCorrespondence}{
  A \textit{correspondence} between
  $\de{E}\subset J^k(E)$ and $\de{F}\subset J^l(F)$,
  is a fibered submanifold $\Phi$ of $J^k(E)\times_M J^l(F)$
  with $\text{cod}(\Phi)(x)\ge \text{min}(\text{dim}(E_x),\text{dim}(F_x))$
  which is almost diagonal to $\de{E}$ and $\de{F}$.
}
Given a natural space to relate two PDEs,
one can define their common part as a set-theoretic intersection.
\gd{implicitIntersection}{
  Given a correspondence $\Phi$ between $\de{E}$ and $\de{F}$, their
  \textit{intersection} $\de{I}$ is defined by
  \begin{equation}
    \mathcal{I}:=\mathcal{E}_J\cap\mathcal{F}_J\cap \Phi.
  \end{equation}
}
Those definitions allow to define shared structure in section \ref{sec:sharedStructure}.

\subsection{Local description}
\label{sec:localDescription}

Given a smooth manifold $X$, we denote its local coordinates (in some suitably adapted chart)
by $C_X$. If $m$
denotes the dimensions of $M$
and if $e$ and $f$ denote the dimensions of the fibers of $E$ and $F$, then the
local coordinates of the manifolds described above are given by
\begin{equation}
  \begin{split}
    &C_M = (x^i),\quad C_E=(x^i,u^j),\quad C_{J^k(E)}=(x^i,u^j_\alpha),
    \quad i\le m,~j\le e,~|\alpha|\le k\\
    &C_{F} = (x^i,v^g),\quad C_{J^l(F)}=(x^i,v^g_\beta),
    \quad i\le m,~g\le f,~|\beta|\le l\\
    &C_{J^k(E)\times_M J^l(F)} = (x^i,u^j_\alpha,v^g_\beta),\quad
    i\le m,~j\le e,~g\le f,~|\alpha|\le k,~|\beta|\le l
  \end{split}
\end{equation}
If $\de{E}\subset J^k(E)$ is a submanifold, then (by proposition \ref{prop:partialConversePreimage}) it can locally always be
described as the kernel of independent functions, i.e.
by equations $F_a(x^i,u^j_\alpha)=0$, where $1\le a\le r$.
In other words, the submanifold $\de{E}$
is locally, in some neighbourhood $U\subset J^k(E)$ defined by those
points $(x^i,u^j_\alpha)$ contained in $U$
that are subject to the conditions $F_a(x^i,u^j_\alpha)=0$.
Instead of $\de{E}=\{~(x^i,u^j_\alpha) \in U\subset J^k(E)~|~F_a(x^i,u^j_\alpha)=0~\}$,
the following shorthand notation is used.
\begin{equation}
  \begin{split}
  \de{E}:\eq{ F_a(x^i,u^j_\alpha)=0 }
  \end{split}
\end{equation}
If we now pull back $\de{E}$ to $\de{E}_J=\pi_E^{-1}(\de{E})\subset J^k(E)\times_M J^l(F)$,
then $\de{E}_J$ is locally described by those equations that define the points in the
inverse image $\pi^{-1}(U\cap \de{E})$.
This inverse image consists of all points $(x^i,u^j_\alpha,v^g_\beta)$
such that $\pi(x^i,u^j_\alpha,v^g_\beta)=(x^i,u^j_\alpha)\in U\cap \de{E}$. But
$(x^i,u^j_\alpha) \in U\cap \de{E}$, precisely iff $F_a(x^i,u^j_\alpha)=0$. Thus, the points
in $\pi^{-1}(U\cap\de{E})$ are described by the same equations as the points
in $U\cap \de{E}$. As a consequence, $\de{E}_J$ is locally described by the conditions
$F_a(x^i,u^j_\alpha)=0$ but now imposed on an open neighbourhood of $J^k(E)\times_M J^l(F)$.\\
Furthermore, if $\de{E}_J$ and $\de{F}_J$ are locally defined by points
fulfilling the equations $F_a^{\de{E}}(x^i,u^j_\alpha)=0$ and $F_b^{\de{F}}(x^i,v^g_\beta)=0$,
then their intersection is necessarily locally defined by those points
that simultaneously
fulfill both equations. In other words, the \ti{intersection}
of $\de{E}_J$ and $\de{F}_J$
is locally described by the \ti{union} of their equations.\footnote{The intuitive reason is that each equation represents a constraint on the space of solutions
  and therefore the intersection, which is smaller than both original
  solution spaces,
must be described by the union of those constraints.}
As a consequence, all local descriptions can be summarized as follows.
\begin{equation}
  \begin{split}
    J^k(E)\supset
    \de{E}:&~\{~ F^{\de{E}}_a(x^i,u^j_\alpha)=0,\quad\qquad 0\le a\le r_{\de{E}}~\},\\
      J^l(F)\supset
      \de{F}
      :&~\{~ F^{\de{F}}_b(x^i,v^g_\beta)=0,\quad\qquad 0\le b\le r_{\de{F}}~\},\\
        J^k(E)\times_M J^l(F)\supset
        \Phi:&~\{~ \phi_c(x^i,u^j_\alpha,v^g_\beta)=0,\qquad 0\le c \le r_\Phi~\},\\
        J^k(E)\times_M J^l(F)\supset
        \de{I}:&~\{~ F^{\de{E}}_a(x^i,u^j_\alpha)=
        F^{\de{F}}_b(x^i,v^g_\beta)=\phi_c(x^i,u^j_\alpha,v^g_{\beta})=0\}.
    \label{eq:localEquations}
  \end{split}
\end{equation}

\ye{ex:intersection}{%
Here, a simple version of the motivating example of the previous subsection
  \ref{sec:correspondenceMotivatingExample} is rephrased in the present terminology.
  The equations are modeled on a flat, Euclidean spacetime $\mathbb{R}^3\times\mathbb{R}$ with
  local coordinates $(x^i,t),~i\in\bbr{1,2,3}$.\\
  For Hydrodynamics, let the
  fibered manifold (which is now a trivial vector bundle)
  be $\pi:E:=M\times \mathbb{R}^3\ra M$ with local coordinates
  $(x^i,t,u^i)$ and dimension $\te{dim}(E)=m+e=4+3$.
  Let $u^{i,j}$ denote the coordinates corresponding to
  $\partial u^i/\partial x^j$ and recall that the sum convention is used.
  Let $p:M\ra\mathbb{R}$ be a given function called pressure and $\rho,\nu\in\mathbb{R}$
  be the constant density and viscocity.
  Denote by $p^{~,i}$ the components of the gradient of $p$.
  Describe $J^2(E)$ with local coordinates
  $(x^i,t,u^i,u^{i,j},u^i_t,u^{i,jk},u^{i,j}_t)$ where $u^{i,jk}=u^{i,kj}$.
  The Navier-Stokes equations described in (\ref{eq:navier-stokes})
  in this setting are then given by
  \begin{equation}
    \de{E}:\eq{~u_t^i + u^ju^{i,j} =-  \frac{1}{\rho} p^{~,i} + \nu u^{i,jj},\qquad u^{i,i}=0}
    \label{eq:example4NavierStokesEq}
  \end{equation}
  Magneto-statics is also modeled on $M$, even though the equations do not
  involve any time-component. (Since a realistic experiment always
  takes place in space and time, even though the fields might not change over time,
  this is not a bad assumption.) Thus, for magneto-statics,
  the vector bundle $\xi:F:=M\times \mathbb{R}^3\to M$ with local coordinates
  $(x^i,t,B^i)$ is defined and the magneto-static equations
  (corresponding to (\ref{eq:m-static}))
  are described
  on $J^1(F)$ with local coordinates $(x^i,t,B^i,B^{i,j})$ via
  \begin{equation}
    \begin{split}
      \de{F}:\eq{~\varepsilon_{ijk}B^{k,j}=I^i,\qquad B^{i,i} = 0}
    \end{split}
    \label{eq:magnetostaticsjetterminology}
  \end{equation}
  (the letter $I$ is used for the current density
  instead of $j$ to avoid confusion with other $j$'s).
  The natural product bundle $J:= J^2(E)\times_MJ^1(F)$
  has local coordinates $(x^i,t,u^{i},B^i,u^{i,j},u^i_t,u^{i,jk},u^{i,j}_t,B^{i,j},B^i_t)$
  one can define $\Phi\subset J$ as the submanifold
  locally given by
  \begin{equation}
    \Phi:\eq{~B^i = \varepsilon_{ijk}u^{k,j}}
    \label{eq:example4correspondenceDef}
  \end{equation}
  Since $\Phi$ does not contain any equations that relate the coordinates of
  $J^{2}(E)$ or $J^1(F)$ among themselves, the projection $\pi_E(U)$, of all of its open subsets
  $U\subset \Phi$,
  is open in $J^2(E)$ and $\pi_F(U)$ is open in $J^1(F)$.
  Hence $\Phi$ is almost diagonal.
  As a consequence, $\pi_E(U)\cap \de{E}$ is also always open in $\de{E}$
  and $\pi_F(U)\cap\de{F}$ is always open in $\de{F}$.
  Hence, $\Phi$ is almost diagonal to $\de{E}$ and $\de{F}$
  and is therefore a correspondence in the sense of definition \ref{implicitCorrespondence}.
  We can thus define a valid intersection $\de{I}$
  by the following equations.
  \begin{equation}
    \begin{split}
      \de{I}=\pi_E^{-1}(\de{E})\cap\pi_F^{-1}(\de{F})\cap\Phi:~
        \eq{
          u_t^i + u^ju^{i,j} =-  \frac{1}{\rho} p^{~,i} + \nu u^{i,jj},~ u^{i,i}=0\\
          \varepsilon_{ijk}B^{k,j}=I^i,\quad B^{i,i} = 0\\
          B^i = \varepsilon_{ijk}u^{k,j} &
        }
    \end{split}
    \label{eq:intersectionMagnetoHydro}
  \end{equation}
  The shared structure contained in this intersection
  can be computed with the methods that are going to be introduced
  in the following sections.
  In subsection \ref{sec:sharedStructureMSHD}, it is shown that this shared structure
  indeed corresponds to the one described in (\ref{eq:comparison}),
  and the assumption of a static fluid flow, guessed in (\ref{eq:assumptions}),
  is the result of the computation
  of the minimal consistency conditions for shared structure to arise.
}

\newpage
\section{Consistency conditions}
\label{sec:constistencyConditionsSection}

The conditions that need to be satisfied in order to be able to speak of a meaningful
intersection of two PDEs are related to at least two areas, namely
transversality theory in differential topology and the theory of formal integrability.
The next subsections, as well as section \ref{sec:formalIntegrability}, provide all corresponding background information in those areas
that are needed to understand the rest of the article.

\subsection{Smoothness conditions}
\label{sec:existenceSmoothInter}
In this subsection is investigated under which circumstances the intersection $\de{I}$ of two
differential equations is actually again a differential equation, that means
a smooth submanifold of a jet space.\\
The intersection theory of differential topology
can answer this question. The remainder of this subsection largely follows
\cite{guillemin2010differential} and those theorems that are needed in the present context are cited.
The starting point is the preimage theorem which is a quite straightforward
consequence of the inverse function theorem and the local submersion theorem.
\gd{def:regularValue}{
  For a smooth map $f:X\ra Y$, a point $y\in Y$ is called a \ti{regular value}
  if the pushforward (or differential) $df_x:T_{x}X\ra T_{f(x)}Y$ is surjective for all $x\in f^{-1}(y)$.
}
\bp{prop:preimage}{%
\textbf{(Preimage theorem)}$~$
  If $y$ is a regular value of $f:X\ra Y$, then $f^{-1}(y)$ is a smooth submanifold
  with dimension $\te{dim}(X)-\te{dim}(Y)$.
}
Note that it is often not hard to check if the pushforward of a smooth map is surjective.
It amounts to checking the rank of the Jacobian matrix.

There is also a partial converse to the theorem, namely
\bp{prop:partialConversePreimage}{%
  If $Z\subset X$ is a smooth submanifold, then it can locally be defined as the kernel
  of independent smooth functions.
}
The following proposition is also useful.
\bp{prop:kerdfIsTangentSpace}{%
  Let $f:X\ra Y$ be smooth and $y$ a regular value of $f$. The tangent space of $Z:=f^{-1}(y)$
  is given by $T_xZ=\ker(df_x)$ for any $x\in Z$.
}
The next step is to consider what happens if one does not only look at the preimage of a single
regular value but at the preimage of a submanifold. Then one can use the definition of
transversality to prove the following theorem.
\gd{def:transversality}{%
  The map $f:X\ra Y$ is said to be transversal to the submanifold $Z\subset Y$,
  abbreviated $f\tv Z$, if the equation
  \begin{equation}
    \te{im}(df_x) + T_{f(x)}Z=T_{f(x)}Y
    \label{eq:transversalityCondition}
  \end{equation}
  holds true at each point $x$ in $f^{-1}(Z)$.
}
\bp{prop:transversalitySubmfd}{%
  If the smooth map $f: X \ra Y$ is transversal to a submanifold
  $Z \subset Y$, then the preimage $f^{-1}(Z)$ is a submanifold of $X$.
  Moreover,
  \begin{equation}
    \te{cod}(f^{-1}(Z)\subset X)=\te{cod}(Z\subset Y)
    \label{eq:codRelation}
  \end{equation}
}
Given the manifold $Y$ and two submanifolds $X\subset Y$ and $Z\subset Y$,
one can apply the above theorem to their intersection $X\cap Z$ as follows:
If $i:X\ra Y$ is the canonical inclusion that embeds $X$ into $Y$
then $X\cap Z = i^{-1}(Z)$. Since $\text{im}(di_x)=T_xX$, and
$T_{i(x)}Z=T_xZ$, one obtains
\bp{prop:transversalIntersect}{%
  If
  $X$ and $Z$ are smooth submanifolds of $Y$, then $X\cap Z$ is a smooth submanifold of
  $Y$ iff $X\tv Z$, that means
  \begin{equation}
    T_xX+T_xZ=T_xY
    \label{eq:intersectCondition}
  \end{equation}
  for all $x\in X\cap Z$. In this case
  $\text{cod}(X\cap Z)=\text{cod}(X) + \text{cod}(Z)$.
}
Condition (\ref{eq:intersectCondition}) can be checked locally.
Indeed, we obtain the following proposition as a consequence.
\bp{prop:computationOfTransversality}{%
  If $X$ and $Z$ are submanifolds of $Y$, locally described
  by equations of the form $F^X_a(y^i)=0$ with $1\le a\le r$ and $F^Z_b(y^i)=0$
  with $1\le b\le q$, then
  $X \tv Z$ iff $dF$ as defined in eq. (\ref{eq:jointsystem}) has full rank.
}
\begin{proof}
If $Y$ has local coordinates $(y^i)$,
then, since $X$ is a smooth submanifold of $Y$,
by proposition \ref{prop:partialConversePreimage},
every local chart $U\subset X$ is described as the kernel of
independent functions $F^X:Y\to \mathbb{R}^r$, i.e. $U=(F^X)^{-1}(0)$, or,
equivalently, we write as before $X:\{F^X_a(y^i)=0\}$ with $1\le a\le r$.
Using proposition \ref{prop:kerdfIsTangentSpace},
we can then compute $T_xX$ as the kernel of $dF^X$.
Similarly, if $Z$ is locally described
by $F^Z_b(y^i)=0$ with $1\le b\le q$, then $T_xZ=\ker(dF^Z)$ and $X\cap Z$ is locally described by the joint system of those equations, i.e. by
\begin{equation}
  0 = F_c(y^i)=
  \begin{cases}
    F_c^X(y^i),& 1\le c\le r\\
    F_{c-r}^Z(y^i),& r+1\le c\le r+q
  \end{cases}
  \label{eq:jointsystem}
\end{equation}
By proposition \ref{prop:transversalIntersect}, $X\tv Z$
iff $X\cap Z=F^{-1}(0)$ is a smooth submanifold,
which, by the preimage theorem, \ref{prop:preimage},
is true if $dF$ is surjective, i.e. has full (row) rank.
\end{proof}
Furthermore, using Sard's theorem, one can prove the transversality theorem
which guarantees that almost all maps of a family of smooth maps are transversal to some submanifold in the codomain.
\bp{prop:Sard}{(\tf{Sard})
  The set of values of a smooth map $f:X\ra Y$ which are not regular has Lebesgue measure zero.
}
This means ``almost all'' points of a smooth map are regular. However,
sets of measure zero can be quite large, for example the subset $\mb{R}^n$
has measure zero in $\mb{R}^{n+1}$.
\bp{prop:transversalityTheorem}{(\tf{Transversality Theorem})
  Suppose that $F : X \times S \ra Y$ is a smooth map
  between smooth manifolds, where only $X$ has boundary, and let $Z$ be any boundaryless
  submanifold of $Y$.
  One can use $F$ to define a smooth family of homotopic maps by $f_s(x):=F(x,s)$.
  If both $F$ and $\partial F$ are transversal to $Z$, then for almost every
  $s \in S$, both $f_s$ and $\partial f_s$, are transversal to $Z$.
}
For a map $f:X\ra \mb{R}^m$ this immediately implies that transversality is a generic feature
because one can simply define $S$ as an open subset of $\mb{R}^m$ and define
$F(x,s):=f(x)+s$. As $S$ is open in $\mb{R}^m$, this means that $F$ is surjective everywhere
and therefore definition \ref{def:transversality} is always fulfilled.
Following this thought further, one can prove the so-called \tf{transversality homotopy theorem}.
\bp{prop:transhomtheorem}{%
  For any smooth map $f: X \ra Y$ and
  any boundaryless submanifold $Z$ of the boundaryless manifold $Y$, there exists
  a smooth map $g : X \ra Y$ homotopic to $f$ such that $g \tv Z$ and $g \tv \partial Z$.
}
Now reconsider two differential equations $\de{E}\subset J^k(E)$ and $\de{F}\subset J^l(F)$.
Using the above, one can show the following.
\bp{prop:EJismanifold}{%
  $\de{EF}=\de{E}_J\cap\de{F}_J$
  as defined in eq. (\ref{eq:EF}) is a PDE, i.e. a smooth submanifold.
}
\begin{proof}
  To check that $\de{EF}$ is a smooth submanifold, it suffices, by
  prop. \ref{prop:transversalIntersect},
  to check that
  $\de{E}_J\tv\de{F}_J$, which in turn, can be checked locally using
  prop. \ref{prop:computationOfTransversality}.
  So if $\de{E}$ and $\de{F}$ are locally described by the independent
  smooth functions
  $F^{\de{E}}_a(x^i,u^j_\alpha)$ and $F^{\de{F}}_b(x^i,v^g_\beta)$
  as in eq. (\ref{eq:localEquations}),
  then we must check if the differential of the joint system of equations $F_c=0$ as defined in
  eq. (\ref{eq:jointsystem}) has full rank. Since
  $\de{E}$ and $\de{F}$ are assumed to be fibered submanifolds of $J^k(E)$ and $J^l(F)$,
  they do not impose any conditions on $M$.
  Furthermore, $F^{\de{E}}$ does not depend on $v^g_\beta$
  and $F^{\de{F}}$ does not depend on $u^j_\alpha$. Hence, $dF^{\de{E}}$ and $dF^{\de{F}}$
  are linearly independent. Since they are both assumed to have full rank
  and they are independent, the joint system $dF$ must also have full rank.
\end{proof}
This theorem implies the following.
\bc{cor:intersectionDiffEq}{
  $\de{I}=\de{EF}\cap\Phi$ is a PDE iff $\de{EF}\tv\Phi$.
}
Whether $\de{EF}\tv\Phi$ or not depends on the definition of $\Phi$
and can not be proven in general.
To check it explicitly in practice for a given $\de{EF}$ and a given $\Phi$,
one can calculate the rank of the joint system as described in prop. \ref{prop:transversalIntersect} and \ref{prop:computationOfTransversality}.
If this rank is locally maximal, then
transversality is guaranteed. If the rank is not locally maximal but locally constant,
then we can restrict the codomain
such that the smooth system becomes locally maximal.
Therefore, the intersection is also a well-defined smooth submanifold at those
points around which the system is locally constant.\\
This is also the reason why the preimage of a differential operator
which has locally constant rank is a smooth submanifold, i.e. a differential equation
(if it is not empty), see
condition (\ref{eq:PDEcriteria}).\\
However, even if $\de{EF}$ is not transversal to $\Phi$,
then the transversality theorem \ref{prop:transhomtheorem} implies that
it suffices to deform $\Phi$ (locally this means to perturb the smooth functions
describing $\Phi$) just ever so slightly in order to obtain an intersection
that is a well-defined object in the category of smooth manifolds.
Furthermore, if some smooth functions $\phi_c$ locally describes our manifold $\Phi$,
one way to make it transversal to $\de{EF}$ is to use $F(x,s):=\phi(x)+s$
for some very small $s$.
Thus, the theorem assures us that taking intersections of $\de{EF}$ and $\Phi$
is not a hopeless endeavor but to the contrary
can always lead to a smooth manifold at least after slight deformations.

\subsection{Differential consistency}
\label{sec:diffConsistency}

In the last subsection was clarified when the intersection of two differential equations
is actually again a differential equation.
As a next step, it is assumed that the intersection \ti{is} a differential equation, i.e. a smooth submanifold,
and it is asked if the PDE has solutions.\\
Ultimately, one is interested in the existence of smooth (or even more general) solutions
but since there is not yet any general theory
that allows to compute whether a solution (in any non-analytic category)
of a PDE exists or not, it is necessary at this point to ask for something weaker.
The next best thing after a general condition that allows
to compute the existence of solutions is
to ask for the existence of so-called formal solutions.
Formal solutions are formal power series that formally solve the PDE
(i.e. the series satisfies all algebraic equations describing the
smooth solution spaces that characterize the PDE and its prolongations)
but is not guaranteed to converge or might converge to something that is not a solution.\\
Formal solutions are tractable because their existence is encoded
in the \ti{differential consequences} of a PDE.

In particular,
if one can prolong an equation infinitely many times in a certain smooth way without obtaining
any contradiction, then
one ``point'' of the infinite prolongation $P^\infty(\de{E})$ can be seen as the sequence
of coefficients for a (not necessarily converging)
taylor expansion that solves the equation locally around the
projection of that point. To understand this better, the reader is encouraged to take a look at example \ref{ex:diffEq}.\\
However, as remarked below eq. (\ref{eq:prolSubmfdJetSpace}),
the prolongations of $\de{E}$
do not necessarily exist.
This means that in order to
check if formal solutions exist,
one needs a general formalism to determine if a PDE is
differentially consistent in the sense that all of its prolongations exist.\\
Furthermore, recall that in the motivating example in subsection \ref{sec:correspondenceMotivatingExample},
we had to make certain physical assumptions (\ref{eq:assumptions}). It would
be beautiful if those assumptions could be obtained in a systematic way.
In general, if one could obtain
the minimal amount of assumptions that must be made to make a system differentially consistent
(if such assumptions exist), then this would be optimal.
Fortunately, one can use the theory of \ti{formal integrability}
for this purpose.
In particular, the ``physical assumptions'' come out of the formalism as ``integrability conditions''
that are needed for consistency.\\
Since the theory is somewhat involved,
the next section provides an introduction to the theory of formal integrability.
In the section after the next, those notions of formal integrability are combined with
the notions of correspondence and intersection defined above
to define what it means for two theories to share structure.

\newpage
\section{Formal Integrability}
\label{sec:formalIntegrability}
Subsequently, the introduction follows
\cite{goldschmidt1967linear},
\cite{goldschmidt1967}
and
\cite{bryantGoldschmidt} (chapter IX)
to introduce the notion of formal integrability.
The first subsection contains the necessary definitions and
the derivation of explicit coordinate expressions which are missing in
Goldschmidt's publications, as well as the derivation of proposition \ref{prop:commutationFormalInt} that can simplify some computations.\\
The second subsection describes the main theorems of the
formal theory. The third subsection discusses integrability conditions
which are especially important for subsequent constructions.
The reader already familiar with formal integrability can
directly proceed with subsection \ref{sec:integrabilityConditions}.
The reader who prefers to learn with examples
is referred to subsection \ref{sec:explicitExampleFormalInt}.

\subsection{Definitions and Preliminaries}
\begin{enumerate}
  \item
    Recall that if $X,Y$ and $N$ are manifolds and $f:Y\ra X$ is a smooth map and $\pi:N\ra X$ is a fiber bundle
    with fibers denoted by $N_x,~x\in X$,
    then $f^*N$ denotes the pullback bundle over $Y$ and it is defined as follows:
    \begin{equation}
      f^*N := \bbr{ N_{f(y)}~|~y\in Y}.
      \label{eq:pullbackBundle}
    \end{equation}
    To each point $y\in Y$, we attach the fiber $N_{f(y)}$ that would usually be attached
    to the point $x=f(y)\in X$.\\
    Suppose we are given the following configuration of smooth maps between smooth manifolds:
    \begin{center}
      \begin{tikzcd}[ampersand replacement = \&, column sep=huge, row sep=tiny]
        \& X_1 \& \ar[']{l}{\pi_1} N^1\\
        Y \ar{ru}{f}\ar[']{rd}{g} \&\&\\
    \& X_2 \& \ar{l}{\pi_2} N^2
    \et
    \end{center}
    where $\pi_i:N^i\ra X_i$ are vector bundles (not just fiber bundles). Then we define
    \begin{equation}
      N^1\otimes_{Y} N^2:= f^*N^1\otimes g^*N^2
      = \bbr{ N^1_{f(y)}\otimes N^2_{g(y)} ~|~y\in Y}.
      \label{eq:pullbackBundleTensor}
    \end{equation}
    which is a vector bundle over $Y$.

  \item
    Now, for any $k\ge 0$, let $V(J^k(E))\ra J^k(E)$ denote the vertical
    subbundle of the tangent bundle $TJ^k(E)$ of $J^k(E)$
    containing those vectors which are tangent to the fibers of $\pi:J^k(E)\ra M$. It is a bundle over $J^k(E)$.
    In a local neighbourhood $U\subset J^k(E)$ with coordinates $(x^i,u^j,u^j_\sigma)$,
    $V(J^k(E))$ is the span of the vector fields
    \begin{equation}
      V(J^k(E)) = \te{span}\br{ \frac{\partial}{\partial u^j},\frac{\partial}{\partial u^j_\sigma}}
    \end{equation}
    and we have
    \begin{equation}
      \pi_{*,\theta}\br{\frac{\partial}{\partial u^j}\bigg|_{\theta}}=0
      =\pi_{*,\theta}\br{\frac{\partial}{\partial u^j_\sigma}\bigg|_{\theta}}
      \in T_{\pi(\theta)}M
        \label{eq:verticalLocalCoord}
    \end{equation}
    at every point $\theta\in U\subset J^k(E)$.
  \item
    If $M$ denotes our base manifold as before, we denote by $T^*$ its tangent bundle,
    by $S^kT^*$ the $k$-th symmetric power of the tangent bundle and by $\Lambda^kT^*$
    the $k$-th anti-symmetric power. \\
    In local coordinates, general elements of those spaces are written
    \begin{equation}
      \begin{split}
        T^* &\ni v= v_i dx^i,~i\in \bbr{1,\cdots,m}\\
        S^kT^* &\ni a = a_{i_1 \ldots i_k} dx^{i_1} \vee \cdots \vee dx^{i_k},~i_j\in\bbr{1,\cdots,m}
        \\
        \Lambda^kT^* &\ni w = w_{i_1 \ldots i_k} dx^{i_1} \wedge \cdots \wedge dx^{i_k},~i_j\in\bbr{1,\cdots,m}
      \end{split}
      \label{eq:localCoordTensorSpaces}
    \end{equation}
    where the sum convention is always used.
    $S^kT^*$ and $\Lambda^kT^*$ are different in that $dx^{i_j}\vee dx^{i_k}=dx^{i_k}\vee dx^{i_j}$
    but $dx^{i_j}\wedge dx^{i_k}=-dx^{i_k}\wedge dx^{i_j}$.
    As a consequence, $\te{dim}\Lambda^kT^*\le \te{dim}(S^kT^*)$. To calculate the dimension,
    note that there are as many symmetric basis elements as there are ways to put $k$ balls
    between $m-1$ sticks. Thus,
    \begin{equation}
      \te{dim}(S^kT^*)=
      \begin{pmatrix}
        m-1+k\\
        k
      \end{pmatrix},\qquad \te{dim}(\Lambda^kT^*) =
      \begin{pmatrix}
        m\\
        k
      \end{pmatrix}
      \label{eq:dimPowerSpaces}
    \end{equation}
    If one has a multi-index $\alpha$ with $|\alpha|=k$, one can define
    $dx^\alpha_\vee := dx^{\alpha_1} \vee \cdots \vee dx^{\alpha_k}$ and
    $dx^\alpha_\wedge := dx^{\alpha_1} \wedge \cdots \wedge dx^{\alpha_k}$ to write more
    concisely
    \begin{equation}
      \begin{split}
        S^kT^* &\ni a = a_\alpha dx^\alpha_\vee,~|\alpha|=k,
        \qquad
        \Lambda^kT^* \ni
        w = w_\alpha dx^\alpha_\wedge,~|\alpha|=k.
      \end{split}
      \label{eq:localCoordTensorSpacesMultiIndexNotation}
    \end{equation}
  \item
    Define the map $\Delta_{l,k}:S^{l+k}T^*\ra S^lT^*\otimes S^kT^*$ as the composition
    \begin{center}
    \bt[huge]
      S^{l+k}T^* \ar{r}{i} \& \otimes^{l+k} T^* \ar{r}{s_{l,k}} \& S^lT^*\otimes S^kT^*
    \et
    \end{center}
    where $i$ is the injection given by
    \begin{equation}
      i(dx^{i_1}\vee \ldots \vee dx^{i_{k+l}}) :=
      \sum_{\sigma \in \mathfrak{S}_{k+l}} dx^{\sigma(i_1)}\otimes \ldots \otimes dx^{\sigma(i_{l+k})}
      \label{eq:injectionOfSk}
    \end{equation}
    where the sum goes over all entries $\sigma$ of the permutation group $\mathfrak{S}$.
    And $s_{l,k}$ is the projection given by
    \begin{equation}
      s_{l,k}(dx^{i_1}\otimes \cdots \otimes dx^{i_{l+k}}):=
      dx^{i_1}\vee \cdots \vee dx^{i_l}\otimes dx^{i_{l+1}} \vee \cdots \vee dx^{i_{l+k}}
      \label{eq:projectionOnSk}
    \end{equation}
    Thus, all in all, we obtain
    \begin{equation}
      \begin{split}
      \Delta_{l,k}(dx^{i_1}\vee \ldots \vee dx^{i_{k+l}}) =
      \sum_{\sigma \in \mathfrak{S}_{k+l}}
      d&x^{\sigma(i_1)}\vee \ldots \vee dx^{\sigma(i_l)}\\
      &\otimes
      dx^{\sigma(i_{l+1})}\vee \cdots \vee dx^{\sigma(i_{l+k})}.
      \end{split}
      \label{eq:coordsDeltaLK}
    \end{equation}
  \item
    Given some smooth manifold $Y$ and maps $\pi:Y\to M$ and $\pi_0:Y\to E$, define
    \begin{equation}
      F^k_Y := S^kT^*\otimes_Y V(E)
      \label{eq:verticalFiber}
    \end{equation}
    Call it $k$-fiber (over $Y$).
    The $k$-fiber is the vector bundle whose
    fibers have as many dimensions (and hence local coordinates)
    as there are local coordinates of order $k$ on $J^k(E)$.
    This can be seen by observing that
    \begin{equation}
      \te{dim}( (F^k_Y)_{p\in Y}) \ov{\ref{eq:dimPowerSpaces}}
      \te{dim}(S^kT^*)\cdot e\ov{\ref{eq:dimJets}}\te{dim}(J^k(E))-\te{dim}(J^{k-1}(E)).
      \label{eq:dim}
    \end{equation}
    In local coordinates, an element $p\in F^k_Y$ can be written $p=(\theta,a)$ where
    $\theta \in Y$ and
    \begin{equation}
        a = a^j_{i_1 \ldots i_k} dx^{i_1} \vee \cdots\vee dx^{i_k}|_{\pi(\theta)}
        \otimes \frac{\partial}{\partial u^j}\bigg|_{\pi_0(\theta)}
        = a_\alpha dx^\alpha_\vee |_{\pi(\theta)}
        \otimes \frac{\partial}{\partial u^j}\bigg|_{\pi_0(\theta)}.
      \label{eq:localCoordKFiber}
    \end{equation}
  \item
      One can show (Proposition 5.1 of \cite{goldschmidt1967}) that for $k \ge 1$, the jet
      bundle $J^k (E)$ is an affine bundle over $J^{k-1} (E)$, modeled on the vector bundle
      $S^kT^* \otimes_{J^{k-1}(E)} V(E)$ over $J^{k-1}(E)$
      ((\ref{eq:dim}) shows that the dimensions match).\\
      As described in chapter IX.§3 of \cite{bryantGoldschmidt},
      if $\theta \in J^{k-1}(E)$,
      the vector space $S^kT^*_{\pi(\theta)} \otimes V_{\pi_0(\theta)} (E)$
      considered as an additive group acts freely
      and transitively on the fiber of $J^k(E)$ over $\theta$.
      As a consequence, for $a \in S^kT^*_{\pi(\theta)} \otimes V_{\pi_0(\theta)}(E)$, we can
      denote by $q+a$ the image of the element $q$
      of the fiber $J^k(E)_\theta$ under the action of $a$.
      If $(x,u^j,u^j_\sigma)$ are the local coordinates of $q$,
      the local coordinates of $q+a$ are $(x,u^j,z^j_\sigma)$ where
      \begin{equation}
        \begin{pmatrix}
          z^j_\sigma = u^j_\sigma,&\te{ if }|\sigma|<k\\
          z^j_\sigma = u^j_\sigma + a^j_{\alpha=\sigma},&\te{ if }|\sigma|=k
        \end{pmatrix}
      \label{eq:epsilon}
    \end{equation}
    (Goldschmidt also provides an intrinsic definition of this map in §5.)
  \item
    The above described action on the fibers of $\pi^k_{k-1}:J^k(E)\to J^{k-1}(E)$ induces a map
    \begin{equation}
      \mu:F^k_{J^k(E)} \ra V(J^k(E)),~ (\theta,a)\mapsto \frac{d}{dt}(\theta+ta)|_{t=0}
      \ov{\ref{eq:epsilon}) and (\ref{eq:verticalLocalCoord}}
      a^j_{\alpha} \frac{\partial}{\partial u^j_\alpha}\bigg|_{\theta}
      \label{eq:mumap}
    \end{equation}
    where $|\alpha|=|i_1 \cdots i_k|=k$.
  \item
    Because of (\ref{eq:verticalLocalCoord}),
    $(\pi^k_{k-1})_{*,\theta}V_\theta(J^k(E))=V_{\pi^k_{k-1}(\theta)}(J^{k-1}(E))$,
    i.e. the pushforward of $\pi^{k}_{k-1}$ restricted to $V_\theta(J^k(E))$ is a
    surjective map whose kernel consists of the vectors tangent to the fibers of $\pi^k_{k-1}:J^k(E)\ra J^{k-1}(E)$.
    Those vectors are precisely those contained in $\mu(F^k_{\theta\in J^k(E)})$.
    Therefore, we have the exact sequence of vector spaces
    \begin{center}
      \bt[normal]
        0\ar{r}\&
        F^k_{\theta\in J^k(E)} \ar{r}{\mu} \&
        V_{\theta}(J^k(E)) \ar{r}{(\pi^k_{k-1})_{*,\theta}} \&
        V_{\pi^k_{k-1}(\theta)}(J^{k-1}(E)) \ar{r} \& 0
      \et
    \end{center}
    which we can pull back to a sequence of vector bundles using
    (\ref{eq:pullbackBundle}) and (\ref{eq:pullbackBundleTensor}):
    \begin{center}
      \bt[normal]
        0\ar{r}\&
        F^k_{J^k(E)} \ar{r}{\mu} \&
        V(J^k(E)) \ar{r}{(\pi^k_{k-1})_{*}} \&
        (\pi^{k}_{k-1})^*V(J^{k-1}(E)) \ar{r} \& 0
      \et
    \end{center}
    This is
    an exact sequence (see also \cite{bryantGoldschmidt} or \cite{goldschmidt1967}) of vector bundles over $J^k(E)$.
  \item
    Given a differential operator $\varphi:J^k(E)\supset J\ra F$, we can restrict its pushforward
    $\varphi_*$
    to the the vertical subbundle $V(J)$ of $TJ$. By definition, a differential operator
    is a morphism of fibered manifolds. That means, we have $\xi \circ \varphi=\pi$
    (where $\pi:J^k(E)\ra M$ and $\xi:F\ra M$ are projections).
    This implies that vertical vectors of $J$ are mapped to vertical vectors of $F$.
    Thus, we obtain a map $\varphi_*:V(J)\ra V(F)$. Now define the
    \ti{symbol} $\sigma(\varphi)$ (of $\varphi$) as the composition
    \begin{equation}
      \begin{split}
        &\sigma(\varphi):=\varphi_*\circ \mu : F^k_{J} \ra V(F)\\
        & p=(\theta,a) \mapsto \varphi_{*,\theta}
        \br{a^j_{\sigma=i_1 \ldots i_k} \frac{\partial}{\partial u^j_\sigma}\bigg|_{\theta}}
        = a^j_{\sigma} \frac{\partial \varphi^h}{\partial u^j_\sigma}
        \frac{\partial}{\partial w^h}\bigg|_{\varphi(\theta)}
      \end{split}
      \label{eq:symbolOfMap}
    \end{equation}
    Here $h\in \bbr{1,\cdots,\te{dim}(F_{\pi(\theta)})}$ and $|\sigma|=k$ because
    of (\ref{eq:dim}), (\ref{eq:epsilon}) and (\ref{eq:mumap}).
  \item
    The $l$-th prolongation $\sigma^l(\varphi)$ of the symbol $\sigma(\varphi)$ of $\varphi$ is defined as
    the composition
    (see \cite{bryantGoldschmidt}, end of chapter IX)
    \begin{center}
      \bt[huge]
      F^{l+k}_J = S^{l+k}T^*\otimes_J V(E) \ar{r}{\Delta_{l,k} \otimes \te{id}} \&
      S^lT^*\otimes F^k_J \ar{r}{\te{id} \otimes \sigma(\varphi)} \& S^lT^*\otimes_F V(F).
    \et
    \end{center}
    In local coordinates, we can express a point $p\in F^{l+k}_J$ as a tuple
    $p=(\theta \in J\subset J^k(E),a \in S^{l+k}T^*\otimes V(E))$ such that
    \begin{equation}
      \begin{split}
      \sigma^l(\varphi)(p)
      &=
      \sum_{\sigma\in \mathfrak{S}_{k+l}}
      a^j_{\sigma(i_1) \ldots \sigma(i_{l+k})}
      dx^{\sigma(i_1)}\vee \ldots \vee dx^{\sigma(i_l)} \\
      &
      \qquad
      \qquad
      \qquad
      \otimes \sigma(\varphi)\br{dx^{\sigma(i_{l+1})}\vee \cdots \vee dx^{\sigma(I_{l+k})}\otimes
      \frac{\partial}{\partial u^j}\bigg|_{\theta}}\\
      &\ov{$a$ is symmetric}
      \sum_{\sigma\in \mathfrak{S}_{k+l}}
      a^j_{i_{1} \ldots i_{l+k}}
      dx^{\sigma(i_1)}\vee \ldots \vee dx^{\sigma(i_l)} \\
      &
      \qquad\qquad
      \qquad~
      \qquad\qquad\qquad
      \otimes
      \frac{\partial \varphi^h}{\partial u^j_{\sigma(i_{l+1}) \ldots \sigma(i_{l+k})}}
    \frac{\partial}{\partial w^h}\bigg|_{\varphi(\theta)}\\
      \end{split}
      \label{eq:localCoordsProlongSym}
    \end{equation}
    $\sigma^1(\varphi)$ is especially important, and is explicitly rewritten as follows.
    \begin{equation}
      \begin{split}
      \sigma^1(\varphi)(p)
      = \sum_{\sigma\in\mathfrak{S}_{1+k}}
      a^j_{i_{1} \ldots i_{1+k}}
      \frac{\partial \varphi^h}{\partial u^j_{\sigma(i_{1}) \ldots \sigma(i_{k})}}
      dx^{\sigma(i_{1+k})} \otimes
    \frac{\partial}{\partial w^h}\bigg|_{\varphi(\theta)}\\
      \end{split}
      \label{eq:localCoordsProlongSym1}
    \end{equation}
  \item
    The following proposition is useful for practical calculations.
    \bp{prop:commutationFormalInt}{
      The following diagram commutes.
      \begin{center}
        \bt
        S^{l+k}T^*\otimes_J V(E) \ar{r}{\sigma^l(\varphi)}
          \ar{d}{\mu}
          \& S^lT^*\otimes_F V(F) \ar{d}{\mu} \\
          V(P^l(J)) \ar{r}{p^l(\varphi)_*}\& V(J^l(F))
        \et
      \end{center}
      Locally, one can thus use
      (\ref{eq:prolongedSymbolOfMap}) with $|\alpha|=l$
      instead of (\ref{eq:localCoordsProlongSym}).
    }
    \begin{proof}
    If one applies $\mu$ to equation (\ref{eq:localCoordsProlongSym}), one obtains
    {\small
    \begin{equation}
      \begin{split}
      \mu(\sigma^l(\varphi)(p))
    &=
      \sum_{\sigma\in \mathfrak{S}_{k+l}}
      a^j_{\sigma(i_1) \ldots \sigma(i_{l+k})}
      \frac{\partial \varphi^h}{\partial u^j_{\sigma(i_{l+1}) \ldots \sigma(i_{l+k})}}
      \frac{\partial}{\partial w^h_{\sigma(i_1) \ldots \sigma(i_l)}}\bigg|_{p^l(\varphi)(\theta)}
      \end{split}
      \label{eq:muOflocalCoordsProlongSym}
    \end{equation}
  }
    One can furthermore define the composition\footnote{
    To recall the definition of $P^l(J)$, see equation (\ref{eq:prolSubmfdJetSpace}).}
    \begin{center}
      \bt[huge]
      F^{l+k}_J = S^{l+k}T^*\otimes_J V(E) \ar{r}{\mu} \&
      V(P^l(J) )\ar{r}{p^l(\varphi)_*} \& V(J^l(F)).
    \et
    \end{center}
    In local coordinates, this means, for $|\sigma|=l+k,~0\le| \alpha |\le l$, that
    {\small\begin{equation}
      \begin{split}
        &p^l(\varphi)_* \circ \mu : F^{l+k}_{J} \ra V(J^l(F))\\
        & p=(\theta,a) \mapsto p^l(\varphi)_{*,\theta}
        \br{a^j_{\sigma=i_1 \ldots i_{l+k}} \frac{\partial}{\partial u^j_{\sigma}}\bigg|_{\theta}}
        =
        a^j_{\sigma} \frac{\partial (D_\alpha \varphi^h)}{\partial u^j_{\sigma}}
        \frac{\partial}{\partial w^h_\alpha}\bigg|_{p^l(\varphi)(\theta)}
      \end{split}
      \label{eq:prolongedSymbolOfMap}
      \end{equation}
    }
    Note that, if $\varphi:J^k(E)\supset J\ra F$ is a differential operator of order $k$,
    then it involves at most coordinates $u^j_\beta$ with $0\le |\beta|\le k$.
    As a consequence $\partial D_\alpha\varphi^h/\partial u^j_\sigma$ for $|\sigma|=l+k$
    must be zero for all $0\le |\alpha| <l$. Hence, to obtain
    the non-zero components of $p^l(\varphi)_*\circ\mu$, it suffices to
    calculate (\ref{eq:prolongedSymbolOfMap}) for $|\alpha|=l$.\\
    As all terms of $D_{\alpha}\varphi^h$ in (\ref{eq:prolongedSymbolOfMap})
    vanish if they are not highest order, let us calculate what is left of
    $D_{\alpha}\varphi^h$ if we only look at its highest order terms.
    Suppose that $\varphi$ is a differential operator of order $k$, then
    {\small
    \begin{equation}
      \begin{split}
      D_\alpha\varphi^h &=
      D_{\alpha_1 \ldots \alpha_{l-1}} D_{\alpha_l}\varphi^h
      \ov{\ref{eq:totalDiffOperator}}
      D_{\alpha_1 \ldots \alpha_{l-1}}
      \br{\frac{\partial \varphi^h}{\partial x^{\alpha_l}} + \cdots +
      \frac{\partial \varphi^h}{\partial u^j_\theta} u^j_{\theta \alpha_l}}
      \te{ with }|\theta|=k\\
      &\overset{\te{(highest order)}}{\longrightarrow}
      \frac{\partial \varphi^h}{\partial u^j_\theta}
      D_{\alpha_1 \ldots \alpha_{l-1}}
      u^j_{\theta \alpha_l}
      =\frac{\partial \varphi^h}{\partial u^j_\theta}
      u^j_{\theta \alpha_1 \ldots \alpha_l}
      =\frac{\partial \varphi^h}{\partial u^j_\theta}
      u^j_{\theta \alpha}
      \end{split}
      \label{eq:equivalenceReason}
    \end{equation}
  }
    This means the calculation of terms of order $k+l$ of $D_{\alpha}\varphi^h$ only involves
    derivatives of $\varphi^h$ of order $k$. Thus,
    \begin{equation}
      \begin{split}
      \frac{\partial (D_{\alpha}\varphi^h)}{\partial u_{\sigma}^j}
      &=
      \frac{\partial \varphi^h}{\partial u^k_\theta}
      \frac{\partial u^k_{\theta \alpha}}{\partial u_{\sigma}^j}
      \end{split}
      \label{eq:equivalenceReason2}
    \end{equation}
    Now
    $\partial u^k_{\theta \alpha}/\partial u_{\sigma}^j=1$ only if $k=j$ and
    $\theta_1 \cdots \theta_k \alpha_1 \cdots \alpha_l=\sigma_1 \cdots \sigma_{l+k}$
    \ti{or any permutation thereof}.\\
    Therefore, when summing over everything, one obtains
    \begin{equation}
      \begin{split}
        &a^j_{\sigma} \frac{\partial (D_\alpha \varphi^h)}{\partial u^j_{\sigma}}
        \frac{\partial}{\partial w^h_\alpha}
        \bigg|_{p^l(\varphi)(\theta)} \\
        &\qquad= \sum_{\sigma\in \mathfrak{S}_{k+l}} a_{\sigma(i_1) \ldots \sigma(i_{k+l})}
        \frac{\partial \varphi^h}{\partial u^j_{\sigma(i_1) \ldots \sigma(i_k)}}
        \frac{\partial}{\partial w^h_{\sigma(i_{k+1}) \ldots \sigma(i_{k+l})}}
        \bigg|_{p^l(\varphi)(\theta)}
      \end{split}
      \label{eq:equivalenceReason3}
    \end{equation}
    As the sum goes through all permutations, this is equivalent to equation
    (\ref{eq:muOflocalCoordsProlongSym}). Thus, we obtain
    $\mu\circ \sigma^l(\varphi) = p^l(\varphi)_* \circ \mu$.
  \end{proof}
     \item
    Given a differential equation $\de{E}$, define
    \begin{equation}
      g^k := V(\de{E})\cap \mu(F^k_{\de{E}})
      \label{eq:symbolOfRkInrinsic}
    \end{equation}
    and also call it the \ti{symbol} (of $\de{E}$).
    It's $l$-th prolongation is defined as
    \begin{equation}
      g^{k+l} := (S^lT^* \otimes_{\de{E}} V(\de{E})) \cap F^{l+k}_{\de{E}}
      \label{eq:symbolProlong}
    \end{equation}
    If a differential operator $\varphi:J\ra F$ is given such that (\ref{eq:PDEcriteria}) holds,
    \cite{goldschmidt1967} shows that
    the symbol of $\de{E}:=\ker_s(\varphi)$ and its $l$-th prolongation are given by
    \begin{equation}
      g^k = \ker (\sigma(\varphi))|_{\de{E}},\qquad g^{k+l}=\ker(\sigma^l(\varphi))|_{\de{E}}.
      \label{eq:symbolsOfRk}
    \end{equation}
    Set $g^{k+l}=F^{k+l}_{\de{E}}$ for $l<0$ and $F^{-1}_{\de{E}}=0$.
  \item
    Define a map
    $\delta : S^{1+k}T^* \ra T^*\otimes S^{k}T^*$ by setting
    $\delta= \Delta_{1,k}$ (see (\ref{eq:coordsDeltaLK})). Then extend this map by letting the same letter $\delta$
    denote the map
    \begin{equation}
      \begin{split}
      &\delta: T^*\otimes S^kT^*\ra \Lambda^2\otimes S^{k-1}T^*\\
      &dx^{h_1}\otimes dx^{i_1} \vee \cdots \vee dx^{i_{k}}
      \mapsto (-1) dx^{i_1}\wedge \Delta_{1,k-1}(dx^{i_1}\vee \cdots \vee dx^{i_k})
      \end{split}
    \end{equation}
    Now let $n$ be any natural number and $w\in \Lambda^j$ and extend the map again as follows:
    {\small\begin{equation}
      \begin{split}
        \delta: \Lambda^j \otimes F^{n}_Y &\to \Lambda^{j+1} \otimes F^{n-1}_Y\\
        w \otimes dx^{i_1}\vee \cdots \vee dx^{i_n}
        \otimes \frac{\partial}{\partial u^l}
        &\mapsto
         (-1)^j
         w \wedge
        \Delta_{1,n-1}\br{dx^{i_1}\vee \cdots \vee dx^{i_n}} \otimes \frac{\partial}{\partial u^l}.
      \end{split}
      \label{eq:spencerDelta}
    \end{equation}
  }
    If we set $S^{l}T^*=0$ for $l<0$, one can now use this map $\delta$ to obtain the sequence
    \begin{center}
    \bt
    0 \ar{r} \&  S^kT^* \ar{rr}{\delta} \& \& T^*\otimes S^{k-1}T^* \ar{d}{\delta} \\
    0 \&
    \Lambda^m\otimes S^{k-m}T^* \ar{l}
    \& \ldots \ar[']{l}{\delta} \& \Lambda^2\otimes S^{k-2}T^* \ar[']{l}{\delta}
    \et
    \end{center}
    (where $m=\te{dim}(M)$.)
    This sequence is exact (see \cite{goldschmidt1967}, Lemma 6.1).\\
    As $g^{k+l}\subset F^{k+l}_{\de{E}}$ and
    $\delta(g^{n})\subset T^*\otimes_{\de{E}} g^{n-1}$,
    the above map (\ref{eq:spencerDelta}) also gives rise to the sequence
    \begin{center}
    \bt
    0 \ar{r} \&  g^n \ar{rr}{\delta} \& \& T^*\otimes_{\de{E}} g^{n-1} \ar{d}{\delta} \\
    \Lambda^{n-k+1}\otimes_{\de{E}} F^{k-1}_{\de{E}}.
    \&
    \Lambda^{n-k} \otimes_{\de{E}} g^k \ar{l}
    \& \ldots \ar[']{l}{\delta} \& \Lambda^2\otimes_{\de{E}} g^{n-2} \ar[']{l}{\delta}
    \et
    \end{center}
    The cohomology groups of this sequence are denoted by
    \begin{equation}
      H^{n,j} := \frac{\ker(\delta: \Lambda^j\otimes_{\de{E}} g^n \ra \Lambda^{j+1}\otimes_{\de{E}} g^{n-1})}
      {\te{Im}(\delta: \Lambda^{j-1}\otimes_{\de{E}} g^{n+1} \ra \Lambda^{j}\otimes_{\de{E}} g^{n})}
      \label{eq:spencerCohomology}
    \end{equation}
    and are called \ti{Spencer cohomology} groups.\\
    One says that 
    \begin{equation}
      \begin{split}
        \text{$g^k$ is \ti{$r$-acyclic} if
        $H^{n,j}=0$ for all $n\ge k$ and $0\le j\le r$}
      \end{split}
      \label{eq:rAcyclic}
    \end{equation}
    and that $g^{k}$ is \ti{involutive} if 
    \begin{equation}
      \begin{split}
        \text{$g^{k}$ is $\infty$-acyclic, i.e.
        if $H^{n,j}=0~ \forall ~n\ge k,~j\ge 0$.}
      \end{split}
      \label{eq:infAcyclic}
    \end{equation}
  \item
    Finally, one needs the notion of a \ti{quasi-regular basis}. To this end, define the space
    \begin{equation}
      S^{k,j}T^* := \bbr{\te{span}(dx^{i_1}\vee \cdots \vee dx^{i_k})|
      j+1\le i_1 \le \cdots \le i_k\le m=\te{dim}(M)}
      \label{eq:quasiregSymSpace}
    \end{equation}
    Its dimension can be calculated as before by counting the number of possibilities of
    putting $k$ balls between $m-(j+1)$ sticks. The result is
    \begin{equation}
      \te{dim} (S^{k,j}T^*) =
      \begin{pmatrix}
        m-j-1+k\\
        k
      \end{pmatrix}
      \label{eq:dimSkj}
    \end{equation}
    Using this definition, define the $k,j$-fiber
    \begin{equation}
      F^{k,j}_Y := S^{k,j}T^* \otimes_{\de{E}} V(E)
      \label{eq:quasiregKFiber}
    \end{equation}
    and use this to define the $k,j$-symbol and its prolongation
    \begin{equation}
      g^{k,j} := g^k \cap F^{k,j}_{\de{E}},\qquad g^{k+l,j} := g^{k+l}\cap F^{k+l,j}_{\de{E}}.
      \label{eq:quasiregKJsymbol}
    \end{equation}
    If $\varphi:J\ra F$ is a differential operator such that (\ref{eq:PDEcriteria}) holds,
    then the $k,j$-symbol of $\varphi$
    and its prolongation are defined as the restrictions of $\sigma(\varphi)$ and
    $\sigma^l(\varphi)$ to $F^{k,j}_{J}$ and $F^{l+k,j}_J$. Explicitly, we have
    \begin{equation}
      \begin{split}
        \sigma(\varphi)^j :&= \sigma(\varphi)|_{F^{k,j}_{J}}: F^{k,j}_{J}\ra V(F),\\
        \sigma^l(\varphi)^j :&= \sigma^l(\varphi)|_{F^{l+k,j}_{J}}: F^{l+k,j}_{J}\ra V(F),\\
        \Ra~ g^{k,j} &= \ker(\sigma(\varphi)^j)|_{R_k},\qquad g^{l+k,j}=\ker(\sigma^l(\varphi)^j)|_{\de{E}}.
      \end{split}
      \label{eq:quasiregkjSymbol}
    \end{equation}
    Now say that a basis $\bbr{\partial_1, \cdots, \partial_m}|_{x\in M}$ of $T_xM$
    is quasi-regular for $g^k$ at $p\in \de{E}$ if\footnote{The condition on $g^k$
    locally imposes a condition on the dual basis and thus also on the basis.}
    \begin{equation}
      \te{dim} (g^{k+1}_{p}) = \te{dim} (g^k_p) + \sum_{j=1}^{m-1} \te{dim} (g^{k,j}_p).
      \label{eq:quasiregBasisCondition}
    \end{equation}
    And say that there is a quasi-regular basis for $g^k$ if there is a quasi-regular basis for $g^k$ at
    every $p\in \de{E}$.
\end{enumerate}
\subsection{Formal theory}
\label{sec:formalTheory}
Now with all definitions at hand, we can proceed with a motivation for the definition of formal integrability.
Given a differential equation $\de{E}$,
one would like to find its solutions.
In general, solutions around a point are difficult to find.
Recall that a horizontal solution can be described by
a section $s:M\supset U \ra E$ such that
$j^k(s)(U)\subset \de{E}$.
If a section fulfills this property and it is smooth,
then its prolongations also fulfill the prolonged
equations, i.e.
$j^{k+l}(s)(U)\subset P^l(\de{E})$.
In particular, this means, if $s$ is a solution and one chooses a fixed $x\in U$, then it holds true that
\begin{equation}
  j^{k+l}(s)(x) \in P^l(\de{E})\te{ for all }l\ge 0.
  \label{eq:necessaryProlongationCondition}
\end{equation}
Thus, (\ref{eq:necessaryProlongationCondition}) is a necessary condition for the existence of
a smooth solution $s:U\ra E$.\\
A point $\theta\in \de{E}$ is called a solution of order $k$ at $x=\pi(\theta)$.
It is called a solution of order $k$ at $x$ because
by Borel's lemma, one can always find a section $s$
that fulfills $j^k(s)(x)=\theta$. However, this section does not necessarily fulfill
the condition (\ref{eq:necessaryProlongationCondition}).\\
Therefore, given a solution $\theta\in \de{E}$ of order $k$, one wishes to check if there exists
a section such that (\ref{eq:necessaryProlongationCondition}) holds.
If this condition holds at $x=\pi(\theta)$, then one says that $\de{E}$ has a formal
solution at $\theta$. If one can find formal solutions at all points of $\theta \in \de{E}$,
then one says that $\de{E}$ is \ti{formally integrable}.\\
As higher derivatives are promoted to
coordinates in the jet bundle approach,
$\de{E}$ is usually the kernel of an algebraic (most often polynomial) equation.
Therefore, \ti{to find solutions of order $k$ is comparatively easy} because it does not involve
any analysis but algebraic operations are sufficient.\\
Finally, suppose that a formal solution consisting of a section
$s$ that fulfills (\ref{eq:necessaryProlongationCondition})
at the point $x=\pi(\theta)$ is given.
Then the section $s$ we have found is precisely the section whose taylor expansion
is equal to the expansion whose coefficients are $j^l(s)(x)$.
This taylor expansion does not necessarily converge.
It may also happen that it does only converge at $x$ and in no
neighbourhood of $x$. Therefore, it is not necessarily a solution of $\de{E}$
in the usual sense. \\
However, suppose that it does converge in a neighbourhood of $x$, then it is a smooth solution of $\de{E}$.
In general, it is possible to show that a formal solution
always converges
if one works in the
analytic category where all functions are locally given by a converging taylor expansion.
Therefore, in this category, formal integrability is also a sufficient condition for the
existence of (local) solutions.
\\
To motivate the precise definition of formal integrability,
note that the requirement that any solution of order $k$ can be extended to a solution of infinite order
can only be fulfilled if the prolongation of any order of the equation does not impose new
constraints on the coordinates of the solution up to order $k$
(``new constraints'' means new equations involving coordinates up to order $k$ which are not equivalent to the equations one started with).
For suppose we started with a solution of order $k$ that did not fulfill those constraints,
then this solution could not be extended to a solution of the order which imposes those constraints.\\
If no new constraints are imposed on the coordinates of order $k$ by the prolongation, this means
geometrically that $P^l(\de{E})$ is a surface which can be given local coordinates
that agree with those of $\de{E}$ up to order $k$. Then,
\begin{equation}
  \pi^{k+l+1}_{k+l}:P^{l+1}(\de{E})\ra P^l(\de{E})\qquad\te{is surjective for all }l\ge 0.
  \label{eq:motivationFormalIntegrability1}
\end{equation}
One might define formal integrability using just this condition.
However, in most cases one would like to work in the smooth category
in order to find out if smooth solutions exist for some equation.
This requires us to impose an additional smoothness condition.
To ask if a smooth solution exists given some $k$-th order solution is equivalent to
asking whether the prolongation is smooth to all orders.
As a solution of order $k+l$ is a section such that $j^{k+l}(s)(U)\subset P^l(\de{E})$ (with $U\subset M$),
smoothness of the section can only be guaranteed if $P^l(\de{E})$ is a smooth submanifold of $J^{k+l}(E)$.
Goldschmidt shows in proposition 7.1 of \cite{goldschmidt1967} that $\pi^{1+k}_k:P^1(\de{E})\ra \de{E}$
is a smooth fibered submanifold of $\pi^{1+k}_k: J^{1+k}(E)|\de{E}
\ra \de{E}$ if and only if
$g^{1+k}$ (defined in \eqref{eq:symbolProlong}) is a vector bundle over $\de{E}$ and $\pi^{1+k}_k:P^1(\de{E})\ra \de{E}$ is surjective.
Those considerations motivate the following definition:
\gd{def:formalIntegrability}{
  A differential equation $\de{E}$ is said to be \ti{formally integrable} if
  \begin{enumerate}
    \item $\pi^{k+l+1}_{k+l}:P^{l+1}(\de{E})\ra P^l(\de{E})$ is surjective,
    \item $g^{k+l+1}$ is a vector bundle
  \end{enumerate}
  for all $l\in\bbr{0,1,2,\cdots}$.
}
The above definition requires to check an infinite amount of conditions.
Goldschmidt proved a theorem that facilitates to determine formal integrability in
a finite amount of steps. It is based on theorem 8.1 of \cite{goldschmidt1967} which we cite here:
\bp{prop:formalIntGoldschmidt2acyclic}{%
  If $\de{E}$ is a differential equation, then it is formally integrable if and only if
  \begin{enumerate}
    \item $\pi^{k+1}_k:P^1(\de{E})\ra \de{E}$ is surjective,
    \item $g^{k+1}$ is a vector bundle over $\de{E}$,
    \item $g^{k}$ is 2-acyclic.
  \end{enumerate}
}
Recall that $g^k$ is 2-acyclic (cf. \eqref{eq:rAcyclic}) if the Spencer cohomology groups $H^{n,j}$ (see (\ref{eq:spencerCohomology}))
vanish for all $n\ge k$ and $0\le j\le 2$.
However, \cite{goldschmidt1967} also proves in Lemma 6.2 that $g^k$ is always 1-acyclic,
i.e. $H^{n,j}=0$ for all $n\ge k$ and $0\le j \le 1$.
Therefore, one can replace the last condition by the requirement that
\begin{equation}
  H^{n,2}=0 \te{ for all $n\ge k$}.
  \label{eq:requirementSpencerCohomologies}
\end{equation}
This still seems to require an infinite number of calculations. However, \cite{goldschmidt1967}
shows in Lemma 6.4
\bp{prop:formalIntfiniteSteps}{%
  If the dimension of $V_{\pi(\theta)}(E)$ does not depend on $\theta \in \de{E} \subset J^k(E)$,
  then there exists an integer $k_0>k$ depending only on $\te{dim}(M)$ and $k$ and
  $\te{dim}V_{\pi(\theta)}(E)$ such that
  $g^{k_0}$ is involutive, i.e. that $g^{k_0}$ is $\infty$-acyclic, i.e. $H^{n,j}=0~ \forall ~n\ge k_0,~j\ge 0$.
}
Similarly, he uses Lemma 6.4 and proposition 7.2 to prove theorem 8.2 which reads
\bp{prop:finiteFormalInt2}{%
  If the dimensions of all components of $E$ are the same and $\de{E}\subset J^k(E)$ is a differential equation,
  then there exists an integer $k_0>k$ depending only on $\te{dim}(M)$ and $k$ and $\te{dim}(E)$ such that
  $\de{E}$ is formally integrable
  if and only if
  \begin{enumerate}
    \item $\pi^{k+1+l}_{k+l}:P^{1+l}(\de{E})\ra P^l(\de{E})$ is surjective,
    \item $g^{k+1+l}$ is a vector bundle over $\de{E}$,
  \end{enumerate}
  for all $0\le l\le k_0-k$.
}
The last two propositions in particular imply that whenever the dimension of $E$ is constant (which is the case in most applications, where one often chooses $E=\mathbb{R}^m\times \mathbb{R}^e$ or some other manifold with constant dimension), then it must be possible to determine whether $\mathcal{E}$ is formally integrable in \textbf{finitely many steps}. This means that one actually does not have to compute the infinitely many cohomology groups appearing in \eqref{eq:requirementSpencerCohomologies}. 

\bigskip

Nevertheless, proposition \ref{prop:finiteFormalInt2} does not tell us how large this finite $k_0$ might be. 
In general, there does not seem to be a simple way to estimate this, which can be problematic. However, it turns out that one can prove stronger statements about the stronger condition of $\infty$-acyclicity / involutivity (defined in \eqref{eq:infAcyclic}). In \cite{bryantGoldschmidt}, theorem 2.14  in Chapter IX (according to them, going back to Serre), states:
\bp{prop:equivQuasiRegBaseInvol}{%
  The following conditions are equivalent:
  \begin{enumerate}
    \item There exists a quasi-regular basis (cf. \eqref{eq:quasiregBasisCondition}) of $g^k$ at $\theta\in \de{E}$,
    \item $g^k$ is involutive at $\theta$, i.e. $H^{n,j}_\theta=0~\forall n\ge k,j\ge 0$.
  \end{enumerate}
}
This means, if there is a quasi-regular basis, then $g^k$ is $\infty$-acyclic and therefore
also $2$-acyclic. Hence, combining proposition \ref{prop:formalIntGoldschmidt2acyclic} with the last proposition,
we obtain
\bp{prop:formalInt}{%
  If $\de{E}$ is a differential equation, then it is formally integrable if
  \begin{enumerate}
    \item $\pi^{k+1}_k:P^1(\de{E})\ra \de{E}$ is surjective,
    \item $g^{k+1}$ is a vector bundle over $\de{E}$,
    \item There exists a quasi-regular basis for $g^k$.
  \end{enumerate}
}
\gd{def:involution}{
  A PDE $\de{E}$ is called \textit{involutive} if and only if it satisfies the conditions of proposition \ref{prop:formalInt}.
}
As a corollary, an involutive equation is also formally integrable but the converse is not true (because $2$-acyclicity does not imply $\infty$-acyclicity). Indeed there are examples of equations that are formally integrable but not involutive. Thus, though the above proposition \ref{prop:formalInt} is more readily used in practice than propositions \ref{prop:formalIntGoldschmidt2acyclic} or \ref{prop:finiteFormalInt2}, it only provides a sufficient but not a necessary condition for formal integrability.
\yr{rem:4}{%
  An extensive treatment including possible subtleties of involution and formal integrability can be found in
  \cite{seiler2009involution}.
}
The above propositions \ref{prop:finiteFormalInt2} and \ref{prop:formalInt} are the central propositions of this subsection. In practice, one can use them to determine formal integrability and involutivity in finitely many steps.
In actual calculations of the rank of $g^k$
(which is necessary for validating
condition 3 of proposition \ref{prop:formalInt}),
it may happen that one must determine the rank of a larger matrix. As written above, the code for a small program computing it can be found in appendix \ref{app:formalIntSolver} but much more sophisticated algorithms are provided in \cite{seiler2009involution}.

\bigskip

Given formal integrability of an equation $\de{E}$, it becomes possible to show
the existence of local solutions in the analytic category as mentioned at the beginning of the subsection.
The precise definition of analyticity is
\gd{def:analyticity}{%
  A map is called analytic if, around any point, it can locally be defined by a convergent power series.
  (Note that this definition can also be applied to real functions. If the condition holds, they are called real-analytic).\\
  A manifold is called analytic if all of its transition functions are analytic.
  The analytic category is defined as the category in which the objects are analytic manifolds
  and the morphisms are analytic maps between them.
}
The existence of local solutions in the analytic category is guaranteed by
theorem 9.1 of \cite{goldschmidt1967} which is here rephrased as follows:
\bp{prop:formalIntImpliesExistenceOfAnalytSol}{%
  Suppose that $\de{E}$ is a formally integrable differential equation which is analytic.
  Then given a point $\theta\in P^l(\de{E})$ (for any $l\in \bbr{0,1,2,\cdots}$),
  it is possible to find an analytic section $s:U\ra E$ where $U$ is a neighbourhood of $x=\pi(\theta)$
  such that $j^{k+l}(s)(x)=\theta$ and $s$ is a local solution of $\de{E}$.
}
One might wonder if it is possible to prove something stronger, for example that smoothness
guarantees existence of local solutions. This is not possible because of "Lewy's example", a well-known counter-example.

\subsection{Integrability conditions}
\label{sec:integrabilityConditions}

When checking for formal integrability
or involutivity of a system of differential equations,
it may happen that the first prolongation $P^1(\de{E})$ does not project
surjectively to $\de{E}$ via $\pi^{k+1}_k$, or that $g^{k+1}$ is not a smooth vector bundle
or that there exists no quasi-regular basis for $g^{k}$
but that the PDE can become formally integrable
if certain integrability conditions $\de{B}(\de{E})$ are \ti{added to} $\de{E}$,
i.e. by defining $\de{B}:=\de{B}(\de{E})\cap \de{E}$,
$\de{B}$ can become formally integrable. This subsection gives a
definition for $\de{B}(\de{E})$ that is useful for identifying minimal consistency conditions when comparing systems of differential equations and field theories.\\
The definition is motivated by the following example in which surjectivity fails to hold.
\ye{ex:intCond}{
  Let $\pi:E:=\mb{R}^2\times\mb{R}$ be a fibered manifold with local coordinates
  $(x,t,u)$. Define $F:=\mb{R}^2\times\mb{R}^2$ and consider the differential operator
  \begin{equation}
    \Phi:J^2(E)\ra F,\qquad (x,t,u,u_x,u_t,u_{xx},u_{tt},u_{xt})\mapsto (x,t,u_x,u_{tt})
    \label{eq:intConEx}
  \end{equation}
  Now the differential equation $\de{E}=\ker_0(\Phi)$ is given by
  \begin{equation}
    \de{E}= \bbr{(x,t,u,0,u_t,u_{xx},0,u_{xt})}
    \label{eq:diffEqConEx}
  \end{equation}
  The first prolongation is
  \begin{equation}
    \begin{split}
      P^1(\de{E}) = \ker(\Phi^1)&=\ker
    \begin{pmatrix}
      u_x,&
      u_{tt},&
      u_{xx},&
      u_{xt},&
      u_{ttx},&
      u_{ttt}
    \end{pmatrix}\\
    &= \bbr{(x,t,u,0,u_t,0,0,0,u_{xxx},u_{xxt},0,0)}
    \end{split}
  \end{equation}
  so that $\pi^3_2 (P^1(\de{E}))=\ker\bbr{(x,t,u,0,u_t,0,0,0)}$ which is much smaller than $\de{E}$.
  The reason is that due to the prolongation, there arise additional constraints
  on coordinates \tf{of the order of $\de{E}$}, here of second order,
  namely on $u_{xx}$ and $u_{tx}$.
  Concretely, they are given by
  \begin{equation}
    \begin{split}
      \de{B}(\de{E}):~\{~u_{xx}=0,~u_{tx}=0~\}.
    \end{split}
  \end{equation}
  $\de{B}(\de{E})$ are the \ti{integrability or consistency conditions}
  of this system $\de{E}$.
  Therefore, to restore surjectivity,
  one can try to include those additional constraints right from the start
  and define
  \begin{equation}
    \begin{split}
      \de{B}:&= \de{E}\cap\de{B}(\de{E}) = \ker
    \begin{pmatrix}
      u_x,&
      u_{tt},&
      u_{xt},&
      u_{xx}
    \end{pmatrix}
    = \bbr{(x,t,u,0,u_t,0,0,0)}
    \end{split}
  \end{equation}
  Now if we prolong \ti{this} system, then the result is similar to $P^1(\de{E})$ except
  for the fact that the additional constraints $u_{xtx}=0$, $u_{xxt}=0$ are additionally imposed.
  However, those are only new constraints on the coordinates of order $3$.
  Therefore, the projection $\pi^3_2|_{P^1(\de{B})}$ is now indeed surjective.\\
  Furthermore, we can read off a solution from $\de{B}$, namely
  \begin{equation}
    u(x,t)=At+B
    \label{eq:solConEx}
  \end{equation}
  which is a meaningful solution because it \ti{also} is a solution of $\de{E}$.
  Indeed, for reasons of consistency just shown above, those are the only solutions of $\de{E}$.
  Therefore, the procedure to define a new system for which surjectivity is guaranteed is meaningful as long as $\de{B}$ is again a PDE (in particular, it must be non-empty).
}
Motivated by the observations above, we define consistency / integrability conditions as follows.
\gd{def:integrabilityConditions}{
  The \textit{integrability condition} of a given PDE $\de{E}\subset J^k(E)$
  is defined to be the biggest smooth submanifold $\de{B}(\de{E})$ of the lowest order jet space $J^l(E)$ (with $l\ge k$) such that $\de{B}:=\de{B}(\de{E})\cap (\pi^{l}_k)^{-1}(\de{E})$ is formally integrable (where $\pi^{l}_k:J^l(E)\to J^k(E)$ is the canonical projection).\\
  If $\de{B}$ is non-empty, smooth, and has a component with dimension bigger zero, it is called the \textit{formal closure} of $\de{E}$. Otherwise the formal closure of $\de{E}$ does not exist and $\de{E}$ is said to be \textit{(formally) non-integrable}.
}
Note that the formal closure (or its non-existence) can always be computed in finitely many steps
because formal integrability can be checked in finitely many steps
using proposition \ref{prop:finiteFormalInt2} which is of practical importance. \\
Furthermore, it can often be useful to attempt to compute the \textit{involutive closure} / \textit{completion} of the PDE $\de{E}$ instead because the conditions of proposition \ref{prop:formalInt} are easier to check. If the PDE in question admits such a completion, one does not need to check formal integrability anymore. If it does not admit such a completion, one can still resort to checking the conditions of proposition \ref{prop:finiteFormalInt2}.

\bigskip

For the intersections of our physical theories, it might occur
quite often that the intersections are formally integrable only
after redefining them as systems that take the consistency conditions into account.
Those consistency conditions that are automatically found when checking formal integrability
are precisely \ti{the minimal amount of assumptions that must be made in order
to make the system consistent.}
Therefore, they are actually really useful for us because
\ti{they can be interpreted as the minimal physical assumptions under which a correspondence
becomes meaningful.}\\
This means that without knowing exactly what assumptions
are reasonable to relate two systems, we can just define a correspondence
and then find it out. 
This happens later in the example where magneto-statics and hydrodynamics are shown to
share an intersection whose consistency conditions had to be guessed
in equation (\ref{eq:assumptions}) in the motivating example in subsection \ref{sec:correspondenceMotivatingExample}.\\
The correspondences themselves still have to be guessed. However, symmetries can provide clues about which correspondences might be especially meaningful as explained in section \ref{sec:equivUpToSym}.

\subsection{Explicit examples of the application of proposition \ref{prop:formalInt}}
\label{sec:explicitExampleFormalInt}

An example where formal integrability fails to hold (if no new constraints are added) was provided above in example \ref{ex:intCond}. In this subsection, involutivity, and thus also formal integrability, is proved for a very simple example using proposition \ref{prop:formalInt}. Despite the simplicity of the equation, the example is very detailed to illustrate the formalism.
The reader not interested in this illustration can directly continue with the next
section. More involved examples are provided in section \ref{chap:applications}.\\
Below, $\de{E}$ and $P^l(\de{E})$ are defined as kernel of a differential operator $\varphi$
and its prolongation, using (\ref{eq:defKer}) and (\ref{eq:prolongRk}). Thus, it is possible
to obtain $g^k$ and $g^{k+l}$ as the kernel of $\sigma(\varphi)$ and its prolongation
using (\ref{eq:symbolsOfRk}). Furthermore, $g^{k,j}$ and $g^{k+l,j}$ can be obtained using (\ref{eq:quasiregkjSymbol}).

\bigskip


First define $\pi:E\ra M$ as follows:
\begin{equation}
  M:=\mathbb{R},~E:=\mathbb{R}\times\mathbb{R},~\pi=\te{pr}_1:E\ra M\te{ is the projection onto the first factor}.
  \label{eq:simpleSetting}
\end{equation}
Let $J:=J^1(E)\simeq\mathbb{R}^3$ with local coordinates $(x,u,u_x)$.
Then define a differential operator $\varphi:J\ra F:=E$ by
\begin{equation}
  \varphi(x,u,u_x)=(x,\varphi^1(x,u,u_x)):=(x,u_x-u)
  \label{eq:example1DE}
\end{equation}
which is a first order linear operator.
Its kernel
\begin{equation}
  \begin{split}
  \de{E} :&= \ker \varphi \ov{\ref{eq:defKer})}
  \bbr{ \theta \in J~|~\varphi(\theta)=0(\pi(\theta))=(x,0)~}\\
  &=\bbr{\theta \in J~|~u_x=u}
  =\bbr{(\rho,\lambda,\lambda)~|~\rho,\lambda\in\mathbb{R}}
  \end{split}
  \label{eq:example1Rk}
\end{equation}
is a first order linear differential equation corresponding to a two-dimensional subspace of $J$.
We know it to have the general solution
\begin{equation}
    u(x):=u(s_E(x))=A\exp(x),~ A\in\mathbb{R}
  \label{eq:example1sol}
\end{equation}
but want to show formal integrability of $\de{E}$ to
illustrate the general methods introduced above.\\
To show all 3 conditions of proposition \ref{prop:formalInt}, we first need to calculate $P^1(\de{E})$ and $g^{k+1}=g^2$.
To this end,
note that the prolongation of $J=J^1(E)$ is $J^2(E)\simeq \mathbb{R}^4$ with local coordinates $(x,u,u_x,u_{xx})$.
Thus, we can
use (\ref{eq:prolongvarphiExplicit}) to prolong $\varphi$ to obtain
\begin{equation}
  p^1(\varphi)(\theta \in J^2(E)) = (\varphi,D_x\varphi)(\theta)=(x,u_x-u,u_{xx}-u_{x}) \in J^1(E).
  \label{eq:example1prolongvarphi}
\end{equation}
such that
\begin{equation}
  \begin{split}
    P^1(\de{E}) &= \ker p^1(\varphi) \ov{\ref{eq:prolongRk}}
    \bbr{ \theta \in J^2(E)~|~(\varphi(\theta),D_x\varphi(\theta))=0(\pi(\theta))}\\
    &= \bbr{ \theta \in J^2(E) ~|~u_x-u=0,~u_{xx}-u_x=0}
    = \bbr{ (\rho,\lambda,\lambda,\lambda) ~|~\rho,\lambda\in \mathbb{R}}.
  \end{split}
  \label{eq:example1prolongRk}
\end{equation}
Now that $P^1(\de{E})$ and $\de{E}$ are explicitly given, one can see that the restriction of $\pi^2_1$ to $P^1(\de{E})$
\tf{surjectively} projects down to $\de{E}$. Explicitly,
\begin{equation}
  \pi^2_1 P^1(\de{E}) = \bbr{ \pi^2_1(\rho,\lambda,\lambda,\lambda) } = \bbr{ (\rho,\lambda,\lambda)}=\de{E}.
  \label{eq:example1surjectivity}
\end{equation}
This means condition 1. of proposition \ref{prop:formalInt} is fulfilled.
In fact, there even is an inverse map sending $(\rho,\lambda,\lambda)$ back to $(\rho,\lambda,\lambda,\lambda)$,
so $P^1(\de{E})\simeq \de{E}$. This continues for higher orders. We have
\begin{equation}
  p^l(\varphi) = (\varphi, D_x\varphi, \cdots, D_x^l \varphi)~
  \Ra~P^l(\de{E}) = \bbr{~(\rho,
  \lambda, \cdots,\lambda
  )~|~\rho,\lambda\in \mathbb{R}~}~
  \simeq ~\de{E}.
\end{equation}
Now let us calculate $g^1$ and $g^2$. To do so, we must first calculate the symbol of $\varphi$.
To do this, we must first clarify how an element $p\in F^1_{\de{E}}$ looks like. This can be done using
(\ref{eq:localCoordKFiber}). Note that our manifold $M=\mathbb{R}$ is one dimensional
and therefore $T^*M$ has basis $dx$ while $V(E)$ has basis $\partial/\partial u$. Thus,
\begin{equation}
  F^1_{\de{E}} \ni p=(\theta,a)=\br{ (x,u,u),~\br{a^1_1 dx \otimes \frac{\partial}{\partial u}\bigg|_{\pi_0(\theta)}}}.
  \label{eq:example1kjFiber}
\end{equation}
As a consequence $\te{dim}(F^1_{\de{E}})_{\theta\in \de{E}}=1$ and we obtain
\begin{equation}
  \sigma(\varphi)(p) \ov{\ref{eq:symbolOfMap}} a^1_1~ \frac{\partial \varphi^1}{\partial u_x}
  \frac{\partial}{\partial u}\bigg|_{\pi_0(\theta)}
  \ov{\ref{eq:example1DE}}
  a^1_1
  \frac{\partial}{\partial u}\bigg|_{\pi_0(\theta)}
  \label{eq:example1symbolvarphi}
\end{equation}
\begin{equation}
  \begin{split}
  g^1 &\ov{\ref{eq:symbolsOfRk}} \ker(\sigma(\varphi))|_{\de{E}}
  \ov{\ref{eq:example1symbolvarphi}}
  \bbr{ p \in F^1_{\de{E}}~\bigg|~\theta\in \de{E}\te{ and }
  a^1_1
\frac{\partial}{\partial u}\bigg|_{\pi_0(\theta)} = 0} \\
&~=~\bbr{ (\theta,0)~|~\theta\in \de{E}} ~\simeq~ \de{E}
  \end{split}
  \label{eq:example1symbol}
\end{equation}
This shows that $g^1$ is the trivial vector bundle over $\de{E}$ whose fibers consist of
the zero-point only. Similarly,
\begin{equation}
  F^2_{\de{E}} \ni p=(\theta,a)=\br{ (x,u,u),~\br{a^1_{11} dx\vee dx \otimes \frac{\partial}{\partial u}\bigg|_{\pi_0(\theta)}}}.
  \label{eq:example1kjFiber1}
\end{equation}
whose fibers are also one-dimensional and therefore
\begin{equation}
  \sigma^1(\varphi)(p) \ov{\ref{eq:localCoordsProlongSym}}
  2 a^1_{11}~ \frac{\partial \varphi^1}{\partial u_{x}}
  \frac{\partial}{\partial u}\bigg|_{\pi_0(\theta)}
  \ov{\ref{eq:example1DE}}
  2 a^1_{11}
  \frac{\partial}{\partial u}\bigg|_{\pi_0(\theta)}
  \label{eq:example1symbol1varphi}
\end{equation}
such that
\begin{equation}
  \begin{split}
  g^2 &\ov{\ref{eq:symbolsOfRk}} \ker(\sigma^1(\varphi))|_{\de{E}}
  \ov{\ref{eq:example1symbol1varphi}}
  \bbr{ p \in F^{2}_{\de{E}}~|~\theta\in \de{E}\te{ and }
    a^1_{11} \frac{\partial}{\partial u}\bigg|_{\pi_0(\theta)} = 0}\\
&~= ~\bbr{ (\theta,0)~|~\theta\in \de{E}}\simeq g^1 ~ \simeq ~ \de{E}
  \end{split}
  \label{eq:example1symbol2}
\end{equation}
As a consequence, $g^2$ is also a trivial \ti{vector bundle over $\de{E}$}. This proves that condition
2. of proposition \ref{prop:formalInt} is fulfilled. In fact, one can see that $g^{1+l}\simeq g^1$ for all $l$.
As a consequence, we do not even need to test condition 3 of proposition 5 because
this together with $P^l(\de{E})\simeq \de{E}$ directly shows that the definition
\ref{def:formalIntegrability} of formal integrability is fulfilled.\\
Nevertheless, let us test condition 3 of proposition \ref{prop:formalInt} explicitly.
To this end, we must check condition (\ref{eq:quasiregBasisCondition})
for all $p\in \de{E}$.
To do so, we must calculate $g^{1,j}$. However,
the definition of $S^{k,j}T^*$ (see (\ref{eq:quasiregSymSpace})) requires
that $j+1\le i_1 \le \cdots \le i_k \le m=1$ which is only possible for $j=0$,
i.e. $S^{k,j>0}T^*=0$. But the sum in (\ref{eq:quasiregBasisCondition}) only
goes from $j=1$ to $j=m-1=0$. As a consequence, we only have to verify that
\begin{equation}
  \te{dim}(g^2)=\te{dim}(g^1).
  \label{eq:example1quasiregCondition}
\end{equation}
This does hold because $g^2\simeq g^1$ as shown above. This shows that all conditions of
proposition \ref{prop:formalInt} are satisfied and our equation
(\ref{eq:example1Rk})
is involutive and thus formally integrable.

\newpage
\section{Shared structure}
\label{sec:sharedStructure}
\subsection{Definition}
Now that the notions of intersection and correspondence are developed and
that the theory of formal integrability has been reviewed, everything can be combined
to define what it means for two theories to share structure.\\
So suppose we are given two fibered manifolds $\pi:E\ra M$ and $\xi:F\ra M$
and would like to compare the differential equations $\de{E}\subset J^k(E)$ and
$\de{F} \subset J^l(F)$.
Consider
$\de{EF}$ as defined in eq. (\ref{eq:jointsystem}) which is a PDE by
proposition \ref{prop:EJismanifold}. However, given a correspondence $\Phi$,
corollary \ref{cor:intersectionDiffEq}
shows that $\de{EF}\cap \Phi$ is only a PDE if $\de{EF}\tv\Phi$.
Thus, the following definition is useful.
\gd{def:shareintersection}{
  $\de{E}$ and $\de{F}$ \ti{share an intersection}
  $\de{I}:=\de{EF}\cap\Phi$ \ti{(under the correspondence $\Phi$)}
  if $\de{EF}\tv \Phi$.
}
Now let us suppose that $\de{E}$ and $\de{F}$ do share an intersection under $\Phi$.
From the discussion in subsection \ref{sec:diffConsistency}, it is clear that
sharing an intersection is not enough for saying that two theories share structure in a meaningful way.
Instead, one should require that the system is differentially consistent / formally integrable
as well.
\gd{def:sharedstructure}{(\tf{Shared Structure})
  Two differential equations $\de{E}$ and $\de{F}$ \ti{share structure} if
  they share an intersection $\de{I}$
  that has a formal closure $\de{B}$
  (in the sense of definition \ref{def:integrabilityConditions}).
}
Note that if only an open subset of $\de{I}$ has a formal closure, then one can always
restrict $\Phi$ such that $\de{I}'$ has a formal closure.\\
This definition is meaningful because formal integrability
guarantees that all $N$-th order solutions on an open subset of $\de{B}$ can be prolonged to formal solutions.
As explained in subsection \ref{sec:formalTheory},
those $N$-th order solutions can be constructed very easily by defining
a Taylor expansion using as coefficients the entries of any point $\theta=(x^i,u^j_\alpha, v^g_\beta)$
in this open subset of $\de{B}$.
So if two differential equations share structure, then this usually means that the
formal closure of their intersection has a lot of formal solutions
and in this case the corresponding theories have quite a lot in common.

\bigskip

Given the geometric theory of shared structure, one can also obtain a natural notion of equivalence of PDEs. Most canonically, equivalence is perhaps defined as follows.
\gd{def:equivalence}{
  Two systems of PDEs $\de{E}\subset J^k(E)$ and $\de{F}\subset J^l(F)$
  are said to be \ti{equivalent} if there exists a diffeomorphism
  $L:\de{E}\to\de{F}$ that preserves the Cartan distribution,
  i.e.
  \begin{equation}
    \begin{split}
      dL_\theta(\de{C}_\theta\cap T_\theta\de{E})=
      \de{C}_{L(\theta)}\cap T_{L(\theta)}\de{F}
      \quad \forall\theta\in \de{E}.
    \end{split}
  \end{equation}
}
This diffeomorphism is very similar to a Lie transformation
that is used to define a symmetry, below in section \ref{sec:equivUpToSym}.
However, to integrate the above definition into the product bundle setting defined above, one could define equivalence also as follows.
\gd{def:equivalenceProductBundle}{
  Two systems of PDEs $\de{E}\subset J^k(E)$ and $\de{F}\subset J^l(F)$
  are said to be \ti{equivalent} if there exists a correspondence $\Phi$
  and intersection $\de{I}=\Phi\cap\de{EF}$ s.t.
  $\pi_E|_{\de{I}}$ and $\pi_{F}|_{\de{I}}$ map $\de{I}$ diffeomorphically
  onto $\de{E}$ and $\de{F}$ and preserve the Cartan distribution.
}
This second definition is equivalent to the first.
It might look somewhat more convoluted
but the product space is then a more suitable setting
for investigating relationships that are weaker than equivalence, such as shared subsystems in the form of shared structure that still allow for the transfer of some shared solutions,
as demonstrated by the numerous constructions of the subsequent subsections.

\subsection{Solution transfer}
\label{sec:solutionTransfer}
In this subsection, we assume that the intersection $\de{I}$ of two differential equations
$\de{E}$ and $\de{F}$
is itself a differential equations with solutions and investigate the relationship
between those solutions and the solutions of $\de{E}$ and $\de{F}$.
Recall that a solution $S$ of $\de{I}$ is a locally maximal $\text{dim}(M)$-dimensional
integral submanifold of $\de{C}$ with $S\subset \de{I}$
as described in item \ref{solution}.
Let $J:=J^k(E)\times_M J^l(F)$ and $\Pi:=\pi^k\times_M \xi^l: J\to M$, the natural projection
to the base space.
A submanifold $S\subset J$ is called \ti{horizontal}
if $d\Pi|_{\theta}:T_\theta S\to T_{\Pi(\theta)}M$ is injective
for all $\theta\in S$.

\bp{prop:solTransfer}{
  If $\de{E}$ and $\de{F}$ share the intersection $\de{I}=\de{EF}\cap\Phi$
  (where $\de{EF}=\pi_E^{-1}(\de{E})\cap \pi_F^{-1}(\de{F})$)
  and $\de{I}$ has a horizontal solution $S$, then $\pi_E(S)$ is a solution of $\de{E}$
  and $\pi_F(S)$ is a solution of $\de{F}$.
}
\pr{
  In order to verify that $S_E:=\pi_E(S)$ is a solution of $\de{E}$,
  we must verify that
  \begin{enumerate}
    \item $S_E\subset \de{E}$,
    \item $S_E$ is a smooth submanifold with dimension $m=\text{dim}(M)$,
    \item $S_E$ is an integral submanifold of $\de{C}$, i.e. $T_\theta S_E \subset \de{C}_\theta$
      for all $\theta \in S_E$,
    \item $S_E$ is a locally maximal integral submanifold.
  \end{enumerate}
  Since $S\subset \de{I}=\de{EF}\cap\Phi$, we have in particular $S\subset \de{EF} = \de{E}_J\cap\de{F}_J$
  and $S\subset \de{E}_J=\pi_E^{-1}(\de{E})=\{\theta\in J^k(E)\times_M J^l(F)~|~\pi_E(\theta)\in\de{E}\}$. Thus, $\pi_E(S)\subset \de{E}$ and the first
  item is verified.\\
  As $S$ is horizontal, it can locally
  be described as the image of the prolongation, $U_S:=\text{im}(j^k(s_E)\times_M j^l(s_F))$
  of a local section
  $s=s_E\times_M s_F:U\subset M \to E\times_M F,~x^i\mapsto (x^i,(s_E)^j(x),(s_F)^g(x))$.
  The prolongation and hence also
  $S$ can locally be described by the tuple $(x^i,(s_E)^j_\alpha(x),(s_F)^g_\beta(x))$.
  As a consequence, $\pi_E(S_U)$
  has the local description $(x^i,(s_E)^j_\alpha(x))$ on $J^k(E)$.
  At each point of $\pi_E(S_U)$,
  one can thus define $m=\text{dim}(M)$ tangent vectors,
  the $n$-th of which is given by
  \begin{equation}
    v_n := \frac{\partial x^i}{\partial x^n} \frac{\partial}{\partial x^i} +
    \sum_{j=1}^e\sum_{|\alpha|< k}\frac{\partial (s_E)^j_\alpha(x)}{\partial x^n} \frac{\partial}{\partial u^j_\alpha}
    \label{eq:tangentVectors}
  \end{equation}
  They are all non-zero and linearly independent because $\partial x^i/\partial x^n=\delta^{in}$.
  Thus, $\text{dim}(\pi_E(S))\ge \text{dim}(M)$.
  However, since $\text{dim}(\pi_E(S))\le \text{dim}(S)=\text{dim}(M)$,
  we obtain $\text{dim}(\pi_E(S_U))=\text{dim}(M)$.
  Hence, $d\pi_E|_S$ is locally a bijection and the vectors defined in (\ref{eq:tangentVectors})
  span the tangent space around a generic point of $S_E$. Furthermore, since $\pi_E$ is smooth,
  $\pi_E|_S$ is also smooth. Therefore, $\pi_E|_S$ is a local diffeomorphism.\\
  Now suppose we have two local neighbourhoods $O, O'\subset S_E=\pi_E(S)$
  which are such that $O\cap O'\ne \varnothing$.
  Then, since $\pi_E$ is a local diffeomorphism, we obtain corresponding
  open subsets $U=\pi_E^{-1}(O)$ and $U'=\pi_E^{-1}(O')$ in $S$.
  Furthermore, $\pi_E^{-1}(O\cap O')=\pi_E^{-1}(O)\cap\pi_E^{-1}(O')=U\cap U'$
  because inverse images always preserve intersections.
  Then, since $S$ is a smooth manifold, we also have a smooth transition map
  $\varphi:U\to U'$. As a consequence, since composition of smooth maps are smooth,
  $\pi_E\circ \varphi\circ \pi_E^{-1}|_{O\cap O'}:O\cap O'\to O'$
  is a smooth transition map on $S_E$. Therefore, all local pieces $\pi_E(U)$ coming
  from the local pieces $U\subset S$ of the solution $S$ piece together
  to form a global smooth, $\text{dim}(M)$-dimensional submanifold $\pi_E(S)$ of $\de{E}$.
  This verifies the second item.\\
  To show that $S_E$ is an integral submanifold of $\de{C}$, it suffices to
  show that the tangent vectors (\ref{eq:tangentVectors}) that
  locally span the tangent space of $S_E$ are annihilated by
  the Cartan forms $w^j_\alpha = du^j_\alpha - u^j_{\alpha i} dx^i$.
  Indeed we immediately obtain $w^j_\alpha|_{(x^i,u^j_\alpha)=(x^i,s^j_\alpha(x))}(v_n)=0$
  which verifies the third item.
  Since $S_E$ is already $\text{dim}(M)$ dimensional, no open subset
  of it can be embedded into a solution of higher dimension which implies
  the fourth item. Thus, $S_E$ is a solution of $\de{E}$.\\
  Since the above did not make any assumptions about
  $\de{E}$ which are not shared by $\de{F}$, the same conclusion also holds for $\de{F}$ and
  $\pi_F(S)$ is a solution of $\de{F}$.
}
In the case of non-horizontal / singular solutions,
one has to be a bit more careful. In that case, not all solutions
are projected to smooth submanifolds via $\pi_E$ and $\pi_F$.
\ye{ex:singularSolutions}{Consider $\pi:E\to M$ with $M:=\mathbb{R}$, $E:=M\times M$ and $\pi$ the
  projection to the first factor and consider
  another, identical bundle $\xi:F\to M$. Assume we are given the ODEs
  $\de{E}\subset J^1(E)$
  and $\de{F}\subset J^1(F)$ described by\\
  \begin{minipage}{0.6\textwidth}
    \begin{equation}
      \begin{split}
        \de{E}:\{ x^2+u_x^2=1\},\qquad
        \de{F}:\{v_x^2=2v+ \frac{1}{2}\}.
      \end{split}
    \end{equation}
    Note that the solution of $\de{E}$ is singular because the smooth
  integral submanifold described by
  \begin{equation}
    \begin{split}
      \tf{x}=
      \begin{pmatrix}
        x=\sin(2t)\\
        u = t+ \frac{1}{4}\sin(4t)\\
        u_x = \cos(2t)
      \end{pmatrix},
      \qquad
      \tf{v}:=
      \frac{\partial \tf{x}}{\partial t}
      =
      \begin{pmatrix}
        2\cos(2t)\\
        1+\cos(4t)\\
        -2\sin(2t)
      \end{pmatrix}
    \end{split}
  \end{equation}
  gives rise to a section $s:\mathbb{R}\to E,~t\mapsto (x(t),s(x(t)))$ with singular points at
  $x=x(t=\pi/4+n\pi),~n\in\mathbb{Z}$ because $d\pi^1_0|_{t=(\pi/4+n\pi)}(\tf{v}) = \tf{0}$ (where $\pi^1_0:J^1(E)\to E$).
  \end{minipage}
  \hspace{0.05\linewidth}
  \begin{minipage}{0.3\textwidth}
    \begin{center}
      \includegraphics[width=.8\textwidth]{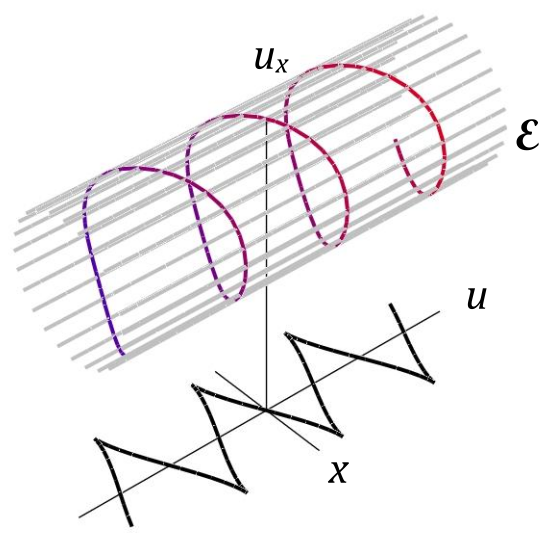}
      \footnotesize Visualization of a singular solution of the equation $\de{E}:\{ x^2+u_x^2=1\}$. Figure taken from a talk of Luca Vitagliano.\normalsize 
    \end{center}
  \end{minipage}\\
  The equation $\de{E}$ and its singular solution, as well as the figure that visualizes the solution, were presented in a talk by Luca Vitagliano, in relation to the publication \cite{Vitagliano_2014}.\\
  In the present example, the aim is to illustrate how such singular solutions relate to the notion of a correspondence.
  To this end, define such a correspondence
  between $\de{E}$ and $\de{F}$
  on $J:=J^1(E)\times_M J^1(F)$ by
  \begin{equation}
    \begin{split}
      \Phi:\{ u_x^2+v_x^2 = 1\}.
    \end{split}
  \end{equation}
  Then one solution $S$ of the submanifold $\de{I}=\de{E}\cap\de{F}\cap \Phi$
  is described by
  \begin{equation}
    \begin{split}
      \tf{x}=
      \begin{pmatrix}
        x=\sin(2t)\\
        u = t+ \frac{1}{4}\sin(4t)\\
        u_x = \cos(2t) \\
        w = - \frac{1}{4}\cos(4t)\\
        w_x = \sin(2t)
      \end{pmatrix},
      ~
      \tf{v}:=
      \frac{\partial \tf{x}}{\partial t}
      =
      \begin{pmatrix}
        2\cos(2t)\\
        1+\cos(4t)\\
        -2\sin(2t) \\
        \sin(4t)\\
        2\cos(2t)
      \end{pmatrix},
      ~
      \tf{v}(t=\pi/4+n\pi)
      =
      \begin{pmatrix}
        0\\
        0\\
        -2\\
        0\\
        0
      \end{pmatrix}
    \end{split}
  \end{equation}
  In the present situation, we obtain
  $d\pi_F|_{t=(\pi/4+n\pi)}(\tf{v})=\tf{0}$.
  This means that $\pi_F(S)$ is not a smooth manifold because it contains
  singular points. However, after removing those, $\pi_F(S)$ becomes smooth
  (but disconnected).\\
  Note that $\pi_E(S)$ is, however, a smooth submanifold even though
  it is, by definition, a singular solution. This means that
  singular solutions might lead to singular points of $\pi_E(S)$ or $\pi_F(S)$
  but not in all cases. The next proposition answers under which conditions it does not.
}
For non-horizontal solutions, the following proposition still holds.
\bp{prop:singularSolutionTransfer}{
  If $\de{E}$ and $\de{F}$ share the intersection $\de{I}=\de{EF}\cap\Phi$
  (where $\de{EF}=\pi_E^{-1}(\de{E})\cap \pi_F^{-1}(\de{F})$)
  and $\de{I}$ has a (possibly singular) solution $S$, then $\pi_E(S)$ is a solution of $\de{E}$
  if $d\pi_E|_S$ is injective and $\pi_F(S)$ is a solution of $\de{F}$
  if $d\pi_F|_S$ is injective.
}
\pr{
  If $d\pi_E|_S$ is injective, then, since $\pi_E$ is smooth, $\pi_E|_S$ is a
  local diffeomorphism onto its image. As shown in the proof of
  proposition \ref{prop:solTransfer}, this implies that $\pi_E(S)$ is a smooth
  submanifold of $\de{E}\subset J^k(E)$.
  To show that it is a solution, it only remains to show that $\pi_E$ preserves the Cartan
  distribution, i.e. $d\pi_E(v\in \de{C}_\theta)\subset \de{C}_{\pi_E(\theta)}~\forall \theta$.
  Since $v\in \de{C}_\theta$ locally lies in the span of the vector fields
  \begin{equation}
    \begin{split}
      D_q &=
      \frac{\partial}{\partial x^q}
      + \sum_{j=1}^e\sum_{|\alpha|<k} u^j_{\alpha q} \frac{\partial}{\partial u^j_\alpha}
      + \sum_{g=1}^f\sum_{|\beta|<l} v^j_{\beta q} \frac{\partial}{\partial v^j_\beta}
      \text{ and }~D^j_\delta:=\frac{\partial}{\partial u^j_\delta},~|\delta|=k, \te{ as well as }~D^g_\kappa:=\frac{\partial}{\partial v^g_\kappa},~|\kappa|=l,
    \end{split}
  \end{equation}
  and since $\pi_E(x^i,u^j_\alpha,v^g_\beta)=(x^i,u^j_\alpha)$, 
  one obtains
  \begin{equation}
    \begin{split}
      d\pi_E = \sum_{i=1}^m\frac{\partial}{\partial x^i} \otimes dx^i
      + \sum_{j=1}^e \sum_{|\alpha|\le k} \frac{\partial}{\partial u^j_\alpha} \otimes du^j_\alpha
    \end{split}
  \end{equation}
  and consequently
  \begin{equation}
    \begin{split}
      d\pi_E(D_q)&=
      \frac{\partial}{\partial x^q}
      + \sum_{j=1}^e\sum_{|\alpha|<k} u^j_{\alpha q} \frac{\partial}{\partial u^j_\alpha},\qquad
      d\pi_E(D^j_\delta) =  D^j_\delta,\qquad d\pi_E(D^g_\kappa) = 0.
    \end{split}
  \end{equation}
  Thus, $d\pi_E(\de{C}_\theta)=\de{C}_{\pi_E(\theta)}$. As a consequence,
  since $S$ was an integral submanifold of $\de{C}$, i.e. $v \in \de{C}_\theta~\forall
  v\in T_\theta S$, and those vectors are mapped to the Cartan distribution
  of $J^k(E)$ by $d\pi_E$, it follows that $\pi_E(S)$ must also be an integral submanifold of the
  Cartan distribution. Since it is a smooth submanifold of dimension $\text{dim}(M)$,
  this implies that it is a (possibly singular) solution of $\de{E}$.
}
In particular, the above proposition yields the following corollary.
\bc{cor:singSolTransferAfterRemovingSingularPoints}{
  If $\de{E}$ and $\de{F}$ share the intersection $\de{I}=\de{EF}\cap\Phi$
  (where $\de{EF}=\pi_E^{-1}(\de{E})\cap \pi_F^{-1}(\de{F})$)
  and $\de{I}$ has a (possibly singular) solution $S$, then if
      $S_{\text{inj}(E)}:=\{\theta\in S~|~d\pi_E|_S \text{ is injective}\}$
  has dimension $\text{dim}(M)$,
  $\pi_E(S_{\text{inj}(E)})$
  is a (possibly singular) solution of $\de{E}$.
  The same holds for $\pi_F(S_{\text{inj}(F)})$.
}

\newpage
\section{Bäcklund correspondences}
\label{sec:BaecklundCorrespondencesSection}

In this section, it is shown how the present framework naturally generalizes Bäcklund
transformations which can sometimes serve to generate non-trivial solutions of non-linear PDEs.\\
Another definition of Bäcklund transformations within the
beautiful theory of coverings can be found in subsection 3.8 of
\cite{nonlocalTrends1989} and also in
subsection 1.11 of chapter 6 of \cite{Krasilshchik1999}.
However, the theory of coverings takes place in the category
of infinitely prolonged differential equations which is not
convenient in the present situation for two reasons:
First, the present setting was developed to
compare two differential equations that might not share enough structure
to be formally integrable which forces us to
stay on the level of finite jets.
Second, singular solutions are more difficult to deal with on infinite jet spaces
because the Cartan distribution becomes purely horizontal.
Therefore, a generalization of Bäcklund transformations on the level of
finite jets is useful for the present purposes.\\
As a first step, the definition of a Bäcklund transformation
described on p. 134-140 in \cite{rogers1982baecklund} is rewritten and somewhat simplified
using the present notation.
As before, let $\pi:E\to M$ be a fibered manifold, $J^k(E)$
the $k$-th order jet space over $E$ and $\xi:F\to M$ another fibered manifold
with the same base space $M$.
If $\psi:J^k(E)\times_M J^0(F)\to J^1(F)$ is a morphism of fibered manifolds,
then $p^1(\psi):J^{k+1}(E)\times_M J^1(F)\to J^1(J^1(F))$
denotes the prolongation of $\psi$.
As already explained around equation (\ref{eq:repeatedJet}),
there is a well-defined inclusion $i_{1,1}:J^2(F)\to J^1(J^1(F))$
that embeds $J^2(F)$ into $J^1(J^1(F))$.
\gd{def:BaecklundMap}{
  A \ti{Bäcklund map} is
  a morphism of fibered manifolds, $\psi:J^k(E)\times_M J^0(F)\to J^1(F)$,
  such that
  \begin{equation}
    \begin{split}
      \xi^1_0\circ\psi=\pi_2,
    \end{split}
    \label{eq:baecklundCondition}
  \end{equation}
  where $\xi^1_0:J^1(F)\to F$ and $\pi_2:J^k(E)\times J^0(F)\to J^0(F)=F$.
  The \textit{Bäcklund compatibility condition}
  \begin{equation}
    \begin{split}
      \text{im}(p^1(\psi))\subset i_{1,1}(J^2(F))
    \end{split}
    \label{eq:intCondBT}
  \end{equation}
  gives rise to a subset $\de{P}(\psi)\subset J^{k+1}(E)\times_M J^1(F)$ given by
  \begin{equation}
    \begin{split}
      \de{P}(\psi) :=
      \{ \theta \in J^{k+1}(E)\times_M J^1(F)~|~p^1(\psi)(\theta) \in i_{1,1}(J^2(F))\}
    \end{split}
    \label{eq:BaecklundB}
  \end{equation}
}
\gd{def:ordinaryBLMap}{
  If $\de{P}(\psi)$
  contains a system that only depends on the coordinates of $J^{k+1}(E)$, i.e. if one has
  \begin{equation}
    \begin{split}
      \de{E} = \pi_E'(\de{P}(\psi))
      \text{ for some PDE }\de{E}\subset J^{k+1}(E),
    \end{split}
    \label{eq:ordinaryCondition}
  \end{equation}
  where $\pi_E':J^{k+1}(E)\times_M J^1(F)\to J^{k+1}(E)$,
  then $\psi$ is called an \ti{ordinary Bäcklund map} for $\de{E}$.
}
To provide a better understanding of this definition, a brief description
of all conditions in local coordinates is given. Let the coordinates
of $J^k(E)\times_M J^0(F)$ be $(x^i,u^j_\alpha,v^g)$, $|\alpha| \le k$ and those
of $J^1(F)$ be $(x^i,w^g,w^g_{b})$, $i,b\in\{1,\cdots,m\}$.
The condition that $\psi$ is a morphism of fibered manifolds locally translates into the description
\begin{equation}
  \begin{split}
    (x^i,u^j_\alpha,v^g)\mapsto
    (x^i,w^g
    = \psi^g(x,u,v), w^g_{b} = \psi^g_{b}(x,u,v))
  \end{split}
\end{equation}
The condition (\ref{eq:baecklundCondition}) then
locally implies
\begin{equation}
  \begin{split}
    w^g = \psi^g(x,u,v)=v^g,
  \end{split}
  \label{eq:localBTcond1}
\end{equation}
and the compatibility condition (\ref{eq:intCondBT})
can locally be understood as follows.
Let $J^{k+1}(E)\times_M J^1(F)$ have local coordinates $(x^i,u^j_\alpha,v^g_\beta)$,
this time with $|\alpha|\le k+1$ and $|\beta|\le 1$
and the local coordinates of $J^1(J^1(F))$ be
$(x^i,w^g,w^g_b,(w^g)_{b},(w^g_b)_{c})$,
$b,c\in\{1,\cdots,m\}$.
Then, for $p^1(\psi):J^{k+1}\times_M J^1(F)\to J^1(J^1(F))$, one obtains
\begin{equation}
  \begin{split}
    p^1(\psi)
    \begin{pmatrix}
      x^i\\
      u^j_{\alpha}\\
      v^g_\beta
    \end{pmatrix}
    =
    \begin{pmatrix}
      x^i\\
      w^g = \psi^g(x,u,v) \ov{\ref{eq:localBTcond1}} v^g\\
      w^g_{b} = \psi^g_{b}(x,u,v))\\
      (w^g)_{b} = D_{b} \psi^g(x,u,v) \ov{\ref{eq:localBTcond1}} D_b v^g = v^g_b \\
      (w^g_b)_{c} = D_c \psi^g_b(x,u,v)
    \end{pmatrix},
  \end{split}
\end{equation}
where $D_b$, as before, is the total differential operator.
\begin{equation}
  \begin{split}
    D_b = \frac{\partial}{\partial x^b} + \sum_{j=1}^e \sum_{|\alpha|<k+1} u^j_{\alpha b} \frac{\partial}{\partial u^j_\alpha}
    +\sum_{g=1}^f \sum_{|\beta|<1} v^g_{\beta b} \frac{\partial}{\partial v^g_\beta}
  \end{split}
\end{equation}
Since the subset $i_{1,1}(J^2(F))$ in $J^1(J^1(F))$ has local coordinates
$(x^i,w^g,w^g_b,(w^g)_{b}=w^g_b,(w^g_b)_{c}=w^g_{bc}=w^g_{cb}=(w^g_c)_b)$,
the local equations describing $\de{B}$ defined in (\ref{eq:BaecklundB})
by the compatibility condition (\ref{eq:intCondBT}) are finally given by
\begin{equation}
  \begin{split}
    \de{P}(\psi):~\{~ v^g_b = \psi^g_b,\quad D_c\psi^g_b=D_b\psi^g_c ~\}.
  \end{split}
  \label{eq:localB}
\end{equation}
This concludes the descriptions of the local coordinates involved
in the definition of a Bäcklund map.

\bigskip

The next step is to use a Bäcklund map to define a Bäcklund transformation.
To this end, note first that, since the restriction of $p^1(\psi)$
to $\de{P}(\psi)$ by construction has an image
that lies in $J^2(F)$, one can define a map
$\psi^1:\de{P}(\psi) \to J^2(F)$, simply given
by $\psi^1(\theta):=p^1(\psi)|_{\de{P}(\psi)}(\theta)$.
This procedure can be iterated to obtain a map $\psi^r:P^{r-1}(\de{P}(\psi))\to J^{r+1}(F)$
where $P^{r-1}(\de{P}(\psi))$ is the $r-1$-th prolongation of $\de{P}(\psi)$.
\gd{def:BaecklundTransformation}{
  If $\psi:J^k(E)\times_M J^0(F)\to J^1(F)$
  is an ordinary Bäcklund map for $\de{E}$
  and if, for some $r$,
  a system of differential equations $\de{F} \subset J^{r+1}(F)$ contains
  the image of $\psi^r:\de{P}(\psi)^{r-1}\to J^{r+1}(F)$, then $\psi$ is called
  a \ti{Bäcklund transformation} between $\de{E}$ and $\de{F}$.
}
The idea behind those definitions is
to reduce the equations locally describing $\de{F}$ to first order equations
with the help of $\de{E}$ and $\psi$. One usually obtains
the following proposition that is reproven in the present terminology, for convenience.
\bp{prop:BaecklundTReduction}{
  Suppose that $\psi$ is a Bäcklund transformation between
  $\de{E}\subset J^{k+1}(E)$ and $\de{F}\subset J^{r+1}(F)$.
  If $s_E:U\subset M \to E$ is a horizontal solution of $\de{E}$
  (i.e. $\text{im}(j^{k+1}(s_E))\subset \de{E}$),
  then a solution $s_F:U\subset M\to F$ of 
  $~\de{F}$ (with $\text{im}(j^{r+1}(s_F))\subset \de{F}$)
  can be obtained by solving the following system of PDEs
  \begin{equation}
    \begin{split}
      j^1(s_F)=\psi(j^k(s_E)\times_M s_F)
    \end{split}
  \end{equation}
  which is first-order in $s_F$ (recall that $s_E$ is already given) and locally described by
  \begin{equation}
    \begin{split}
      \frac{\partial s_F^g(x)}{\partial x^b} =
      \psi_b^g \br{x^i,\partial_\alpha s_E^j(x),s_F^{h}(x)},
    \end{split}
    \label{eq:firstOrderSystem}
  \end{equation}
  where $i,q\in\{1,\cdots,m\}$, $j\in\{1,\cdots,e\}$, $g,h\in\{1,\cdots,f\}$
  and $0 \le |\alpha|\le k+1$.
}
\pr{
  A horizontal solution of $\de{P}(\psi)$ is
  described by a section $s = s_E \times_M s_F : U\subset M\to E\times_M F$
  such that
  \begin{equation}
    \begin{split}
      \te{im}(j^{k+1}(s_E) \times_M j^1(s_F))
      \subset \de{P}(\psi).
    \end{split}
    \label{eq:solB}
  \end{equation}
  Since $s$ is assumed to be smooth, (\ref{eq:solB}) holds if
  \begin{equation}
    \begin{split}
      \text{im}(j^{k+r}(s_E)\times_M j^{r}(s_F))\subset P^{r-1}(\de{P}(\psi)).
    \end{split}
    \label{eq:solB2}
  \end{equation}
  Since by assumption $\psi^r(P^{r-1}(\de{P}(\psi)))\subset \de{F}$,
  (\ref{eq:solB2}) in turn implies
  $\text{im}(\psi^r(j^{k+r}(s_E)\times_M j^{r}(s_F)))
  \subset \de{F}$. At the same time,
  \begin{equation}
    \begin{split}
      \psi^r(j^{k+r}(s_E)\times_M j^r(s_F))
      &= p^{r}(\psi)(j^{k+r}(s_E)\times_M j^r(s_F))\\
      &\ov{\ref{eq:prolDiffOp}}
      j^r(\psi(j^k(s_E)\times_M s_F))
    \end{split}
  \end{equation}
  Thus, if $s=s_E\times_M s_F$ is a solution of $\de{P}(\psi)$ and one can find a section $s_{F'}:U\subset M\to F$ such that
  \begin{equation}
    \begin{split}
      j^1(s_{F'})=\psi(j^k(s_E)\times_M s_F),
    \end{split}
    \label{eq.w}
  \end{equation}
  then $s_{F'}$ is a solution of $\de{F}$.
  Since $\pi_2=\xi^1_0 \circ \psi$ by (\ref{eq:baecklundCondition}), we also have
  \begin{equation}
    \begin{split}
      s_F = \pi_2(j^k(s_E)\times_M s_F) =
      \xi^1_0(\psi(j^k(s_E)\times_M s_F)),
    \end{split}
    \label{eq:firstorder}
  \end{equation}
  and since $s_F$ is holonomic, this implies
  \begin{equation}
    \begin{split}
      j^1(s_F)=\psi(j^k(s_E)\times_M s_F)
    \end{split}
    \label{eq:w2}
  \end{equation}
  In other words, $s_{F'}=s_F$ always solves (\ref{eq.w}).
  As a conclusion, whenever $s=s_E\times_M s_F$ solves $\de{P}(\psi)$,
  then $s_F$ itself is such that it solves $\de{F}$.\\
  Hence, if a solution $s_E$ of $\de{E}$ is given,
  a solution $s=s_E\times_M s_F$
  of $\de{P}(\psi)$ can be found by finding $s_F$ s.t. (\ref{eq:solB}) holds.
  As we also assume that $\psi$ is ordinary for $\de{E}$,
  equation (\ref{eq:ordinaryCondition}) holds,
  which implies that $\pi_E^{-1}(\text{im}(s_E))$
  contains the image of a section $s=s_E\times_M s_F$
  which is contained in $\de{P}(\psi)$.
  Therefore
  given a solution $s_E$, we get $s_F$ by solving the remaining equation
  describing $\de{P}(\psi)$, $v^g_b=\psi^g_b$ (cf. (\ref{eq:localB})) that is eq.
  (\ref{eq:w2}), which in local coordinates
  is described by the system (\ref{eq:firstOrderSystem}).
}
As a next step, Bäcklund transformations are identified as a special case
of the present framework.
\bp{prop:GeneralizationOfBäcklundMap}{
  Every Bäcklund transformation $\psi: J_B:=J^k(E)\times_M J^0(F)\to J^1(F)$
  between $\de{E}\subset J^{k+1}(E)$ and $\de{F}\subset J^{r+1}(F)$
  gives rise to an intersection $\de{I}=\de{EF}\cap \Phi$
  where $\de{EF}$ is constructed as in eq. (\ref{eq:EF})
  on the natural product bundle $J:=J^{k+1}(E)\times_M J^{r+1}(F)$ of $\de{E}$ and $\de{F}$ and $\Phi$ is completely determined by $\psi$.\\
  $\Phi$ fulfills a condition equivalent to (\ref{eq:baecklundCondition}) and
  the projection of the prolongation $\pi^{k+2,~r+2}_{k+1,~1}(P^1(\Phi))$ corresponds to the compatibility condition 
  $\de{P}(\psi)$ defined in (\ref{eq:BaecklundB}).
}
\pr{
  As before, given $\de{E}\subset J^{k+1}(E)$ and $\de{F}\subset J^{r+1}(F)$,
  one can form the natural product bundle
  $J:=J^{k+1}(E)\times_M J^{r+1}(F)$
  and pull $\de{E}$ and $\de{F}$ back to $\de{E}_J$ and $\de{F}_J$
  via $\pi_E:J\to J^{k+1}(E)$ and $\pi_F:J\to J^{r+1}(F)$,
  i.e. $\de{EF}:=\pi_E^{-1}(\de{E})\cap\pi_F^{-1}(\de{F})$ as in eq. (\ref{eq:EF}).
  Next, one can define a correspondence $\Phi$ as follows
  \begin{equation}
    \begin{split}
      \Phi :~\{ \xi^{r+1}_1 \circ \pi_F = \psi \circ \pi_B \}
    \end{split}
    \label{eq:BaecklundCorrespondence}
  \end{equation}
  where $\xi^{r+1}_1:J^{r+1}(F) \to J^1(F)$
  and $\pi_B: J\to J_B$.
  \\
  Recall that $\pi_2:J_B\to J^0(F)$. Since
  $\pi_2 \circ \pi_B = \xi^{r+1}_0\circ\pi_F$ and
  $\xi^1_0\circ \xi^{r+1}_1=\xi^{r+1}_0$,
  applying $\xi^1_0$ to both sides
  of the equation (\ref{eq:BaecklundCorrespondence}) defining $\Phi$
  results in
  \begin{equation}
    \begin{split}
      \pi_2 \circ \pi_B = \xi^{r+1}_0\circ \pi_F
      = \xi^1_0 \circ \xi^{r+1}_1 \circ \pi_F \ov{\ref{eq:BaecklundCorrespondence}}
      \xi^{1}_0\circ\psi\circ \pi_B,
    \end{split}
  \end{equation}
  which is equivalent to condition (\ref{eq:baecklundCondition})
  but this time imposed on $\Phi$ on $J$ instead of on $\psi$ on $J_B$.
  Note that the condition here is trivially fulfilled
  because we are only considering a submanifold $\Phi$ on one product bundle with one set
  of coordinates $(x,u,v)$ instead of a morphism $\psi$ between two different
  fibered manifolds with two different sets of coordinates
  $(x,u,v)$ and $(x,w)$ that required the additional condition
  $v=w$. This is an indication that the present approach is more natural.\\
  If the coordinates of $J$ are $(x^i,u^j_\alpha,v^g_\beta)$
  and of $J^1(F)$ are $(x^i,w^g,w^g_b)$,
  then $\psi\circ \pi_B$ and $\xi^{r+1}_1 \circ \pi_F$ are locally given by
  \begin{equation}
    \begin{split}
      \begin{pmatrix}
        x^i \\
        v^g \\
        v^g_b
      \end{pmatrix} = \xi^{r+1}_1(\pi_F(x^i,u^j_\sigma,v^g_\lambda))
      \ov{\ref{eq:BaecklundCorrespondence}}
      \psi(\pi_B(x^i,u^j_\sigma,v^g_\lambda)) =
      \begin{pmatrix}
        x^i \\
        \psi^g(x^i,u^j_\delta,v^g) \\
        \psi^g_b(x^i,u^j_\delta,v^g)
      \end{pmatrix}
    \end{split}
    \label{eq:BäcklundEquationsPhi}
  \end{equation}
  which correspond to the equations described in eq. (\ref{eq:localBTcond1}) and the left equation in \eqref{eq:localB}. (Note that, in the eq. above, $|\sigma|\le k+1$, $|\lambda|\le r+1$ but $|\delta|\le k$.) \\
  Condition \eqref{eq:intCondBT} is a projected version of the compatibility condition that is enforced by the intersection in the definition of a prolongation, cf. eq. \eqref{eq:prolSubmfdJetSpace},
  \begin{equation}
    \begin{split}
      P^1(\Phi)=p(~J^1(\Phi)~\cap~ (i_{k+1,1}\times_M i_{r+1,1}(P^1(J)))~),
    \end{split}
  \end{equation}
  where 
  \begin{equation}
    \begin{split}
      P^n(J)=J^{k+1+n}(E)\times_M J^{r+1+n}(F).
    \end{split}
    \label{eq:prolongJ}
  \end{equation}
  Indeed, by eq. \eqref{eq:prolongRk} and eq. \eqref{eq:BäcklundEquationsPhi},
  \begin{equation}
    \begin{split}
      P^1(\Phi) &: \bbr{ \theta \in P^1(J)~\big|~D_b \psi^g(\kappa)=v^g_b=\psi^g_b(\kappa),~D_a\psi^g_b(\kappa) = v^g_{ab}= D_b\psi^g_a(\kappa)}
    \end{split}
    \label{eq:P1phi}
  \end{equation}
  where $\kappa=\pi^{k+2,~r+2}_{k+1,~1}(\theta)$ and $\pi^{a,~b}_{c,~d}:J^{a}(E)\times_M J^b(F)\to J^c(E)\times_M J^d(F)$ is the canonical projection. Since $\kappa \in J^{k+1}(E)\times_M J^1(F)$, those equations (apart from the condition $v^g_{ab}=D_b\psi^g_a$) are preserved under projection, and one obtains 
  \begin{equation}
    \begin{split}
      \pi^{k+2,~r+2}_{k+1,~1}(P^1(\Phi))&=\bbr{\theta \in J^{k+1}(E)\times_M J^1(F)~|~v^g_b=\psi^g_b,~D_a\psi^g_b= D_b\psi^g_a} \\
      &\ov{\ref{eq:localB}} \de{P}(\psi)
    \end{split}
    \label{eq:PpsiPphi}
  \end{equation}
  As a result, the compatibility conditions of a Bäcklund map can be understood as the equations arising upon prolongation of the correspondence $\Phi$.
  }
  Prolonging \eqref{eq:PpsiPphi}, one obtains 
  \begin{equation}
    \begin{split}
      \pi^{k+1+n,~r+1+n}_{k+n,~n}(P^n(\Phi))&=P^{n-1}(\de{P}(\psi))
    \end{split}
    \label{eq:prolPpsiPphi}
  \end{equation}
\bp{prop:GeneralizationOfBäcklundTransformations}{
  $\de{I}$ as defined in proposition \ref{prop:GeneralizationOfBäcklundMap}
  allows to transfer solutions from $\de{E}$ to $\de{F}$ in the sense of proposition \ref{prop:BaecklundTReduction}. 
}
\pr{
  To show that $\Phi$ facilitates to transfer solutions from $\de{E}$ to $\de{F}$ by solving a first-order system, one can proceed as follows.
  By proposition \ref{prop:solTransfer}, we know that any solution $S$ of $\de{I}=\de{EF}\cap \Phi$ can be projected to solutions $\pi_E(S)$ and $\pi_F(S)$ of $\de{E}$ and $\de{F}$ respectively. What's special about Bäcklund transformations, is that solving $\Phi$ alone is actually sufficient. The reason is that the differential consequences of $\Phi$ contain the equations describing $\de{E}$ and $\de{F}$. To show that, we will show that $\pi_J^r(P^r(\Phi))\subset \de{EF}$ where $\pi^r_J:P^r(J)\to J$ (and $P^r(J)$ is given by eq. \eqref{eq:prolongJ}).\\
  Since $\psi$ is a Bäcklund transformation between $\de{E}$ and $\de{F}$, definition
  \ref{def:ordinaryBLMap} holds, i.e. $\pi_E'(\de{P}(\psi))=\de{E}$. Recall that $\pi_E=\pi_E'\circ \pi^{k+1,~r+1}_{k+1,~1}$. Then $\pi_E'(\de{P}(\psi))=\de{E}$ implies that
  \begin{equation}
    \begin{split}
      \pi_E^{-1}(\pi_E'(\de{P}(\psi))) &\subset \pi_E^{-1}(\de{E}) = \de{E}_J \quad\text{ where } \\
      \pi_E^{-1}(\pi_E'(\de{P}(\psi)))
      &= (\pi^{k+1,~r+1}_{k+1,~1})^{-1}\circ (\pi_E')^{-1}\circ \pi_E' (\de{P}(\psi))\\
      & = (\pi^{k+1,~r+1}_{k+1,~1})^{-1}(\de{P}(\psi))\\
      &\ov{\ref{eq:PpsiPphi}}(\pi^{k+1,~r+1}_{k+1,~1})^{-1}\big(\pi^{k+2,~r+2}_{k+1,~1}(P^1(\Phi))\big)
    \end{split}
    \label{eq:BcEJ}
  \end{equation}
  Note also that apart from $v^g_{ab}=D_a\psi^g_b$, the eqs describing $P^1(\Phi)$ are first order in $v$ (cf. eq. \eqref{eq:P1phi} and \eqref{eq:PpsiPphi}). Therefore,
  \begin{equation}
    \begin{split}
      (\pi^{k+1,~r+1}_{k+1,~1})^{-1}\big(\pi^{k+2,~r+2}_{k+1,~1}(P^1(\Phi))\big) ~\cap~\bbr{v^g_{ab}=D_a\psi^g_b} &=
      \pi^{k+2,~r+2}_{k+1,~r+1}(P^1(\Phi)) \\
      & = \pi_J^1(P^1(\Phi)) 
    \end{split}
  \end{equation}
  and thus
  \begin{equation}
    \begin{split}
      \pi^1_J(P^1(\Phi))\subset \big(~\de{E}_J~\cap~\bbr{v^g_{ab}=D_a\psi^g_b}\big) \subset \de{E}_J
    \end{split}
  \end{equation}
  Since projections of further prolongations can only increase the number of constraints / equations, we can conclude 
  \begin{equation}
    \begin{split}
      \pi^n_J(P^n(\Phi))\subset \pi^1_J(P^1(\Phi))  \subset \de{E}_J,\qquad \forall ~n\ge 1.
    \end{split}
    \label{eq:containedInEJ}
  \end{equation}
  Next, we want to show that we also have $\pi^r_J(P^r(\Phi))\subset \de{F}_J$. To do so, we use 
  definition \ref{def:BaecklundTransformation} that guarantees that a Bäcklund transformation satisfies
  $\text{im}(\psi^r)\subset \de{F}$ which implies
  \begin{equation}
    \begin{split}
      \pi_F^{-1}(\text{im}(\psi^r))&\subset \pi_F^{-1}(\de{F}) = \de{F}_J
    \end{split}
    \label{eq:contain42}
  \end{equation}
  where
  \begin{equation}
    \begin{split}
      \text{im}(\psi^r) &= \text{im}\left(p^r(\psi)|_{P^{r-1}(\de{P}(\psi))}\right) = p^r(\psi)(P^{r-1}(\de{P}(\psi)) \\
      &\ov{\ref{eq:prolPpsiPphi}} p^r(\psi)(\pi^{k+1+r,~r+1+r}_{k+r,~r}(P^r(\Phi)))\\
      &= p^r(\psi\circ \pi^{k+1,~r+1}_{k,~0})(P^r(\Phi))\\
      &\ov{\ref{eq:BaecklundCorrespondence}} p^r(\xi^{r+1}_1\circ \pi_F)(P^r(\Phi)) = \xi^{r+1+r}_{1+r}\circ p^r(\pi_F)(P^r(\Phi))
    \end{split}
    \label{eq:con42}
  \end{equation}
  Since
  \begin{equation}
    \begin{split}
      \bt J^{k+1+r}(E)\times_M J^{r+1+r}(F) \ar{r}{p^r(\pi_F)}\& J^{r+1+r}(F)\ar{r}{\xi^{r+1+r}_{1+r}}\& J^{r+1}(F) \et
    \end{split}
  \end{equation}
  commutes with 
  \begin{equation}
    \begin{split}
      \bt J^{k+1+r}(E)\times_M J^{r+1+r}(F) \ar{r}{\pi_J^r}\& J^{k+1}\times_M J^{r+1}(F)\ar{r}{\pi_F}\& J^{r+1}(F) \et,
    \end{split}
  \end{equation}
  we obtain 
  \begin{equation}
    \begin{split}
      \text{im}(\psi^r)&\ov{\ref{eq:con42}}\pi_F\circ\pi^r_J(P^r(\Phi))\\
      \Rightarrow \pi^r_J(P^r(\Phi)) &~\overset{(\ref{eq:contain42})}{\subset} F_J
    \end{split}
  \end{equation}
  Together with \eqref{eq:containedInEJ}, we thus finally obtain 
  \begin{equation}
    \begin{split}
      \pi^r_J(P^r(\Phi))\subset \de{EF}:=\de{E}_J\cap\de{F}_J
    \end{split}
    \label{eq:baecklundEssentialProperty}
  \end{equation}
  which expresses the essential property of a Bäcklund transformation:
  The equations $\de{EF}$ are differential consequences of $\Phi$.
  (Since $\pi^r_J(P^r(\Phi))$, which includes the differential consequence of $\Phi$ up to order $k+1$ in $u$ and $r+1$ in $v$, is contained in $\de{EF}$,
    the equations that locally describe $\de{EF}$ are in turn a subset of
  the equations of the (smaller) space $\pi^r_J(P^r(\Phi))$).\\
  Since we assume that solutions are smooth, every solution of $\Phi$ must also be a solution of any prolongation $P^n(\Phi)$. Since the prolongation $P^r(\Phi)$ of $\Phi$ contains both, the equations describing $\de{E}$ and those describing $\de{F}$, the solution of $\Phi$ must also solve $\de{E}$ and $\de{F}$. 
  By proposition \ref{prop:solTransfer}, solutions of $\de{I}$ can be projected
  to solutions of $\de{E}$ and $\de{F}$ via $\pi_E$ and $\pi_F$
  (even singular ones if the conditions in prop. \ref{prop:singularSolutionTransfer}
  are fulfilled).\\
  As a final step, let us show that solving a first-order system is sufficient if a general solution to $\de{E}$ is given.
  Suppose that $\de{E}$ has a general family of solutions $S_E^\alpha$, parameterized by $\alpha$, that is
  locally described by sections $s_E^\alpha$.
  Since $\pi_J^r(P^r(\Phi))\subset\de{E}_J$, the pullback of the family of solutions $\pi_E^{-1}(\text{im}(s_E^\alpha))\subset \de{E}_J$
  should intersect solutions of $P^r(\Phi)$
  that can be found by looking for a section $s_F$
  such that the prolongation of $s_E^\alpha \times_M s_F$ is contained in $\Phi$ for some $\alpha$. The resulting system of equations
  is then first order in $s_F$, namely locally described by (\ref{eq:BäcklundEquationsPhi}).
  This solution can then be mapped to $\de{F}$ as explained above,
  by prop. \ref{prop:solTransfer}.
  Hence, given $s_E$, it suffices to solve the first order PDE \eqref{eq:firstOrderSystem}
  to obtain a solution of $\de{F}$
  which concludes an alternative proof of proposition \ref{prop:BaecklundTReduction}
  in a more general setting.
}
The proof above makes it clear that the exact form of $\Phi$
is not really essential for transferring solutions and reducing
the order of equations as long as $\de{E}_J$ and $\de{F}_J$
are differential consequences of $\Phi$, i.e.
as long as (\ref{eq:baecklundEssentialProperty}) is satisfied. In particular,
staying in the natural product bundle makes it unnecessary to impose conditions
like (\ref{eq:baecklundCondition}) or to require that the codomain
of $\psi$ is a first order jet space. Thus, the following
generalization seems appropriate.
\gd{def:BaecklundCorrespondence}{
  A correspondence $\Phi \subset J:=J^{k}(E)\times_M J^l(F)$ between
  two differential equations $\de{E}\subset J^k(E)$ and $\de{F}\subset J^l(F)$
  is said to be a \textit{Bäcklund correspondence} or to \textit{have the Bäcklund
  property} if, for some $r\ge 1$,
  \begin{equation}
    \begin{split}
      \de{P}(\Phi):=\pi_J^r(P^r(\Phi))\subset \de{EF}.
    \end{split}
    \label{eq:mild}
  \end{equation}
  where $\de{P}(\Phi)$ is the projection of the prolongation of $\Phi$.
  It is said to be a \textit{strict Bäcklund correspondence} if
  \begin{equation}
    \begin{split}
      \pi_E(\de{P}(\Phi))=\de{E}\text{ and }\pi_F(\de{P}(\Phi))=\de{F}.
    \end{split}
    \label{eq:strict}
  \end{equation}
}
Note that (\ref{eq:strict}) implies (\ref{eq:mild}) because
\begin{equation}
  \begin{split}
  (\ref{eq:strict})~&\Rightarrow~\de{P}(\Phi)\subset
  \begin{cases}
    \pi_E^{-1}(\pi_E(\de{P}(\Phi))) = \pi_E^{-1}(\de{E})=\de{E}_J\\
    \pi_F^{-1}(\pi_F(\de{P}(\Phi))) = \pi_F^{-1}(\de{E})=\de{F}_J
  \end{cases}\\
  &~\\
  &\Rightarrow~\de{P}(\Phi)\subset \de{E}_J\cap\de{F}_J=\de{EF}.
  \end{split}
\end{equation}
With this definition, the following proposition holds.
\bp{prop:BaecklundGeneralization}{
  Whenever a correspondence $\Phi$ between two PDEs $\de{E}$ and $\de{F}$
  is Bäcklund, every solution of $\Phi$ is a solution of both,
  $\de{E}$ and $\de{F}$.
}
\pr{
  Since $\de{P}(\Phi) \subset \de{EF}$, and, since $\de{P}(\Phi):=\pi_J^r(P^r(\Phi))$, also $\de{P}(\Phi) \subset \Phi$,
  we obtain $\de{P}(\Phi) \subset \de{I}=\de{EF}\cap \Phi$.
  Hence, what solves $\de{P}(\Phi)$ also solves $\de{I}$.
  But $\de{P}(\Phi)$ is solved when $\Phi$ is solved because $\de{P}(\Phi)$ is the projection of differential consequences of $\Phi$. Hence, a solution of $\Phi$ solves $\de{I}$ and then this solution can be mapped to $\de{E}$ and $\de{F}$ by proposition \ref{prop:solTransfer}.
}
Again, given the solution of one of the equation might allow to reduce the order of the other:
\bp{prop:baecklundTransfer}{
  If $\Phi$ is a strict Bäcklund correspondence between $\de{E}$ and $\de{F}$,
  and a solution $S$ of $\de{E}$ is given, then a solution of $\de{F}$
  can be found by finding a solution of
  $\pi_E^{-1}(S)\cap \Phi$.
}
\pr{
  Since $\Phi$ is a strict Bäcklund correspondence, one
  has $\pi_E(\de{P}(\Phi))=\de{E}$. This means that, apart from $\de{E}_J$, the prolongation of $\Phi$ does not impose additional equations, purely in terms of coordinates of $J^k(E)$, on $J$. 
  Hence $\pi_E^{-1}(S)=S\times_M J^l(F)$ intersects the solution space of $\Phi$.
  If a solution in this intersection can be found, it also solves $\de{F}$
  by proposition \ref{prop:BaecklundGeneralization}.
}
The present approach
generalizes the usual definition of a Bäcklund transformation
because one can now define a correspondence of any order
and the dependence on the coordinates of $J^l(F)$ can be arbitrary
apart from the requirement that $\Phi$ should be
an almost diagonal fibered submanifold of $J^k(E)\times_M J^l(F)$. Despite the increased generality, solutions can still be transferred in a similar way to the simpler case.

\ye{ex:liouville}{
  A very classical example that illustrates Bäcklund transformations is the
  one involving the Liouville equation $u_{12}=\exp(u)$.
  It is briefly rephrased in the present terminology to illustrate the general ideas above.
  Consider $\pi:E:=\mathbb{R}\times\mathbb{R}\to\mathbb{R}=:M$ with local coordinates
  $(x,y,u)$ and $\xi:F\simeq E\to M$ with local coordinates $(x,y,v)$ and the equations
  $\mathcal{E}:~\{~u_{12}=e^u~\}\subset J^2(E)$ and $\mathcal{F}:~\{~v_{12}=0~\} \subset J^2(F)$.
  We relate them on $J^2(E)\times_MJ^2(F)$ by a correspondence $\Phi$ determined by the equations
  \begin{equation}
    \begin{split}
      \Phi:\eq{
        v_1 = u_1 + \beta \exp\br{ \frac{u+v}{2} }, &\quad
        v_2 = -u_2 - \frac{2}{\beta}\exp\br{ \frac{u-v}{2} }
  }
    \end{split}
    \label{eq:implicitCorrBäcklundLiouville}
  \end{equation}
  First, we check that this $\Phi$ is indeed a correspondence. Since it is defined by two independent equations, it has codimension $2\ge 1=\text{min}(\text{dim}(E_x,F_x))$.
  Its projection to both, $J^2(E)$ and $J^2(F)$ does not impose any conditions
  and thus, it is almost diagonal to $\de{E}$ and $\de{F}$.
  The prolongation $P^1(\Phi)$ of $\Phi$ is described by the equations
  describing $\Phi$ and additionally by the following ones.
  \begin{equation}
    \begin{split}
      \eq{
        v_{11} = u_{11} + \beta \exp\br{ \frac{u+v}{2} } \frac{u_1+v_1}{2} \\
        v_{12} = u_{12} + \beta \exp\br{ \frac{u+v}{2} } \frac{u_2+v_2}{2} \\
        v_{21} = -u_{21} - \frac{2}{\beta}\exp\br{ \frac{u-v}{2} }\frac{u_1-v_1}{2} \\
        v_{22} = -u_{22} - \frac{2}{\beta}\exp\br{ \frac{u-v}{2} }\frac{u_2-v_2}{2}
      }
    \end{split}
    \label{eq:prolongLiouville}
  \end{equation}
  The compatibility conditions $v_{12}=v_{21}$ and $u_{12}=u_{21}$,
  that must be imposed when taking the prolongation, result in the following two equations, which,
  together with eq. (\ref{eq:implicitCorrBäcklundLiouville}) and (\ref{eq:prolongLiouville})
  describe $\de{P}(\Phi)$:
  \begin{equation}
    \begin{split}
      \eq{ u_{12}=\exp(u),\qquad v_{12}=0 }
    \end{split}
  \end{equation}
  As can be seen, the correspondence was designed such that its differential consequences are included in both of the intersected equations, i.e.
  \begin{equation}
    \begin{split}
      \de{P}(\Phi) \subset \de{EF}\cap\Phi=:\de{I} \subset \de{EF},
    \end{split}
  \end{equation}
  Thus, $\Phi$ is a Bäcklund correspondence.
  Furthermore, since $\Phi$ is almost diagonal, we obtain $\pi_E(\de{P}(\Phi))=\de{E}$ and
  $\pi_F(\de{P}(\Phi))=\de{F}$. Hence, $\Phi$ is a strict
  Bäcklund correspondence.\\
  Thus, by proposition \ref{prop:baecklundTransfer},
  solutions an be transferred between the PDEs.
  The general solution of $v_{12}=0$ is given by $v(x,y)=A(x)+B(y)$
  and plugging this into (\ref{eq:implicitCorrBäcklundLiouville}) results in a PDE for $u(x,y)$
  that can be integrated
  (though it is not completely trivial), and one obtains the solution
  \begin{equation}
    \begin{split}
      u(x,y)=2\ln\br{ \frac{\exp\br{ \frac{A(x)-B(y)}{2}} }{ \frac{\beta}{2}\int_{x_0}^x \exp\br{A(x')}dx'+ \frac{1}{\beta}\int_{y_0}^y \exp\br{-B(y')}dy'} }
    \end{split}
  \end{equation}
  As mentioned by \cite{rogers1982baecklund},
  this encouraging result was an important motivation for the search of
  Bäcklund transformations.
}

\newpage
\section{Equivalence up to symmetry and quotient equations}
\label{sec:equivUpToSym}

When comparing two theories in mathematically different formulations
that only differ up to a symmetry which is physically not relevant,
then one would like to find a way to compare the two theories
after removing this symmetry.
For example, classical electrodynamics can be formulated in terms
of gauge potentials and in terms of Faraday tensors.
At least classically, those two theories are physically equivalent
because only the fields are measurable quantities.
To formalise this physical equivalence mathematically,
Weatherall
invented
the solution-Category approach described in \cite{weather2014} and \cite{weather2015}
which was already mentioned in the introduction \ref{sec:previousAttempts}.
The idea behind this formalism was, among other things, to show
that those mathematical structures in which the morphisms
between the objects of the solution categories are induced (via the pushforward or pullback)
by the diffeomorphisms of the underlying manifold
are more natural than those in which those symmetries have to be ``added by hand''
in order to achieve an equivalence to other physically equivalent formulations.\\
The aim of the present section is to show how one can approach those ideas
in the category of smooth manifolds.\\
The section describes the general idea how to ``quotient out'' a symmetry of an equation
and how to obtain the corresponding invariant equation.
Basically, the invariant equation is realised by replacing the variables in the equation by
the invariants of the symmetry. So the real work consists in finding all functionally independent invariants.
Though the present approach was developed somewhat independently,
quotient equations are a well-known concept
(cf. \cite{Krasilshchik1999} (chapter 3.6), \cite{svinolupovSokolov1992},
  \cite{kruglikov2015global}, \cite{schneider2020solutions}, also \cite{Valiquette2015}
is related).
\\~\\
We start with
the geometric definition of a symmetry of a PDE (taken from \cite{Krasilshchik1999})
\gd{def:lieTrf}{
  A \ti{Lie transformation} is
  a diffeomorphism  $L: J^k (E) \ra J^k (E)$
  such that $dL_{\theta} (\mathcal{C}_\theta) = C_{L(\theta)}~\forall \theta\in J^k (E)$
  (where $\de{C}$ is the Cartan distribution on $J^k(E)$).
    A vector field $X$ on the manifold $J^k(E)$ is called a \ti{Lie field},
    if shifts along its flow
    are Lie transformations.
}
\gd{def:symm}{
  A Lie transformation $S$
  which is such that $S(\mathcal{E}) = \mathcal{E}$
  is called a \ti{symmetry} of the differential equation $\mathcal{E} \subset J^k(E)$.
  A Lie field $X$ is called an \ti{infinitesimal symmetry} of the
    equation $\mathcal{E}\subset J^k(E)$, if it is tangent to $\mathcal{E}$.
}
Having defined Symmetries, we can proceed to define the concept of an invariant of a symmetry
(taken from \cite{collon2012}).
\gd{def:invariant}{
  Given a Lie transformation $S$ on $J^k(E)$, an \ti{invariant} of this transformation
  is a map $I:J^k(E)\ra \mathbb{R}$ such
  that $S^*I=I$, i.e. $I(\theta)=I(S(\theta))~\forall\theta\in J^k(E)$.
}
Now suppose that $S$ is a symmetry of the equation $\mathcal{E}$,
i.e. $S(\mathcal{E})=\mathcal{E}$.
If the
equation is given as the kernel of a differential operator $\Phi:J\subset J^k(E)\ra F$,
where $\pi':F\ra M$ is another fibered manifold,
i.e. $\de{E}=\ker_s(\Phi)$, where $s:M\ra F$ is a suitable section, then this implies that
$\Phi(\theta)=s(\pi^k(\theta))$ iff $S^*\Phi(\theta)=\Phi(S(\theta))=s(\pi^k(S(\theta)))$.\\
Observe that $\Phi$ itself does not have to be invariant but the condition $\Phi(\theta)=s(\pi^k(\theta))$
only holds for $\theta\in \de{E}$ which is invariant.
But this means that it should be possible to perform algebraic operations on the equation
$\ker_s(\Phi)$ which facilitate to
reformulate the equation in terms of
invariants of the symmetry, at least at all those points where those algebraic operations are well-defined.
In other words,
it should be possible to
find a $\Phi':J'\subset J^k(E)\ra F$ such that $\de{E}=\ker_{s'}(\Phi')$
and $S^*\Phi'=\Phi'$, at least at all those points making up $J'$
where the algebraic operations on $\ker_s(\Phi)$ do not
lead to a division by zero.\\
To find out how to find this $\Phi'$, let us suppose that we have a Lie group $G$ that acts on $J^k(E)$.
We write this action as $g\cdot \theta:=S_g(\theta)$ where $S_g:J^k(E)\ra J^k(E)$ is the
Symmetry on our bundle corresponding to the action of $g\in G$.
Given such a symmetry group, we can try to find
the generating functions of all $S_g$-Invariants on $J^k(E)$. They can be found in a
systematic way using the following proposition (also taken from \cite{collon2012}):
\bp{prop:1}{
  If $G$ is a group of symmetries acting on $J^k(E)$,
  then all invariants $I$ of this symmetry group fulfill the equations
  \begin{equation}
    \boxed{
      X(I)=0
    }
    \label{eq:invariantFinder}
  \end{equation}
  where $X$ are the infinitesimal symmetries corresponding to the action of the Lie algebra of $G$.
}
\pr{
  For an invariant $I$ of a group it is true by definition that
  $S^*_gI=I,~\forall g\in G$. As we assume a Lie group, we can write $S_g=\exp(a X_g)$
  where $X_g$ is the infinitesimal generator corresponding to the action of $g$. Thus,
  \begin{equation}
    \begin{split}
    0&=\frac{d}{da}I(\theta)\bigg|_{a=0}
    =\frac{d}{da}I(S_g(\theta))\bigg|_{a=0}\\
    &=\frac{d}{da}I(\exp(aX_g)\theta)\bigg|_{a=0}
    = I'(\theta) X_g|_\theta = X_g(I).
    \end{split}
    \label{eq:invariantFinderProof}
  \end{equation}
  This is true for all $g$ and thus for all $X$ in the Lie algebra.
}
This means that if we have a finite number of generators for our symmetry group,
then it becomes possible to find all functionally independent invariants by finding
the most general solution of a finite number of equations of the form (\ref{eq:invariantFinder}).
\\
Now suppose we have found out that any invariant of a given group action on a given bundle must be a function of the
functionally independent invariants $(I_1,\cdots,I_r)$. Furthermore, suppose that
the equation $\mathcal{E}$ on $J^k(E)$ is also invariant under the group action. Then, as explained before,
it must be possible to express $\Phi'$, whose kernel is $\mathcal{E}$, almost everywhere
as a function of $I_1,\cdots,I_r$.
To formalize this idea, one can create a new fibered manifold using those invariants
on which this quotient equation emerges.
To do so, one must choose $\text{dim}(M)$ functionally independent invariants
that act as coordinates of the base space $N$ of this new fibered manifold.
The remaining invariants can then serve to indicate how many
dimensions the fibers $F_\theta$
of the new manifold $\xi:F\to N$ should have.
In general, the base coordinates do not agree with those of $M$ and
then one needs to invoke Tresse derivatives to construct a jet space over $F$
or modify the Cartan distribution.
However, in the following, the simpler special case, in
which the coordinates of $M$ are invariant under the symmetry, is assumed
because the main purpose is to illustrate how quotient equations naturally fit
into the present setting involving correspondence and intersection.
There are quite a number of symmetries like translations and dilations of the dependent
coordinates that are included in this special assumption.
The more general case is also compatible with the present approach
and might be described more explicitly in future work.\\
Thus, for now we assume $(I_1,\cdots,I_m)=(x^1,\cdots,x^m)$ and therefore set $N=M$
and create a new fibered manifold
    $\xi:F\ra M$
where the fibers $F_x$ are chosen as the spaces where the invariants live
and consist of $l=r-m$ dimensions (i.e. locally they are isomorphic to $\mathbb{R}^l$)
where $r> m$ is the number of the functionally independent invariants found in the previous step
and $m=\text{dim}(M)$.
Then denote the corresponding
local coordinates of the fibers by $(v^g)=(v^1,\cdots,v^l)$.
Now the invariants $(I_1,\cdots,I_r)$
naturally determine a \ti{correspondence} $\Phi(I)$ on the product bundle
\begin{equation}
  \begin{split}
    J(I):= J^k(E)\times_M J^0(F),
  \end{split}
  \label{eq:invariantTotalSpace}
\end{equation}
namely
\begin{equation}
  \Phi(I): ~\{~v^1 = I_{m+1}(x^i,u^j_\alpha),\cdots, v^l = I_{r=m+l}(x^i,u^j_\alpha)~\}
  \label{eq:correspondenceOfQuotientEq}
\end{equation}
If one computes the prolongations $P^l(\de{Q}(I))$ of the intersection
\begin{equation}
  \begin{split}
    \de{Q}(I):=(\pi_E')^{-1}(\de{E})\cap\Phi(I),
  \end{split}
  \label{eq:quotientIntersection}
\end{equation}
where $\pi_E':J^k(E)\times_M J^0(F)\to J^k(E)$,
then, since $\de{E}$ is invariant with respect
to the symmetry used to construct the invariances expressed by
the correspondence $\Phi(I)$ which relates
the equation to the coordinates $(v^1,\cdots,v^l)$,
$\de{E}$ must necessarily give rise to an equation $(\de{F}_{P^l(J(I))} \supset P^l(\de{Q}(I)) ) \subset P^l(J(I))=J^{k+l}(E)\times_M J^l(F)$, for some $l$,
whose local description solely involves $(x^i,v^g_\beta),~|\beta|\le l$. This equation thus reflects a differential consistency condition and could therefore be called a \ti{differential syzygy}, in analogy to syzygies arising in algebra. (The exact number $l$ is determined by the minimal amount of prolongations needed to arrive at such an expression for $\de{F}_{P^l(J(I))}$.)\\
Since the expression describing $\de{F}_{P^l(J(I))}$ only depends on coordinates of $J^l(F)$, this local description is preserved under the projection $\pi^{k+l,~l}_{k,~l}(\de{F}_{P^l(J(I))})=:\de{F}_J \subset J^k(E)\times J^l(F)=:J$. 
Finally, $\de{F}:=\pi_F(\de{F}_J)$ is then called the \ti{quotient equation}.\\
Note that one can take the pullback of $\Phi(I)$ to arrive at the usual notion of a correspondence
\begin{equation}
  \begin{split}
    \Phi:=(\pi^{k,~l}_{k,~0})^{-1}(\Phi(I))\subset J
  \end{split}
  \label{eq:symcorPhi}
\end{equation}
on $J$, between the two equations $\de{E}$ and $\de{F}$. Furthermore, defining
\begin{equation}
  \begin{split}
    \de{Q}:=(\pi^{k,~l}_{k,~0})^{-1}(\de{Q}(I))=\de{E}_J\cap \Phi,\qquad \de{E}_J:=\pi_E^{-1}(\de{E}),\quad \pi_E:J\to J^k(E),
  \end{split}
  \label{eq:BaecklundSymmetrieCorrespondence}
\end{equation}
one can also express $\de{F}$ as the projection of $\de{P}(\de{Q}):=\pi^l_J(P^l(\de{Q}))$, i.e.
\begin{equation}
  \begin{split}
    \de{F}:=\pi_F(\de{P}(\de{Q})),\qquad \de{F}_J:=\pi_F^{-1}(\de{F}),
  \end{split}
  \label{eq:quotientDef}
\end{equation}
(where, as usual, $\pi_J^l:P^l(J)=J^{k+l}(E)\times_M J^{l+l}(F)\to J$.)\\
The quotient equation
can be understood as the system which one obtains after quotienting out the action
of the Group $G$ because locally it represents $\de{E}$ in terms of
coordinates that were constructed from the invariants of this group.
Those ideas are summarized in the following definition.
\gd{def:quotientCorrespondence}{
  If there is a symmetry group $G$ acting on $J^k(E)$ such that the action $S_g$
  is a symmetry of the PDE $\de{E}\subset J^k(E)$ for all $g\in G$, then
  the correspondence $\Phi(I)$ defined in (\ref{eq:correspondenceOfQuotientEq})
  (on the product bundle $J(I)$ defined in (\ref{eq:invariantTotalSpace})),
  determined by the functionally independent invariants
  $I=(I_1,\ldots,I_r)$ (which can be computed by solving (\ref{eq:invariantFinder})),
  is called a \ti{quotient correspondence} for $\de{E}$.
}
At this point, it is important to notice that the symmetry completely
determines the correspondence. This means that
\tf{symmetries
can help to find meaningful correspondences}.

The explanations above then show that the following corollary holds.
\bc{def:quotientEquation}{
  Given a quotient correspondence $\Phi(I)\subset J^k(E)\times_M F$ for $\de{E}$,
  the prolongations $P^l(\de{Q}(I))$ of the intersection $\de{Q}(I)$,
  defined in (\ref{eq:quotientIntersection}),
  for sufficiently high $l$, give rise to an equation on $J^l(F)$, called \textit{quotient equation}, defined as in \eqref{eq:quotientDef},
  and, defining $\Phi$ as in \eqref{eq:symcorPhi}, a \ti{quotient intersection}
  \begin{equation}
    \begin{split}
      \de{I}:=\de{E}_J\cap\de{F}_J\cap \Phi \subset J.
    \end{split}
  \end{equation}
}
Thus, the definition of $\de{I}$ is in harmony with the usual notion of an intersection,
cf. Definition \ref{implicitIntersection}.
\bigskip 

The present framework allows to show that 
a quotient correspondence gives always rise to
a special kind of Bäcklund correspondence.
\bp{prop:quotientCorAndBaeckCor}{
  A quotient correspondence $\Phi(I)$ for some equation $\de{E}$
  determines a strict Bäcklund correspondence $\de{Q}$ where $\de{Q}$ is defined as in \eqref{eq:BaecklundSymmetrieCorrespondence}.
}
\pr{
  By construction, we already have $\pi_F(\de{P}(\de{Q}))=\de{F}$, cf. equation \eqref{eq:quotientDef}. What remains to be shown is that $\pi_E(\de{P}(\de{Q}))=\de{E}$.\\
  Since $\Phi(I)$ is locally explicitly defined by \eqref{eq:correspondenceOfQuotientEq}, always relating $v$-coordinates to $u$-coordinates, it is almost diagonal and since there are no other equations involving $v$-coordinates, all additional conditions that arise upon prolongation of
  $\de{Q}=\de{E}_J\cap \Phi$, apart from the differential consequences of $\de{E}_J$ (which we assume here not to impose conditions of lower order on $u$-coordinates, i.e. $\de{E}$ is assumed to be in involutive form), can always be written as expressions also involving $v$-coordinates and thus do not impose additional equations involving only $u$-coordinates. 
  Hence $\de{P}(\de{Q})=\pi_J^l(P^l(\de{Q}))$ is almost diagonal to $\de{E}$ and $\de{F}$ and such that $\pi_E(\de{P}(\de{Q}))=\de{E}$ and $\pi_F(\de{P}(\de{Q}))=\de{F}$. Thus, $\de{Q}$ is a strict Bäcklund correspondence.
}
If $\Phi(I)$ is understood to contain the information
about the symmetry group $G$, then this last proposition
demonstrates that \textit{Bäcklund correspondences are generalized symmetries}.

\bigskip 

As usual, a Bäcklund transformation allows to transfer solutions between $\de{E}$ and $\de{F}$. However, because of the specific nature of $\Phi(I)$, one can even give an explicit description of the transferred solution, as described by the following proposition.
\bp{prop:symmetrySolutionTransfer}{
  If $S_E$ is a solution of $\de{E}$ and $\de{F}$ is a quotient equation
  of $\de{E}$, then $S_F=\pi_F(\de{P}(\pi_E^{-1}(S_E) \cap \Phi ))$ is a solution of $\de{F}$.
}
\pr{
  Since $\de{Q}$ is a strict Bäcklund correspondence, $\pi_E^{-1}(S_E)=S_E\times_M J^l(F)$ intersects the solution space of $\de{Q}=\de{E}_J\cap \Phi$. 
  Since the constraints imposed by $\de{E}_J$ are described by the same equations as those describing $\de{E}$, which are already solved by $S_E$, one only needs to find a solution of $\pi_E^{-1}(S_E)\cap \Phi$.\\
  At the same time, $\Phi$, described by equations of the form \eqref{eq:correspondenceOfQuotientEq}, explicitly
  and uniquely defines the values of $v^g$ as functions of $(x^i,u^j_\alpha)$. Thus, the prolongation $P^l(\Phi)$ of $\Phi$ determines, without solving any equations, the values of $v^g_\sigma,~|\sigma|\le l$ in terms of $(x^i,u^j_\delta)$ with $|\delta|\le k+l$. However, when considering $P^l(\pi_E^{-1}(S_E)\cap \Phi)$, all coordinates $u^j_{\delta}$ are locally expressible as functions of $x^i$ because $S_E$ is an $m$-dimensional integral submanifold. Hence, one can solve $v^g_\sigma$ for those $x^i$ and project $P^l(\pi_E^{-1}(S_E)\cap \Phi)$ back to $J$, and then to $J^l(F)$, i.e. taking $\pi_F(\pi^l_J(P^l(\pi_E^{-1}(S_E)\cap \Phi)))=\pi_F(\de{P}(\pi_E^{-1}(S_E)\cap \Phi))$, preserving those solutions.
}
As a result, the following definition becomes meaningful.
\gd{def:equivalenceUpToSymmetr}{
Two differential equations $\de{E}\subset J^k(E)$ and $\de{F}\subset J^l(F)$
are said to be \ti{equivalent up to the action of the symmetry Group $G$ on $J^k(E)$}
if $\de{F}$ is the quotient equation of $\de{E}$ with respect to a quotient correspondence determined by $G$.
}
An extended example is given in section \ref{sec:EquivGaugeSym} where
Maxwell's equations formulated in terms of Faraday tensors are shown to be
a quotient equation of Maxwell's equations formulated in terms of gauge potentials.\\
A brief example that is supposed to illustrate the general formalism is given below:

\ye{ex:ColeHopfHeatBurger}{
  On the bundle $\pi:E:=\mathbb{R}^2\times \mathbb{R}\to\mathbb{R}^2=:M$
  with coordinates $(x,y,u)$,
  consider, on $J^2(E)$, the heat equation $\mathcal{E}:~\{~u_2=\beta u_{11}~\}$.
  It is invariant under prolongations of dilations $X = u \frac{\partial}{\partial u}$
  along $u$. The prolongation of $X$
  is given by
  \begin{equation}
    \begin{split}
      X^{(2)}=
      \sum_{|\beta|\le 2}
      u_\beta \frac{\partial}{\partial u_\beta}
    \end{split}
  \end{equation}
  The generators of the differential algebra of all invariants of $X^{(2)}$ are
  given by the solution of (\ref{eq:invariantFinder}).
  \begin{equation}
    \begin{split}
      I_1=x,\quad I_2=y,\quad I_3=\frac{u_1}{u},\quad I_4=\frac{u_2}{u}
    \end{split}
  \end{equation}
  For later convenience, we renorm $I_3$ and write $I_3=-2\beta \frac{u_1}{u}$.
  According to the general procedure above, we now construct a new
  bundle, $\xi:F:=\mathbb{R}^2\times M\to M$
  with coordinates $(x,y,v,w)$.
  On the product bundle $J^2(E)\times_M F$,
  we can define the correspondence $\Phi(I)$ by
  \begin{equation}
    \begin{split}
      \Phi(I):\eq{v = - 2\beta \frac{u_1}{u},\quad w = \frac{u_2}{u}}
    \end{split}
    \label{eq:heatBurgerCorr}
  \end{equation}
  We now want to find the compatibility conditions $\de{P}(\de{Q})$
  where $\de{Q}(I):=\Phi(I)\cap(\pi_E')^{-1}(\de{E})$ (with $\pi_E':J^2(E)\times F\to J^2(E)$). $\de{Q}(I)$ is locally given by
  \begin{equation}
    \begin{split}
      \de{Q}(I):\eq{
        u_2=\beta u_{11},\quad
        v = - 2\beta \frac{u_1}{u},\quad
        w = \frac{u_2}{u}
      }
    \end{split}
  \end{equation}
  Note that the equations imply $w=\beta u_{11}/u$.
  The prolongation of $\de{Q}(I)$ imposes the following additional conditions
  \begin{equation}
    \begin{split}
      \eq{
        u_{12} = \beta u_{111},\quad u_{22}=\beta u_{112},\\
        v_1 = - 2\beta \frac{u_{11}}{u} + 2\beta \br{ \frac{u_1}{u} }^2,\quad
        v_2 = - 2\beta \frac{u_{12}}{u} + 2\beta \frac{u_1u_2}{u^2},\\
        w_1 = \frac{u_{12}}{u} - \frac{u_1u_2}{u^2},\quad
        w_2 = \frac{u_{22}}{u} - \br{ \frac{u_2}{u} }^2
      }
    \end{split}
    \label{eq:prolQ}
  \end{equation}
  The equations imply
  \begin{equation}
    \begin{split}
      v_1 = -2w+ \frac{v^2}{2\beta}~\Rightarrow~ \frac{v^2}{4\beta} - \frac{v_1}{2} = w = \frac{u_2}{u}
    \end{split}
    \label{eq:wu}
  \end{equation}
  Thus, $w$ is a function of $v$ which implies that, on $(\pi_E')^{-1}(\de{E})$, the second generator in \eqref{eq:heatBurgerCorr} depends on the first one, i.e. on $(\pi_E')^{-1}(\de{E})$ there is only one independent generator of the symmetry.
  We can thus expect to find one quotient equation of $\de{E}$ purely in terms of $v$.
  Indeed, the differential consequences of (\ref{eq:wu}) reveal the following relations.
  \begin{equation}
    \begin{split}
      w_1 = \frac{vv_1}{2\beta} - \frac{v_{11}}{2},\quad w_2 = \frac{vv_2}{2\beta} -\frac{v_{12}}{2}
    \end{split}
  \end{equation}
  Furthermore, we can rewrite the eq for $v_2$ in \eqref{eq:prolQ} to obtain a 2nd condition on $w_1$:
  \begin{equation}
    \begin{split}
      v_2 = -2\beta\left(\frac{u_{12}}{u} - \frac{u_1u_2}{u^2}\right)=-2\beta w_1
    \end{split}
  \end{equation}
  Combining the last two expressions for $w_1$, we obtain the following quotient equation
  on $J:=J^2(E)\times_M J^2(F)$, purely in terms of $v$ and its derivatives:
  \begin{equation}
    \begin{split}
      \de{F}_J:\eq{
        v_2 = \beta v_{11} - vv_1
      }
    \end{split}
    \label{eq:BurgersAsResult}
  \end{equation}
  This is Burger's equation, i.e. we computed the well-known Hopf-Cole reduction.\footnote{
    Note that (\ref{eq:wu}) can be seen as a derivation of a correspondence
  $\Phi':~\{~v = -2\beta u_1/u,\quad v^2/(4\beta)-v_1/2=u_2/u~\}$.
  which is a Bäcklund correspondence with diff. consequences $\de{E}_J$ and $\de{F}_J$, that, in contrast to $\de{Q}$, does not require the coordinate $w$ anymore.)}\\
  By proposition \ref{prop:symmetrySolutionTransfer}, solutions of
  $\de{E}$ can be transferred to the quotient equation
  $\de{F}:=\pi_F(\de{F}_J)$ (where $\pi_F:J\to J^2(F)$). Note that the coordinate $w$ is not involved and we could thus also consider $\de{F}$ as an equation on $J^2(G)$ where $\rho:G\to M$ has local coordinates $(x,y,v)$.
  One can e.g. solve the following boundary value problem.
  On the ($(x,y)=(x,t)$) plane:
  \begin{equation}
    \begin{split}
      \begin{cases}
        v = A(x),& t=0,\\
        \mathcal{F}:\{v_2+vv_1 - \beta v_{11} = 0\},& t > 0.
      \end{cases}
    \end{split}
  \end{equation}
  The correspondence $v = - 2\beta \frac{u_1}{u} ~\Ra~u(x,t)=\exp(-1/(2\beta)\int dx~v(x,t))$, transforms this into an initial value problem for $\mathcal{E}$:
  \begin{equation}
    \begin{split}
      \begin{cases}
        u = \exp\br{ -\frac{1}{2\beta} \int_{x_0}^x d\sigma~ A(\sigma)},& t=0,\\
        \mathcal{E}:\{u_2-\beta u_{11}=0\},& t>0
    \end{cases}
    \end{split}
    \label{eq:initialHeat}
  \end{equation}
  The general solution of the heat equation given the initial condition $u(x,0)=g(x)$ is the convolution
  \begin{equation}
    \begin{split}
      u(x,t) &= \int_{-\infty}^\infty dz~ f(x-z,t)g(z)\quad\text{ where }\\
      f(x,t) &= \frac{1}{\sqrt{4\pi\beta t}}\exp\br{-\frac{x^2}{4\beta t}}\text{ is the fundamental solution.}
    \end{split}
  \end{equation}
  In the present case where $g(x)$ is given by (\ref{eq:initialHeat}), this leads to
  \begin{equation}
    \begin{split}
    u(x,t):= \frac{1}{\sqrt{4\pi \beta t}} \int_{-\infty}^\infty dz~ \exp\br{ -\frac{1}{2\beta}\Br{ \frac{(x-z)^2}{2t} + \int^z_{z_0} d\sigma~ A(\sigma)}}
    \end{split}
  \end{equation}
  and using the correspondence $v= -2\beta u_1/u$ again, we obtain, without solving any further equations (as described in proposition \ref{prop:symmetrySolutionTransfer}), the quite general solution of Burgers' equation:
  \begin{equation}
    \begin{split}
      v(x,t) = -2\beta \frac{u_x(x,t)}{u(x,t)} =
      \frac{
      \int_{-\infty}^\infty dz~\frac{x-z}{t} \exp\br{
        -\frac{1}{2\beta}\Br{
          \frac{(x-z)^2}{2t} + \int^z_{z_0} d\sigma~ A(\sigma)
        }
      }
    }{
      \int_{-\infty}^\infty dz~ \exp\br{
        -\frac{1}{2\beta}\Br{
          \frac{(x-z)^2}{2t} + \int^z_{z_0} d\sigma~ A(\sigma)
        }
      }
    }
    \end{split}
  \end{equation}
  This well-known result also appears as a Bäcklund transformation in \cite{rogers1982baecklund}.
  The example is supposed to show how it arises in the present framework as a special case of a solution transfer,
  relating symmetries / quotient equations to correspondences which in turn can give rise to generalized notions of Bäcklund transformations.
}

\newpage
\section{Application to electrodynamics and hydrodynamics}
\label{chap:applications}
In this section, the framework is applied to
study some aspects of electrodynamics and hydrodynamics
in order to illustrate the general aspects outlined in the last sections.
\begin{enumerate}
  \item
    In the first subsection, formal integrability of Maxwell's equations is shown.
    This is a well-known result but provided for completeness.
  \item
    In the second subsection,
    the shared structure of Maxwell's equations in vacuum and the wave equations is computed
    and Maxwell's equations in vacuum are identified as an auto-Bäcklund correspondence of the wave equation.
  \item
    In the third subsection,
    it is shown that
    electrodynamics, formulated in terms of gauge potentials, is equivalent up to gauge symmetries
    to electrodynamics, formulated in terms of Faraday tensors,
    in the precise sense of definition \ref{def:equivalenceUpToSymmetr}.
  \item
    The fourth subsection
    picks up the motivating example
    of subsection \ref{sec:correspondenceMotivatingExample}
    and the shared structure of magneto-statics and the incompressible,
    viscous Navier-Stokes equation.
    It is shown that the integrability conditions coming out of the formalism
    are exactly those physical assumptions that had to be guessed in the motivating example.
\end{enumerate}

~\\For the interested reader, an axiomatic derivation of Maxwell's equations (along the lines of \cite{Zirnbauer1998}) is given in appendix \ref{app:axForm} that the author considers to be rather beautiful. Furthermore, in subsection \ref{sec:phenomenaED} of appendix \ref{app:axForm}, the empirical limits of electrodynamics are discussed in order to show how difficult it is to formalize such considerations, even though they should in fact be a part of a (meta-)theory that compares theories.

\subsection{Formal integrability of Maxwell's equations}
\label{sec:formIntMW}

Let $M$ be our spacetime with local coordiantes $(x^0,\cdots,x^3)$ and
$E=TM$ an 8-dimensional bundle, $\pi:E\to M$,
which locally has the form $U\times\mathbb{R}^4$, $U\subset M$
with local coordinates $(x^0,\cdots,x^3,A^0,\cdots,A^3)$.
We abbreviate those local coordinates with $(x^\mu,A^\mu)$. $A^\mu$ are the local
coordinates of the \textit{gauge potential} of electrodynamics.
In the present context, they are coordinate functions $A^\mu:J^0(\pi)\ra \mathbb{R}$
and they should not be confused with
sections $A^\mu:M\ra J^0(\pi),~ \bo{x} \mapsto A^\mu(\bo{x})$.
One can prolong $J^0(\pi)$ to $J^2(\pi)$ to obtain the local coordinates
\begin{equation}
  (x^\mu,A^\mu,A^{\mu,\nu},A^{\mu,\nu\lambda}).
  \label{eq:localCoordsOnJ2expl}
\end{equation}
As second derivatives commute, the relation $A^{\mu,\nu\lambda}=A^{\mu,\lambda\nu}$ holds for the
corresponding coordinate functions of the prolongation. Thus,
$J^2(\pi)=J^2(4,4)$ is a space with $4+4+4^2+4\cdot4\cdot(4+1)/2=24+40=64$ dimensions.
Furthermore, we
let
$g:TM\otimes TM\ra C^{\infty}(M)$ be the Lorentzian metric of our spacetime $M$.
It is an element of $T^*M\otimes T^*M$.
In local coordinates, it can be written $g=g_{\mu\nu}dx^\mu\otimes dx^\nu$.
If one assumes that the metric is given (e.g. as solution of the Einstein equations)
and that the sources $J^\nu:M\ra\mathbb{R}$ are also given,
one can locally describe Maxwell's equations as the kernel of the differential operator\footnote{
  The notation $a_{[\mu_1 \ldots \mu_n]}$ means antisymmetrisation of the indices,
  e.g. $F_{[\nu\lambda,\mu]}=
   F_{\nu\lambda,\mu}
  -F_{\mu \lambda,\nu}
  +F_{\mu\nu,\lambda}
  -F_{\lambda\nu,\mu}
  +F_{\lambda\mu,\nu}
  -F_{\nu \mu,\lambda}$
  or
  $g_{\mu\lambda}A^{[\nu,\mu]\lambda}=
  g_{\mu\lambda}A^{\nu,\mu\lambda}
  -g_{\mu\lambda}A^{\mu,\nu\lambda}$.
  The Einstein sum convention is used.
}
\begin{equation}
  \varphi:J^2(E) \ra E,\qquad(x^\mu,A^\mu,A^{\mu,\nu},A^{\mu,\nu\lambda})\mapsto
  (x^\mu,g_{\nu\lambda}A^{[\nu,\mu]\lambda} - J^\mu).
  \label{eq:mw-gaugeJetexpl}
\end{equation}
\bp{prop:formalIntOfMaxwellsEq}{
  $\de{E}=\ker(\varphi)$ is involutive and thus formally integrable.
}
\pr{
The prolongation $P^1(\mathcal{E})$
only involves new constraints on 3rd order coordinates.
As a result, $\pi^3_2:P^1(\mathcal{E})\ra\mathcal{E}$ is surjective.
Let us check if the other two conditions of proposition \ref{prop:formalInt} are fulfilled.
\begin{equation}
  \begin{split}
    \sigma(\varphi) &=
    a^{\rho,\kappa\theta} \frac{\partial \varphi^h}{\partial A^{\rho,\kappa\theta}}\frac{\partial}{\partial w^h} \\
    &=
    a^{\rho,\kappa\theta} \frac{g_{\nu\lambda} \partial A^{\nu,\mu\lambda} }{\partial A^{\rho,\kappa\theta}}
      \frac{\partial}{\partial w^\mu}-
    a^{\rho,\kappa\theta} \frac{g_{\nu\lambda} \partial A^{\mu,\nu\lambda} }{\partial A^{\rho,\kappa\theta}}
      \frac{\partial}{\partial w^\mu}
    \\
    &=
    \br{a^{\nu,\mu\lambda} g_{\nu\lambda}
    -a^{\mu,\nu\lambda} g_{\nu\lambda}}
      \frac{\partial}{\partial w^\mu}
    =
    g_{\nu\lambda}
    a^{[\nu,\mu]\lambda}
      \frac{\partial}{\partial w^\mu}
  \end{split}
\end{equation}
Over pairs of indices is summed and thus,
those are in total 4 equations.
When calculating the rank of the symbol,
those 4 equations impose 4 constraints.
This means (recall that $\te{dim}(E)=m+e=4+4$)
\begin{equation}
  \te{dim}(g^2) \ov{\ref{eq:dimPowerSpaces}}
  e
  \begin{pmatrix}
    m-1+2\\
    2
  \end{pmatrix}-e
  \ov{$m=4=e$}
  4
  \begin{pmatrix}
    5\\
    2
  \end{pmatrix}-4 = 4\cdot 10 - 4 = 36.
  \label{eq:example7dimg2}
\end{equation}
Next, calculate the prolongation:
\begin{equation}
  \begin{split}
    \mu\circ\sigma^1(\phi)&\ov{\ref{eq:prolongedSymbolOfMap}}
    a^{\nu,\mu\lambda\theta} \frac{\partial(D_o \phi^h)}{\partial A^{\nu,\mu\lambda\theta}}\frac{\partial}{\partial w^h_o}
    =
    g_{\nu\lambda}a^{[\nu,\mu]\lambda\theta} \frac{\partial}{\partial w^{\mu}_{\theta}}
  \end{split}
  \label{eq:example7symbolsigma1}
\end{equation}
Those are in total 16 equations. However, the rank of the system might be lower if some of them
are functionally dependent.
A small program was implemented that generates the corresponding matrix and calculates the rank.
The code of this program is given in \appendix \ref{app:formalIntSolver}.
The program delivers the rank 15 for the system above for 4 dimensions
This means, one of the functions depends on the others. Therefore, we obtain
\begin{equation}
  \te{dim}(g^3)
  = 4
  \begin{pmatrix}
    6\\
    3
  \end{pmatrix}-15
  = 80-15 = 65.
  \label{eq:example7dimg3}
\end{equation}
The dimension is constant for every local neighbourhood
and thus $g^3$ is a smooth vector bundle over $\mathcal{J}^2$.\\
We can use (\ref{eq:dimSkj}) to obtain the dimensions of
$F^{2,j}_{\mathcal{J}^2}$ .
\begin{equation}
  \te{dim}(F^{2,j}_{\mathcal{J}^2})=
  e\cdot
  \begin{pmatrix}
    m-1-j+2\\
    2
  \end{pmatrix}
\end{equation}
Let us give an explicit basis for them
\begin{equation}
  F^{2,j}_{\mathcal{J}^2} =
  \bbr{
  \te{span}
  \begin{pmatrix}
    dx^{l}\vee dx^n\otimes \partial_u^k
  \end{pmatrix}
  ~|~j+1\le l \le n \le m}
\end{equation}
We can obtain the intersection by restricting $\sigma(\phi)$ to $F^{2,j}_{\mathcal{J}^2}$:
\begin{equation}
  \begin{split}
  g^{2,j} &=
  g^2\cap F^{2,j}_{\mathcal{J}^2}
  = \ker(\sigma(\phi)|_{F^{2,j}_{\mathcal{J}^2}})\\
  &= ~\bbr{
    \begin{pmatrix}
      0=g_{\nu\lambda}(a^{\mu,\nu\lambda}-a^{\nu,\mu\lambda})
      ~|~\te{ last two indices }\in\bbr{j+1,\cdots,m}
    \end{pmatrix}}
  \end{split}
\end{equation}
For $j < m-1$, the above equation always gives rise to $e=4$ different conditions on the components
$a^{\mu,\nu\lambda}$ because the last two indices can be chosen differently.
However, for $j=m-1$
one obtains the equation
\begin{equation}
  g_{mm}a^{\mu,mm}
  - g_{\nu m}a^{\nu,\mu m}\delta^{\mu,m}
\end{equation}
And this means that for $\mu=m$, the last term of the matrix
of derivatives of the equation above with respect to $a^{\mu,\nu\lambda}$
(whose rank corresponds to the rank of the system)
vanishes.
Then they impose one condition less.\\
In accordance with this, the computer program delivers:
\begin{equation}
  \begin{split}
    &\te{rank} \sigma(\phi)|_{F^{2,1}}= 4
    ,\qquad
    \te{rank} \sigma(\phi)|_{F^{2,2}}= 4
    ,\qquad
    \te{rank} \sigma(\phi)|_{F^{2,3}}= 3
  \end{split}
\end{equation}
All in all, we obtain
\begin{equation}
  \begin{split}
  \te{dim}(g^2) + \sum_{j=1}^{3}\te{dim}(g^{2,j})
  &=
  36 +
  4
  \begin{pmatrix}
    4\\
    2
\end{pmatrix}-4+
  4
  \begin{pmatrix}
    3\\
    2
\end{pmatrix}-4+
  4
  \begin{pmatrix}
    2\\
    2
\end{pmatrix}-3
  \\
  &=\te{dim}(g^3).
  \end{split}
\end{equation}
Thus, the system is formally integrable.
}
Note that formal integrability of the Yang-Mills-Higgs equations
was shown for arbitrary dimensions in 1996 by
\cite{formalIntYangMills1996}.

\subsection{Embedding of vacuum electrodynamics in wave equations}
\label{sec:embeddingWave}

As is well-known, when considering Maxwell's equations in flat spacetime in vacuum
(without sources and in Gaussian units)
\begin{equation}
  \nabla\cdot \bo{E} = 0,\qquad
  \nabla\times \bo{E} = - \frac{1}{c} \partial_t\bo{B},\qquad
  \nabla\times\bo{B} = \frac{1}{c} \partial_t\bo{E},\qquad
  \nabla\cdot\bo{B}=0,
  \label{eq:MWvacuum}
\end{equation}
one can derive wave equations as follows
\begin{equation}
  \begin{split}
    \partial_t^2\bo{B}&=-c~ \partial_t(\nabla\times\bo{E})=-c^2~\nabla\times(\nabla\times\bo{B})
    =-c^2~\nabla\cdot(\nabla\cdot\bo{B})+c^2\Delta\bo{B}=c^2\Delta\bo{B}\\
    \partial_t^2\bo{E}&=-c~\partial_t(\nabla\times\bo{B})=-c^2~\nabla\times(\nabla\times\bo{E})
    =-c^2~\nabla\cdot(\nabla\cdot\bo{E})+c^2\Delta\bo{E}=c^2\Delta\bo{E}
  \end{split}
  \label{eq:waveeq}
\end{equation}
In the following is shown how Maxwell's equations in vacuum can
be understood as a Bäcklund correspondence for the wave equations.\\
As can be seen, the wave equations are differential consequences
of Maxwell's equations.
Furthermore, the consequences separate into
constraints imposed solely on $\tf{B}$ and $\tf{E}$.
This suggests to understand (\ref{eq:MWvacuum}) as a correspondence
between the two wave equations.\\
Indeed, if one defines the bundle $\pi:E:=M\times\mathbb{R}^3\to M$ where $M=\mathbb{R}^4$ in this case, with local coordinates $(t,x^i,E^j)$, $i,j\in\{1,\cdots,3\}$
and the bundle $\xi:F\simeq E\to M$ with coordinates $(t,x^i,B^j)$, then
one can define Maxwell's equations in vacuum as a correspondence $\Phi$
on the product space $J:=J^2(E)\times_M J^2(F)$ by
\begin{equation}
  \begin{split}
    \Phi:\eq{
      E^{i,i}=0,\qquad
      \varepsilon_{ijk}E^{k,j} =- \frac{1}{c} B^i_t\\
      B^{i,i}=0,\qquad
      \varepsilon_{ijk}B^{k,j} = \frac{1}{c} E^i_t
    } 
  \end{split}
\end{equation}
The compatibility conditions $\de{P}(\Phi)=\pi_J^1(P^1(\Phi))$ for $\Phi$ are given by
\begin{equation}
  \de{P}(\Phi) = \ker \begin{pmatrix}
  E^{i,i},&
  E^{i,ij},&
  E^{i,i}_t \\
  c~\varepsilon_{ijk}E^{k,j} + B^i_t,&
  c~\varepsilon_{ijk}E^{k,jl} + B^{i,l}_t,&
  c~\varepsilon_{ijk}E^{k,j}_t + B^i_{tt}\\
  c~\varepsilon_{ijk}B^{k,j} - E^i_t,&
  c~\varepsilon_{ijk}B^{k,jl} - E^{i,l}_t,&
  c~\varepsilon_{ijk}B^{k,j}_t - E^i_{tt}\\
  B^{i,i},&
  B^{i,il},&
  B^{i,i}_t
\end{pmatrix},\qquad
  \label{eq:waveIntersect}
\end{equation}
Now, as already shown above,
the entries [2,2] and [3,2] of the matrix can be inserted into the entries
[3,2] and [3,3] to obtain the wave equations via the $\varepsilon_{ijk}$-identities.
\begin{equation}
  \begin{split}
    \de{E}_J:\eq{ E_{tt}^i = c^2~E^{i,jj}},\quad
    \de{F}_J:\eq{ B_{tt}^i = c^2~B^{i,jj}}.
  \end{split}
\end{equation}
Together, $\de{E}_J\cap\de{F}_J=\de{EF}\supset \pi_J^1(P^1(\Phi))=\de{P}(\Phi)$.
Furthermore, since no other equations purely in terms of $E$ or $B$ coordinates are imposed, we have $\pi_E(\de{P}(\Phi))=\de{E}$ and $\pi_F(\de{P}(\Phi))=\de{F}$. Hence $\Phi$
is a strict Bäcklund correspondence.
\\
Furthermore, there is a diffeomorphism $\de{E}\simeq \de{F}$
and therefore this Bäcklund correspondence is actually an auto-Bäcklund correspondence.
This is a useful fact because (it is well-known that) auto-Bäcklund correspondences allow
to generate an infinite amount of solutions.
Indeed, by proposition \ref{prop:baecklundTransfer}, solutions can be transferred
from $\de{E}$ to $\de{F}$ by solving $\Phi$. Since the process involves solving
$\Phi$, the solution obtained for $\de{F}$ is in general different to the solution
coming from $\de{E}$. However, once such a solution of $\de{F}$ is obtained,
one can repeat the process because $\de{E}\simeq \de{F}$ and obtain a new solution
of $\de{F}$ and so on.\\
Another aspect that is shown quite clearly in this geometric product bundle setting,
is that the space of all differential solutions of $\de{E}_J$ and $\de{F}_J$ contain the the space of all differential solutions of $\Phi$ (because they are a differential consequence of $\Phi$, i.e. $\de{P}(\Phi)\subset \de{EF}$).
Thus, one could say that the solution space of electrodynamics in vacuum
is \textit{embedded} into the solution spaces of the wave equations.
Hence, once the most general solution of the wave equations is found
(possibly by utilizing the auto-Bäcklund correspondence),
one can restrict this general solution to the subspace of solutions of Maxwell's equations (that can be obtained simply by inserting the solutions into those equations) to obtain the
general solution of Maxwell's equations in vacuum.

\subsection{Equivalence up to gauge symmetry}
\label{sec:EquivGaugeSym}
In this subsection, the aim is to derive Maxwell's equations in terms of Faraday tensors\footnote{
  As already mentioned in the footnote above eq. \eqref{eq:mw-gaugeJetexpl},
  the notation $a_{[\mu_1 \ldots \mu_n]}$ means antisymmetrisation of the indices,
  e.g. $F_{[\nu\lambda,\mu]}=
   F_{\nu\lambda,\mu}
  -F_{\mu \lambda,\nu}
  +F_{\mu\nu,\lambda}
  -F_{\lambda\nu,\mu}
  +F_{\lambda\mu,\nu}
  -F_{\nu \mu,\lambda}$
  or
  $g_{\mu\lambda}A^{[\nu,\mu]\lambda}=
  g_{\mu\lambda}A^{\nu,\mu\lambda}
  -g_{\mu\lambda}A^{\mu,\nu\lambda}$.
  The Einstein sum convention is used.
}
\begin{equation}
  \de{F}:~\{~g_{\nu\lambda}F^{\nu\mu,\lambda} = J^\mu,\quad F^{[\nu\lambda,\mu]}=0~\}
  \label{eq:mw-gaugefaraday}
\end{equation}
as a quotient
equation by quotienting out gauge symmetries from the equations in terms
of vector potentials
\begin{equation}
  \de{E}:~\{~g_{\nu\lambda}A^{[\nu,\mu]\lambda} = J^\mu~\}
  \label{eq:mw-gaugeJetagain}
\end{equation}
using the methods introduced in section \ref{sec:equivUpToSym}.
Among other things, this shall illustrate that the framework is versatile enough
to answer the questions that the solution-Category approach described in
\cite{weather2014}
answers -
though the way the answer is obtained is quite different.\\
The first equation above can be modeled on the jet bundle $J^1(F)$ where
$F$ is the total space of the bundle $\xi:F:=TM\otimes TM\ra M$
with local coordinates $(x^\mu,F^{\mu\nu})$.
$M$ is a Lorentzian spacetime, equipped with a Lorentzian metric $g\in T^*M\otimes T^*M$.
Its local description reads $g=g_{\mu\nu} dx^\mu\otimes dx^\nu$.\\
The second equation can be modeled as submanifold $\de{E}\subset J^2(E)$
over the bundle $\pi:E:=TM\ra M$ with local coordinates $(x^\mu,A^\mu)$.
$J^2(E)$ has local coordinates $(x^\mu,A^\mu,A^{\mu,\nu},A^{\mu,\nu\lambda})$.
As second derivatives commute, the corresponding relation
$A^{\mu,\nu\lambda}=A^{\mu,\lambda\nu}$
also holds for the
jet bundle coordinates.\\
The differential equation $\de{E}$ is invariant under so called \ti{gauge transformations}
\begin{equation}
  x^\mu\ra x^\mu,\qquad A^\mu \ra A^\mu + \chi^{~,\mu}
  \label{eq:gauge-trf}
\end{equation}
which prolonged to $J^2(E)$ take the form
\begin{equation}
  A := \begin{pmatrix}
    x^\mu\\
    A^\mu\\
    A^{\mu,\nu}\\
    A^{\mu,\nu\lambda}
  \end{pmatrix}
  \qquad\ra\qquad
  A' :=
  \begin{pmatrix}
    x^\mu\\
    A^\mu\\
    A^{\mu,\nu}\\
    A^{\mu,\nu\lambda}
  \end{pmatrix}
  +
  \begin{pmatrix}
    0\\
    \chi^{~,\mu}\\
    \chi^{~,\mu\nu}\\
    \chi^{~,\mu\nu\lambda}
  \end{pmatrix}.
  \label{eq:gauge-trfOnJ2}
\end{equation}
Note that because of the prolongation, we have $\chi^{~,\mu\nu}=\chi^{~,\nu\mu}$
and therefore, if we contract it with some tensor $T_{\mu\nu}$, we obtain
\begin{equation}
  \chi^{~,\mu\nu}T_{\mu\nu}
  =\chi^{~,\mu\nu}\br{\frac{T_{\mu\nu}+T_{\nu\mu}}{2}+\frac{T_{\mu\nu}-T_{\nu\mu}}{2}}
  =\chi^{~,\mu\nu}\frac{T_{\mu\nu}+T_{\nu\mu}}{2}
  \label{eq:symmetrisation}
\end{equation}
because the anti-symmetric part vanishes upon contraction. Similarly,
\begin{equation}
  \begin{split}
  \chi^{~,\mu\nu\lambda}T_{\mu~,\nu\lambda}
  &= \chi^{~,\mu\nu\lambda}\br{
    \frac{T_{\mu~,\nu\lambda}+T_{\lambda~,\mu\nu}+T_{\nu,~\lambda\mu}
  +T_{\mu~,\lambda\nu}+T_{\lambda~,\nu\mu}+T_{\nu,~\mu\lambda}}{3!}}= \chi^{~,\mu\nu\lambda}
    \frac{T_{\mu~,\nu\lambda}+T_{\lambda~,\mu\nu}+T_{\nu,~\lambda\mu}}{3}
  \end{split}
  \label{eq:2symmetrisaton}
\end{equation}
The gauge transformation can be rewritten as the action
of group elements on $A$ to extract the generators $X$.
\begin{equation}
  \begin{split}
  A' &=
  \begin{pmatrix}
    x^\mu\\
    A^\mu\\
    A^{\mu,\nu}\\
    A^{\mu,\nu\lambda}
  \end{pmatrix}
  +
  \begin{pmatrix}
    0\\
    \chi^{~,\mu}\\
    \chi^{~,\mu\nu}\\
    \chi^{~,\mu\nu\lambda}
  \end{pmatrix}
  \\&=
  \exp\left[\chi^{~,\mu}\frac{\partial}{\partial A^{\mu}}
    + \chi^{~,\mu\nu}\frac{\partial}{\partial A^{\mu,\nu}}
    + \chi^{~,\mu\nu\lambda}\frac{\partial}{\partial A^{\mu,\nu\lambda}}
  \right]
  \begin{pmatrix}
    x^\mu\\
    A^\mu\\
    A^{\mu,\nu}\\
    A^{\mu,\nu\lambda}
  \end{pmatrix}
  \\
  &\ov{\ref{eq:2symmetrisaton}}
  \exp\left[
    \chi^{~,\mu}
      \un{\frac{\partial}{\partial A^\mu}}{=:X^\mu}
      + \frac{1}{2} \chi^{~,\mu\nu}
      \un{
        \br{
          \frac{\partial}{\partial A^{\mu,\nu}}
          + \frac{\partial}{\partial A^{\nu,\mu}}
        }
     }{=:X^{\mu\nu}}
        + \frac{1}{3} \chi^{~,\mu\nu\lambda}
      \un{
        \br{
            \frac{\partial}{\partial A^{\mu,\nu\lambda}}
          + \frac{\partial}{\partial A^{\nu,\mu\lambda}}
          + \frac{\partial}{\partial A^{\lambda,\mu\nu}}
        }
      }{=:X^{\mu\nu\lambda}}
  \right]
  \begin{pmatrix}
    x^\mu\\
    A^\mu\\
    A^{\mu,\nu}\\
    A^{\mu,\nu\lambda}
  \end{pmatrix}
  \end{split}
  \label{eq:Lie-AlgebraMW}
\end{equation}
If $\chi^{~,\mu\nu}$ were different from $\chi^{~,\nu\mu}$,
then $\partial/\partial A^{\mu,\nu}$ and $\partial/\partial A^{\nu,\mu}$ would be
two different generators but because $\chi^{~,\mu\nu}=\chi^{~,\nu\mu}$, we obtain the
generator $X^{\mu\nu}$. Similarly for $\chi^{~,\mu\nu\lambda}$ and $X^{\mu\nu\lambda}$.\\
As a consequence, to obtain a functionally independent set of Invariants of gauge transformations, we
use equation (\ref{eq:invariantFinder}) and obtain
\begin{equation}
  X^\mu(I)=0,\qquad X^{\mu\nu}(I)=0,\qquad X^{\mu\nu\lambda}(I)=0.
  \label{eq:mw-invariants}
\end{equation}
\bp{prop:maxwellInvariantproof}{
  This system can only be solved if $I$ is a function of $x^\mu$ and
  \begin{equation}
      I^{\mu\nu}:=A^{\mu,\nu}-A^{\nu,\mu} = A^{[\mu,\nu]}
    \label{eq:quotientSolution}
  \end{equation}
  and its prolongations
  $I^{\mu\nu,\lambda}=A^{[\mu,\nu]\lambda}, \cdots ,
  I^{\mu\nu,\lambda_1 \ldots \lambda_k} = A^{[\mu,\nu]\lambda_1 \ldots \lambda_k}$ and so on.
}
\pr{
  That the generators annihilate $x^\mu$ is trivial because they only
  contain derivatives w.r.t. the dependent variables.
  For deriving (\ref{eq:quotientSolution}), let us consider the equations order by order:
\begin{enumerate}
  \item $0=X^\mu(I)=\partial I/\partial A^\mu$ implies that $I$ does not depend on $A^\mu$.
  \item Now we have $0=X^{\mu\nu}(I)=\partial I/\partial A^{\mu,\nu}+\partial I/\partial A^{\nu,\mu}$.
    The general dependence of $I$ can be found by a coordinate transformation.
    First, let us fix some indices $\mu,\nu,\lambda$ and
    then define $x_1:=A^{\mu,\nu}$ and $x_2:=A^{\nu,\mu}$
    such that the above equation takes the form
    $0=\partial I/\partial x_1 + \partial I/\partial x_2$.
    Now, we introduce the transformation
    \begin{equation}
      \begin{pmatrix}
      x_1':=x_1\\
      x_2':=x_2-x_1
      \end{pmatrix}~\Ra~
      \frac{\partial}{\partial x_i}
      = \frac{\partial x_1'}{\partial x_i} \frac{\partial}{\partial x_1'}
      + \frac{\partial x_2'}{\partial x_i} \frac{\partial}{\partial x_2'}
      \label{eq:coordTrans}
    \end{equation}
    Thus, $\partial/\partial x_1 = \partial/\partial x_1' - \partial/\partial x_2'$ and
    $\partial/\partial x_2 = \partial/\partial x_2'$. Therefore
    \begin{equation}
      0 = \br{\frac{\partial}{\partial x_1}+\frac{\partial}{\partial x_2}}I
      = \frac{\partial I}{\partial x_1'}.
    \end{equation}
    This implies that $I$ can only be any function of $I^{\mu\nu}:=x_2'=x_2-x_1 = A^{\mu,\nu}-A^{\nu,\mu}$.
    This goes through for any choice of $\mu,\nu,\lambda$.
  \item
    $0=X^{\mu\nu\lambda}(I)$ implies
    \begin{equation*}
      0 =
          \br{\frac{\partial}{\partial A^{\mu~,\nu\lambda}}
          + \frac{\partial}{\partial A^{\nu~,\mu\lambda}}
        + \frac{\partial}{\partial A^{\lambda~,\mu\nu}}}I
    \end{equation*}
    We employ the same method as above. We define
    $x_1 = A^{\mu,\nu\lambda}, \cdots, x_3 = A^{\lambda,\mu\nu}$ and the transformation
    \begin{equation}
      \begin{split}
        \begin{pmatrix}
        x_1' = x_1,\\
        x_2'=x_2-x_1,\\
        x_3'=x_3-x_1
        \end{pmatrix}
        ~\Ra~
        \begin{pmatrix}
        \frac{\partial}{\partial x_1} =
        \frac{\partial}{\partial x_1'}-
        \frac{\partial}{\partial x_2'}-
        \frac{\partial}{\partial x_3'},\\
        \frac{\partial}{\partial x_2}=
        \frac{\partial}{\partial x_2'},\\
        \frac{\partial}{\partial x_3}=
        \frac{\partial}{\partial x_3'}.
        \end{pmatrix}
      \end{split}
    \end{equation}
    Thus,
    \begin{equation}
      \sum_i \frac{\partial I}{\partial x_i} = \frac{\partial I}{\partial x_1'}
    \end{equation}
    implying that $I$ is a function of $x_2'=x_2-x_1$ and $x_3'=x_3-x_1$.
    Observe that $x_2'-x_3'=x_2-x_3$ which means that
    this system is linearly equivalent to the system $x_i-x_j$, $i,j\in\bbr{1,2,3}$.\\
    Thus, we can say $I$ to this order only depends on
    \begin{equation}
      I^{\mu\nu\lambda}:= A^{\mu,\nu\lambda}-A^{\nu,\lambda\mu}=A^{[\mu,\nu]\lambda}
    \end{equation}
    or any permutation thereof in $\mu,\nu,\lambda$.

  If $A^\mu(\bo{x})$ is a section, then $A^{[\mu,\nu]\lambda}(\bo{x})=\nabla^\lambda A^{[\mu,\nu]}(\bo{x})$
  and therefore $I^{\mu\nu\lambda}=I^{\mu\nu,\lambda}$ as desired.
\end{enumerate}
    If we prolong the bundle further, this idea continuous for higher orders. For order $n$,
    the equation $X(I)=0$ gives
    \begin{equation}
      0 =\br{ \frac{\partial}{\partial A^{\lambda_1,\lambda_2 \ldots \lambda_n}}
    +\frac{\partial}{\partial A^{\lambda_2,\lambda_3 \ldots \lambda_{1}}}
    + \cdots +
    \frac{\partial}{\partial A^{\lambda_n,\lambda_1 \ldots \lambda_{n-1}}}
  }I =: \sum_{i=1}^n \frac{\partial I}{\partial x_i}
      \label{eq:induction}
    \end{equation}
    Thus, with the transformation
    \begin{equation}
      \begin{split}
        \begin{pmatrix}
        x_1' = x_1,\\
        x_2'=x_2-x_1,\\
        \cdots \\
        x_n'=x_n-x_1
        \end{pmatrix}
        ~\Ra~
        \begin{pmatrix}
        \frac{\partial}{\partial x_1} =
        \frac{\partial}{\partial x_1'}-
        \frac{\partial}{\partial x_2'}-
        \cdots -
        \frac{\partial}{\partial x_n'},\\
        \frac{\partial}{\partial x_2}=
        \frac{\partial}{\partial x_2'},\\
        \cdots \\
        \frac{\partial}{\partial x_n}=
        \frac{\partial}{\partial x_n'}.
        \end{pmatrix}
      \end{split}
    \end{equation}
    we obtain $\partial I/\partial x_1'=0$ and therefore $I$ only depends on
    $x_i-x_1$ or, equivalently, on
    \begin{equation}
      x_i-x_j =
      A^{[\lambda_i, \lambda_j]\lambda_1 \ldots \lambda_{i-1}\lambda_{i+1}\ldots
      \lambda_{j-1}\lambda_{j+1}\ldots \lambda_n} = I^{\mu\nu,\lambda_1 \ldots \lambda_k}
    \end{equation}
    proving the claim.
}
Thus, apart from $x^\mu$, the $I^{\mu\nu}$
are our only functionally and differentially independent Invariants.
Their degree is $n=1$ because they only involve functions from $J^1(E)$.
Define $N=\te{max}(k,n)=k=2$, $L=N-n=1$.
As described in section \ref{sec:equivUpToSym},
one can now create a new bundle $\xi:Q\ra M$
with the same base space $M$ and
where $Q$ is the bundle on which the $I^{\mu,\nu}$ live,
i.e. $TM\otimes TM$.
It is given the local coordinates $(x^\mu,F^{\mu\nu})$
whose number coincides with
the number
of the $I^{\mu\nu}$.
Next, the quotient correspondence $\Phi(I)$ is defined on $J^2(E)\times_M J^0(Q)$,
\begin{equation}
  \Phi(I):~\{~F^{\mu\nu} = I^{\mu\nu} = A^{[\mu,\nu]} ~\}
  \label{eq:newBundleMW}
\end{equation}
By our general theory, prolonging the equation
$\de{Q}(I):=(\pi_E')^{-1}(\de{E})\cap \Phi(I)$ should
give rise to compatibility conditions only involving the $F^{\mu\nu}$-coordinates.
Indeed, a prolongation of $\Phi(I)$ results in
\begin{equation}
  \begin{split}
    P^1(\Phi(I)):~\{~F^{\mu\nu} = A^{[\mu,\nu]},~
    F^{\mu\nu,\lambda} = A^{[\mu,\nu]\lambda} ~\}
  \end{split}
  \label{eq:prolong}
\end{equation}
Thus, intersection with the prolongation of $(\pi_E')^{-1}(\de{E})$ (cf eq. (\ref{eq:mw-gaugeJetagain}))
results in the compatibility condition
\begin{equation}
  \begin{split}
    P^1(\de{Q}(I)) ~:~\{&F^{\mu\nu}=A^{[\mu,\nu]},~F^{\mu\nu,\lambda} = A^{[\mu,\nu]\lambda}\\
    &~g_{\nu\lambda}A^{[\nu,\mu]\lambda} = J^\mu,~g_{\nu\lambda}A^{[\nu,\mu]\lambda \theta} = J^{\mu,\theta},\\
    \Rightarrow&~g_{\nu\lambda}F^{\nu\mu,\lambda}
    = g_{\nu\lambda}A^{[\nu,\mu]\lambda}
  = J^\mu,\\&
~F^{[\nu\lambda,\mu]}=
A^{[\nu,\lambda\mu]}=0~\} ~\subset J^3(E)\times J^1(Q),
  \end{split}
\end{equation}
where $A^{[\nu,\lambda\mu]}=0$ always holds because
$A^{\mu,\nu\lambda}=A^{\mu,\lambda\nu}$.\\
Equations purely in terms of coordinates of $Q$ thus arise already after one prolongation. The natural product bundle is thus $J:=J^2(E)\times_M J^1(Q)$, and defining $\de{Q}:=(\pi^{2,~1}_{2,~0})^{-1}(\de{Q}(I))$ as in equation \eqref{eq:BaecklundSymmetrieCorrespondence}, one obtains the following equation $\de{F}\subset J^1(Q)$ from the compatibility conditions $\de{P}(\de{Q})=\pi_J^1(P^1(\de{Q}))$:
\begin{equation}
  \begin{split}
    \de{F}\ov{\ref{eq:quotientDef}}\pi_F(\de{P}(\de{Q})):~\{~g_{\nu\lambda}F^{\nu\mu,\lambda}
      = J^\mu,~
  ~F^{[\nu\lambda,\mu]}=
  0~\}~\subset J^1(Q).
  \end{split}
\end{equation}
Therefore, one indeed obtains Maxwell's equations in terms of Faraday tensors.
Hence, as defined in definition \ref{def:equivalenceUpToSymmetr},
$\de{E}$ and $\de{F}$ are equivalent up to symmetry and $\de{F}=\pi_F(\de{P}(\de{Q}))$ is the quotient equation of $\de{E}$.
\\
Thus,
``adding Morphisms of some group'' in the solution-Category
can be compared with
``finding the invariant equation with respect to some group'' in the category
of smooth manifolds where differential equations are submanifolds of jet spaces.
The procedure in the category of smooth manifolds might be computationally more involved but in contrast to the
solution-Category approach, it delivers all invariants of the symmetry
and it produces the corresponding quotient equation without the need to know
it before. Furthermore, it enables to see connections and find solutions of many systems of PDEs that
result from solution transfer to the quotient as detailed in proposition \ref{prop:symmetrySolutionTransfer} and also from the quotient back to the original equation (here, for example, the quotient equation is a system of lower order).

\subsection{Shared structure of magneto-statics and hydrodynamics}
\label{sec:sharedStructureMSHD}
In this subsection, the motivating example in subsection \ref{sec:correspondenceMotivatingExample}
is picked up.
In particular, the assumption of a static fluid flow, guessed in (\ref{eq:assumptions}),
arise as the result of the computation of the minimal integrability
conditions for shared structure under the given correspondence.\\
The notation that is used in the following computations is the
one introduced in example \ref{ex:intersection}.
In particular, $\de{E}$ is given by (\ref{eq:example4NavierStokesEq}), $\de{F}$ by (\ref{eq:magnetostaticsjetterminology}),
the correspondence by (\ref{eq:example4correspondenceDef}) and the intersection by (\ref{eq:intersectionMagnetoHydro}), copied here for convenience:
\begin{equation}
  \begin{split}
    \de{I}=\pi_E^{-1}(\de{E})\cap\pi_F^{-1}(\de{F})\cap\Phi:~
      \eq{
        u_t^i + u^ju^{i,j} =-  \frac{1}{\rho} p^{~,i} + \nu u^{i,jj}, u^{i,i}=0\\
        \varepsilon_{ijk}B^{k,j}=I^i,~ B^{i,i} = 0\\
        B^i = \varepsilon_{ijk}u^{k,j} &
      }
  \end{split}
  \label{eq:intersectionMagnetoHydroAgain}
\end{equation}
The first prolongation of $\Phi$ leads to
\begin{equation}
  \begin{split}
    P^1(\Phi):\eq{
        B^i = \varepsilon_{ijk}u^{k,j}
        \quad\bigg|\quad
        \begin{matrix}
          B^{i,l} = \varepsilon_{ijk}u^{k,jl}\\
          B^{i}_t = \varepsilon_{ijk}u^{k}_t
        \end{matrix}
      }
  \end{split}
\end{equation}
With the additional relations, all equations in (\ref{eq:intersectionMagnetoHydroAgain})
can be expressed in terms of the coordinates of $J^2(E)$. In particular, we obtain for the middle row of (\ref{eq:intersectionMagnetoHydroAgain}),
\begin{equation}
  \begin{split}
      \varepsilon_{ijk}B^{k,j}=
      \varepsilon_{ijk}\varepsilon_{klm}u^{m,lj}
      &=
      (\delta^i_l\delta^j_m-\delta^i_m\delta^j_l)
      u^{m,lj}=
      u^{j,ij}-u^{i,jj}-I^i\\
      B^{i,i} &=
      \varepsilon_{ijk}u^{k,ji}
      =
      -\varepsilon_{ijk}u^{k,ji} =
      0
  \end{split}
  \label{eq:example4pullbackEqExplicit}
\end{equation}
where we used that $\varepsilon_{ijk}$ is antisymmetric and thus annihilates $u^{k,ji}$ because it is
symmetric in $ji$.\\
Since all relations in (\ref{eq:intersectionMagnetoHydroAgain})
are now expressed in terms of coordinates of $J^2(E)$
(and $B^{i,i}=0$ is trivially fulfilled), formal integrability
of the whole system amounts to formal integrability of the following system on $J^2(E)$.
\begin{equation}
  \mathcal{I}^2:
  \eq{
    u_t^i + u^ju^{i,j} + \frac{1}{\rho} p^{~,i} = \nu u^{i,jj},\qquad
      u^{i,i}=0\\
      u^{j,ij}-u^{i,jj}=I^i
    }
  \label{eq:example4intersection}
\end{equation}
On $\mathcal{I}^2$, $u^{i,jj}=u^{j,ij}-I^i$. Thus,
we can rewrite the first line as
$u_t^i + u^ju^{i,j} + \frac{1}{\rho} p^{~,i} + \nu I^i-\nu u^{j,ij}$.
To simplify the problem, let us assume that
\begin{equation}
  \nu I^i = -p^{~,i}/\rho,
  \label{eq:example4PhysicalAssumption1}
\end{equation}
corresponding to the first of the two assumptions in (\ref{eq:assumptions}).
Then the above system is equivalent to the system
\begin{equation}
  \mathcal{I}^2 = \ker
  \begin{pmatrix}
      u_t^i + u^ju^{i,j} - \nu u^{j,ij}
      \\
      u^{i,i}\\
      u^{j,ij}-u^{i,jj}-I^i
  \end{pmatrix}
  \label{eq:example4intersectionSimplified}
\end{equation}
\bp{prop:magnetoHydroFormallyIntegrable}{
  The system (\ref{eq:example4intersectionSimplified})
  is not formally integrable without adding the integrability condition
  \begin{equation}
    \de{B}(\de{I}^2):\eq{u^{i,ik}=u^{j,kj}=0}\quad\Ra\quad \frac{d\tf{u}}{dt}=0.
    \label{eq:consistencyConFluidImplication}
  \end{equation}
  and thereafter, for $u^i\ne0$, becomes involutive and thus formally integrable. 
}
\pr{
Consider the first prolongation
\begin{equation}
  \mathcal{I}^3 = \ker
  \begin{pmatrix}
      u_t^i + u^ju^{i,j} - \nu u^{j,ij}
      &\bigg|&
      \begin{matrix}
        u_t^{i,k} + u^{j,k}u^{i,j}+u^ju^{i,jk} - \nu u^{j,ijk} \\
        u_{tt}^i + u^j_tu^{i,j}+u^ju^{i,j}_t - \nu u^{j,ij}_t
      \end{matrix}
      \\
      u^{i,i}
      &\bigg|&
      \begin{matrix}
        u^{i,ik}\\
        u^{i,i}_t
      \end{matrix}
      \\
      u^{j,ij}-u^{i,jj}-I^i
      &\bigg|&
      \begin{matrix}
        u^{j,ijk}-u^{i,jjk}-I^{i,k}\\
        u^{j,ij}_t-u^{i,jj}_t-I^i_t
      \end{matrix}
  \end{pmatrix}
  \label{eq:example4intersectionProlong}
\end{equation}
Due to the term $u^{i,ik}$ on the right side, which is set to $0$ when considering the kernel, constraints on coordinates of order 2 are imposed. Furthermore, the third equation simplifies to $u^{i,jj}=I^i$.
As a consequence, $\pi^3_2:\mathcal{I}^3\ra \mathcal{I}^2$ is not surjective, violating the first condition of proposition \ref{prop:finiteFormalInt2}. Thus, the system is not formally integrable without adding those integrability conditions to $\de{I}^2$.\\
As explained in detail in subsection \ref{sec:integrabilityConditions},
those new constraints can be understood as the \ti{minimal conditions
under which the intersection is differentially consistent}.
The conditions are
\begin{equation}
    u^{i,ik}=u^{j,kj}=0
  \label{eq:consistencyConFluid}
\end{equation}
and thus $du^i/dt\cor u_t^i + u^ju^{i,j}=0$. This means the consistency conditions
induce the constraint of a static fluid flow.
\\
Let us therefore define a new system (as explained in subsection \ref{sec:integrabilityConditions})
which takes those consistency conditions up to order two into account:
\begin{equation}
 \Ra \mathcal{J}^2 = \ker
  \begin{pmatrix}
    \begin{matrix}
      u^{j,j},&
      u_t^{j,j},&
      u^{j,ji}
    \end{matrix}\\
    \begin{matrix}
      u_t^i + u^ju^{i,j},&
      (u_t^i + u^ju^{i,j})^{,k},&
      (u_t^i + u^ju^{i,j})_t
    \end{matrix}\\
    u^{i,jj}+I^{i,k}
  \end{pmatrix}
  \label{eq:example4intersectionprolong2}
\end{equation}
The prolongation $\mathcal{J}^3$ now by construction
either does not lead to equations not contained in $\mathcal{J}^2$
or the prolonged terms
always involve at least one 3rd order coordinate.
For example, the term $(u_t^i + u^ju^{i,j})^{,kl}$ can be solved for $u_t^{i,kl}$
and is thus only turned into a constraint on a coordinate of order 3.\\
As a result, $\pi^3_2:\mathcal{J}^3\ra\mathcal{J}^2$ is surjective.
To verify involutivity, let us check if the other two conditions of proposition \ref{prop:formalInt} are fulfilled.
\begin{equation}
  \begin{split}
    \sigma(\phi) &=
    a^{j,kl} \frac{\partial \phi^h}{\partial u^{j,kl}}\frac{\partial}{\partial w^h}
    +a^{j,k}_t \frac{\partial \phi^h}{\partial u^{j,k}_t}\frac{\partial}{\partial w^h}
    +a^{j}_{tt} \frac{\partial \phi^h}{\partial u^{j}_{tt}}\frac{\partial}{\partial w^h} \\
    &= a^{j,j}_t \partial_w^2+ a^{j,ji} \partial_w^3
    + \br{a^{i,k}_t+u^ja^{i,jk}}\partial_w^5 \\
    &
    \qquad\qquad\qquad
    \qquad\qquad\qquad
    + \br{a^{i}_{tt}+u^ja^{i,j}_t}\partial_w^6
    +a^{i,jj}\partial_w^7
  \end{split}
\end{equation}
This and the system (\ref{eq:example4symbolsigma1}) below are quite high dimensional systems.
Thus, a small computer program was implemented
to determine their rank.
The code is given in appendix \ref{app:formalIntSolver}.
It facilitates to generate the matrix corresponding to the tensor equations automatically.\\
When counting all components of the above equations, one obtains 19 but
calculating the rank with the program gives us 18 constraints (i.e. there is one linear dependence).
Note that even though $a^{i,k}_t$ and $a^{i}_{tt}$ depend on $u^j$ due to the non-linearity,
they depend on it in a smooth way and thus $g^2$ has the same dimension everywhere and
is a smooth vector bundle over $\mathcal{J}^2$.\\
Next, we have to calculate the prolongation:
\begin{equation}
  \begin{split}
    \ker\sigma^1(\phi)&\ov{\ref{eq:prolongedSymbolOfMap}}
    \ker\left(a^{j,klm} \frac{\partial(D_n \phi^h)}{\partial u^{j,klm}}\frac{\partial}{\partial w^h_n}
    + a^{j,k}_{tt} \frac{\partial(D_t \phi^h)}{\partial u^{j,k}_{tt}}\frac{\partial}{\partial w^h_t} \right.
    \\&
    \qquad\qquad~
    \left.
    + a^{j,kl}_t \frac{\partial(D_t \phi^h)}{\partial u^{j,kl}_t}\frac{\partial}{\partial w^h_t}
  + a^{j,kl}_t \frac{\partial(D_n \phi^h)}{\partial u^{j,kl}_t}\frac{\partial}{\partial w^h_n}\right)\\
  &=\ker
  \begin{pmatrix}
    \begin{matrix}
      a^{j,jn}_t,&
      a^{j,j}_{tt},&
      a^{j,jin}
    \end{matrix}\\
    \begin{matrix}
      a^{i,kn}_t+u^ja^{i,jkn},&
      a^{i,k}_{tt}+u^ja^{i,jk}_t,&
      a^{i}_{ttt}+u^ja^{i,j}_{tt}
    \end{matrix}\\
    \begin{matrix}
      a^{i,jjn},&
      a^{i,jj}_t
    \end{matrix}
  \end{pmatrix}
  \end{split}
  \label{eq:example4symbolsigma1}
\end{equation}
If all equations of this system are taken to be independent, then
this imposes $3+1+3!+3\cdot 3!+ 3\cdot 3+3+3\cdot 3+3=52$ constraints.
However, the program computes the rank to be 44 (i.e. there are 8 linear dependencies).
If one sets $u^i=0$, the program still returns 8 in accordance to what was said before
(in particular this constancy means that $g^{k+1}$ is a smooth vector bundle everywhere).
Thus, so far we obtain
\begin{equation}
  \te{dim}(g^2) \ov{\ref{eq:dimPowerSpaces}} 3
  \begin{pmatrix}
    4-1+2\\
    2
  \end{pmatrix}-18=12,~
  \te{dim}(g^3)
  = 3
  \begin{pmatrix}
    4-1+3\\
    3
  \end{pmatrix}-44
  =16.
  \label{eq:example4dimg3}
\end{equation}
If we want to show that the system is formally integrable, then it remains to show that
$\te{dim}(g^{2,1})+ \te{dim}(g^{2,2}) + \te{dim}(g^{2,3}) = 4$.
To calculate this, we consider the kernel of $\sigma(\phi)$ restricted to $F^{2,j}_{\mathcal{J}^2}$.
For $g^{2,1}$, this means that all $a_t$'s fall away. Thus, we obtain
\begin{equation}
  \begin{split}
    \sigma(\phi)|_{F^{2,1}_{\mathcal{J}^2}} &=
    a^{j,ji} \partial_w^3
    +u^ja^{i,jk}\partial_w^5
    +a^{i,jj}\partial_w^7
    \label{eq:example4criticalvelocity}
  \end{split}
\end{equation}
Using the program again, one obtains the rank $14$. Thus,
\begin{equation}
  \te{dim}(g^{2,j})
  = \te{dim}(F^{2,j}_{\mathcal{J}^2})-e=
  e\cdot
  \begin{pmatrix}
    m-j+1\\
    2
  \end{pmatrix}-e
\end{equation}
\begin{equation}
  \te{dim}(g^{2,1})
  = \te{dim}(F^{2,1}_{\mathcal{J}^2})-14=
  e\cdot
  \begin{pmatrix}
    m-j+1\\
    2
  \end{pmatrix}-14
  =
  3\cdot
  \begin{pmatrix}
    4\\
    2
  \end{pmatrix}-14
  = 18 - 14 = 4.
\end{equation}
Note, however, that the rank changes to $6$ if one sets $u^i=0$ above in
equation (\ref{eq:example4criticalvelocity}).
This means that the system is \ti{not} involutive for $u^i=0$.\\
Now, for $u^i\ne 0$, it remains to show that $\te{dim}(g^{2,2})=0=\te{dim}(g^{2,3})$.
For them, we obtain the same system as above but the range of the derivatives
now only covers the coordinates $2$ and $3$.
For $\ker(\varphi)|_{F^{2,2}_{\mathcal{J}^2}}$, the program gives us the rank $9$
(and the rank $5$ for $u^i=0$).
For $\ker(\varphi)|_{F^{2,3}_{\mathcal{J}^2}}$, it delivers rank $3$
(and also rank $3$ for $u^i=0$). Thus,
\begin{equation}
  \begin{split}
  \te{dim}(g^{2,2}) &
  = \te{dim}(F^{2,2}_{\mathcal{J}^2})-9=
  e\cdot
  \begin{pmatrix}
    m-j+1\\
    2
  \end{pmatrix}-9
  =
  3\cdot
  \begin{pmatrix}
    3\\
    2
  \end{pmatrix}-9
  = 9 - 9 = 0.\\
  \te{dim}(g^{2,3}) &
  = \te{dim}(F^{2,3}_{\mathcal{J}^2})-3=
  e\cdot
  \begin{pmatrix}
    m-j+1\\
    2
  \end{pmatrix}-3
  =
  3\cdot
  \begin{pmatrix}
    2\\
    2
  \end{pmatrix}-3
  = 3 - 3 = 0.
  \end{split}
\end{equation}
Therefore, for $u^i\ne 0$, the system is involutive and thus formally integrable.
}
\yr{rem:consistencyCon}{
  As explained in subsection \ref{sec:integrabilityConditions},
  the integrability conditions (\ref{eq:consistencyConFluidImplication}),
  can be interpreted as the minimal amount of physical assumptions
  that have to be made in order to reach consistency.
  Observe how, at this point, the consistency conditions emerge from the formalism
  without the need to guess them as in the motivating example
  \ref{sec:correspondenceMotivatingExample} (second assumption of (\ref{eq:assumptions})).
}
Now using the definitions introduced in section \ref{sec:sharedStructure} about shared structure,
one can make the following conclusions.
Hydrodynamics of an incompressible fluid and magneto-statics
\ti{share structure} under a \ti{linear correspondence of first order} $\Phi$
in case that the fluid flow strength $\bo{u}$ is not zero and condition (\ref{eq:example4PhysicalAssumption1}) holds.
(The formal closure is then $\de{B}:=\de{I}\cap\de{B}(\de{I})$.)
As was explained already in the motivating example, $\bo{u}$ in that case takes the
role of $\bo{A}$ in a fixed gauge in magneto-statics.\\
All solutions of $\de{I}$ are solutions of both the Navier-Stokes equation
and, via the correspondence $\Phi$, of magneto-statics by propostion
\ref{prop:solTransfer}.
Finally, note that this correspondence might not be the only one under which
those two theories share structure.

\newpage
\section{Discussion}
\label{sec:conlusion}

In this section, some conclusions are presented that are supposed to show that the aims, that were described in the introduction (section \ref{sec:introMethods}), were reached, and an outlook to possible future research directions is given.

\subsection{Conclusion}

A geometric framework was developed to compare classical field theories, or more generally, any two systems of PDEs in the category of smooth manifolds,
in a mathematically precise sense.
For every two theories there might be multiple
correspondences relating them, enabling a very versatile comparison,
both of subtheories of a single theory with themselves
and with subtheories of other theories.\\
The methods developed in this contribution allow to give an answer to all requirements
\ref{req:1} - \ref{req:4} described in subsection \ref{sec:requirements} in the following way.
\begin{enumerate}
  \item 
    A geometric answer to (Q.1) ("Are two systems of PDEs equivalent?") is given by Definition \ref{def:equivalenceProductBundle} in section \ref{sec:sharedStructure}.
  \item
    (Q.2) ("Do two PDEs share any subsystem?")
    can be answered by computing the shared structure described in
    definition \ref{def:sharedstructure}, using the methods from
    differential topology and formal integrability introduced in section \ref{sec:constistencyConditionsSection} and \ref{sec:formalIntegrability}.
  \item
    (Q.3) ("When are two systems equivalent up to a symmetry?")
    was answered by definition \ref{def:equivalenceUpToSymmetr}
    via the introduction of quotient equations (corollary \ref{def:quotientEquation})
    which can be computed using (\ref{eq:invariantFinder})
    in combination with differential consequences of $\de{Q}(I)$ as defined in \eqref{eq:quotientIntersection}.
  \item
    Finally, (Q.4) ("How to transfer solutions from one system to another?")
    was answered by propositions \ref{prop:solTransfer}, \ref{prop:singularSolutionTransfer}
    and corollary \ref{cor:singSolTransferAfterRemovingSingularPoints},
    with the generalization of Bäcklund transformations in proposition
    \ref{prop:GeneralizationOfBäcklundTransformations}, definition
    \ref{def:BaecklundCorrespondence} and propositions \ref{prop:BaecklundGeneralization},
    \ref{prop:baecklundTransfer} and the proposition
    about the transfer of solutions to quotient equations \ref{prop:symmetrySolutionTransfer}.
\end{enumerate}
Hence, theoretical analogies of similar systems can now be analyzed, new analogies can be found
using symmetries, and
methods to solve systems can be transferred with a generalization of Bäcklund transformations, that can help to solve some otherwise barely tractable non-linear PDEs. 

\subsection{Outlook}
\label{sec:outlook}

It would be interesting to apply the framework
to the comparison of more complex theories, for example to understand the relations between
general relativity, hydrodynamics and electrodynamics.\\
Perhaps the description of analogue experiments can be made more transparent with the present approach.
\\~\\
Something that is still missing in the present framework is a way to
find the best possible correspondence
(e.g. the ones that maximizes the solution space of the intersection)
between two given theories. A starting point for making progress in this direction might be the relationship between correspendences and symmetries as outlined in section \ref{sec:equivUpToSym}.
\\~\\
An interesting endeavor might be to study how Bäcklund transformations from eq. $\de{E}$ to $\de{F}$ and from $\de{F}$ to $\de{G}$ could give rise to Bäcklund transformations between $\de{E}$ and $\de{G}$ and if those could be used to build up chains of generalized relations between multiple equations that facilitate to map solutions of rather simple equations to ever more complex ones. 
\\~\\
Another future aim would be to describe transitions between theories and approximations of theories
in a mathematically precise way.
They are important both for conceptual reasons  - namely, to identify
how one theory prepares the rise of another - and for
practical purposes - namely, in order to be able to understand
how one should approximate a complicated equation by a simpler one.
\\
In the geometric framework, an equation is a submanifold of a jet bundle
which locally is the kernel of some system of equations. Therefore,
a slight approximation to this system would correspond to a slight deformation of the
submanifold.
Thus, deformation and homotopy theory might serve to describe such transitions.
\\~\\
A natural question is whether it would be possible to extend
the framework to compare quantum theories.
To a certain extend, it can be applied to quantum mechanics
because the Schrödinger equation is also a PDE.
However, in quantum field theory it would perhaps be necessary to
consider functional equations because the Dyson-Schwinger equations,
whose solution is the path integral,
is a functional differential equation.
At some points, \cite{vinogradov2001} points out that cohomology theory could
be used to study problems usually approached by functional analysis.
The advantage would be that cohomology theory directly connects with all areas
of geometry, topology, homological algebra, abstract algebra and would provide
many tools to study quantum field theoretical problems in new ways.
However, it is not yet clear how to set up such a theory.

\backmatter
\appendix
\chapter{}
\section{Introduction to the geometric theory of PDEs}
\label{sec:introgeomethod}
For everything that follows, the following definition will be needed.\footnote{
  The reader not familiar with those geometric notions
  is referred to \cite{kobayashi1996foundations}.
  For an extensive treatment of a wide variety of geometric topics see
  \cite{alekseevski1991geometry}. If a recent treatment is desired, 
  \cite{tu2017differential} is recommended.
  }
\gd{def:fiberedsmooth manifold}{
  Let $M$ denote a smooth manifold with dimension $m$.
  A \ti{fibered smooth manifold} $\pi:E\ra M$ is a differentiable smooth manifold $E$
    together with a differentiable surjective submersion $\pi$ called projection.\footnote{
      A \ti{surjective submersion} is a differentiable surjective map
    such that its pushforward $\pi_*$ is also surjective at each point.}\\
    A \ti{fiber bundle} is a fibered smooth manifold with a local trivialization.\\
  A \ti{vector bundle} is a fiber bundle in which the fibers are vector spaces
  and whose transition maps are linear.
}
Now, before starting to introduce the geometric view on partial differential equations,
it would perhaps be useful to summarise briefly the usual analytic view.
In the analytic view, a differential equation is generally the kernel of a differential
operator imposed on the sections of some fiber bundle (see e.g. \cite{geroch1996}).
If $\pi:E\ra M$ and $\pi':F\ra M$ are fiber bundles and
$\Gamma(E)$ and $\Gamma(F)$ are their spaces of sections
respectively, then a differential operator $\Delta$ of order $k$ is a map
$\Delta:\Gamma(E)\ra\Gamma(F),~s\mapsto \Delta(s)$ that involves
derivatives up to order $k$
(and satisfying certain other conditions). In a diagram, this might be visualised as follows.
\begin{equation}
\bt
  E \& \& F \\
   \&\& \\
   \& M
   \ar[""{name=E}]{uul}{\Gamma(E)}
   \ar[',""{name=F}]{uur}{\Gamma(F)}
   \ar[from=E, to=F, bend left=50, shorten <=6pt,shorten >=6pt]{r}{\Delta}
\et
  \label{eq:diffOpAnalyticView}
\end{equation}
\ye{example:diffOpAnalytic}{
  If $M:=\mathbb{R}^3$ is euclidean 3-space, then we could
model a static fluid flow by specifiying for each point $x\in M$,
a vector $\bo{u}(x)$ in the tangent space $T_xM$ at the point $x$ describing the strength
and direction of the fluid flow. Therefore, $\bo{u}$ itself is a
section from $M$ to $TM$, i.e. $\bo{u}:M\ra TM,~x\mapsto (x,\bo{u}(x))$.
The physical condition of incompressibility of a fluid amounts to the requirement
that the divergence $\varphi(\bo{u}) := \nabla \cdot \bo{u}$ vanishes.
In this case, $\varphi=\nabla\cdot$ is our differential operator and $\varphi(\bo{u})$ is yet another section,
namely the section $\nabla\cdot\bo{u}:M\ra M\times \mathbb{R},~x\mapsto (x,\nabla\cdot \bo{u}(x))$ of
the bundle $\pi':M\times\mathbb{R}\ra M$. The kernel of this operator is thus
the differential equation $\nabla\cdot\bo{u}=0$.
}
In the geometric view on differential equations, the operator $\varphi$ will become a map
between smooth manifolds (and not between spaces of functions or sections).
To facilitate this, the notion of a \ti{jet space} must be introduced.
However, before providing the definition in its beautiful coordinate invariant generality,
the next subsubsection is supposed to give the reader an intuition by providing him
with a very explicit and simple example of a simple equation in local coordinates.
\subsubsection{A simple example to provide intuition}
\label{sec:chap3utu}
\begin{figure}
  \centering
  \includegraphics[width=.49\textwidth]{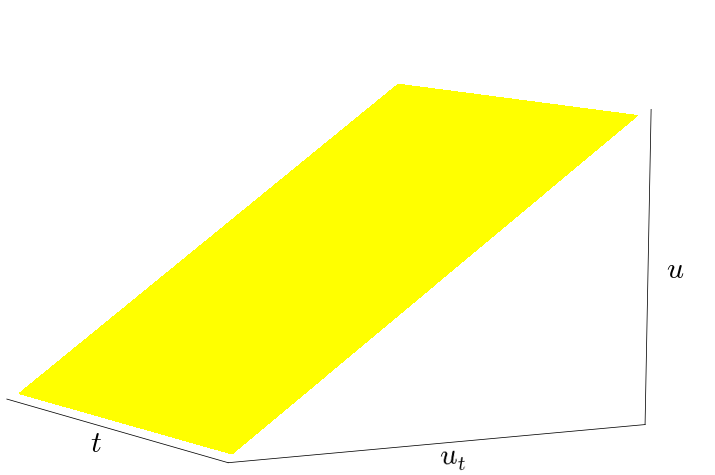}
  \includegraphics[width=.49\textwidth]{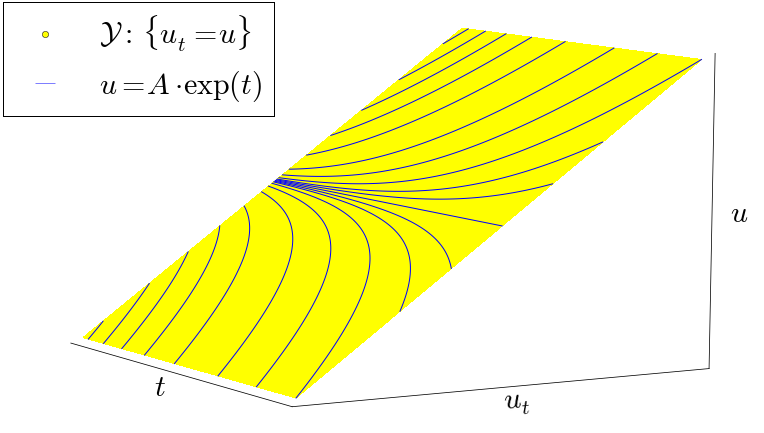}
  \caption{The algebraic solution surface of $u_t=u$ is foliated by differential solutions $u(t)=A\exp(t)$}
  \label{fig:ut1u}
\end{figure}

\begin{figure}[H]
  \centering
  \includegraphics[width=.49\textwidth]{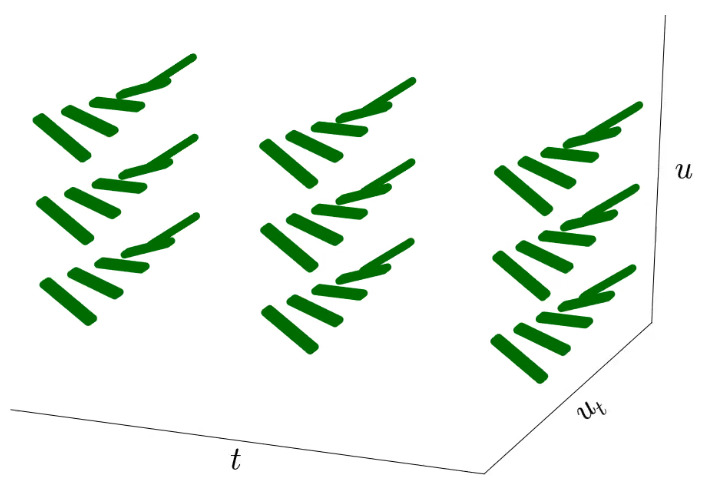}
  \includegraphics[width=.49\textwidth]{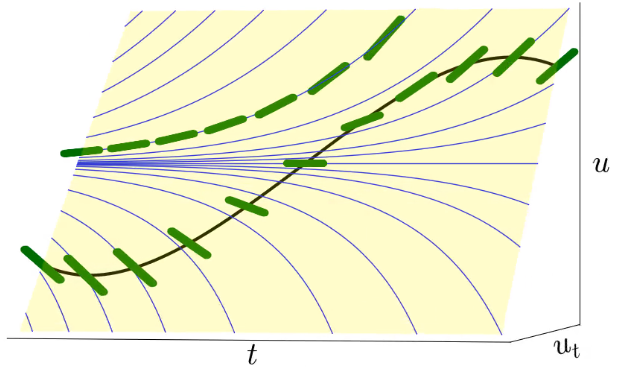}
  \caption{Cartan distribution and curves in relation to it.}
  \label{fig:cartanDisutu}
\end{figure}

Consider the equation
\begin{equation}
  u_t = u
  \label{eq:favourite}
\end{equation}
Here $u:\mathbb{R}\ra\mathbb{R}$ is a function of $t\in\mathbb{R}$
and $u_t:=du/dt$. The solution is $u(t)=A\exp(t),~A\in\mathbb{R}$.
Now the question is how to transform this equation into a geometric object, into a hypersurface
in some space. To obtain some inspiration, we could look at an algebraic equation like
\begin{equation}
  x^2+y^2=1
  \label{eq:algEq}
\end{equation}
imposed on the euclidean plane $\mathbb{R}^2$ with coordinates $x,y$. The natural geometric object
related to this equation is its solution space which in this case is a circle.
In the case of the differential equation (\ref{eq:favourite}) above,
one could consider the space of solutions as well.
However, though we do know that the space of solutions 
consists of functions of the form $u(t)=A\exp(t)$, in general
we do not know the solutions of systems of differential equations. In fact,
what we would like to do is to investigate the differential equation itself in a geometric way,
precisely in order to obtain an answer to questions about solutions, symmetries and so forth.\\
To resolve this dilemma, one can do the very simple but far-reaching step to
regard the differential equation (at first) as an algebraic equation
by promoting all derivatives to new coordinates. In the case above, this would mean
that we create a new space $M:=\mathbb{R}^3$ but instead of giving its (local) coordinates the names
$x,y,z$, we call those coordinates $t,u,u_t$. Observe how $u$ and $u_t$ are now \ti{coordinate functions},
that means maps from $\mathbb{R}^3$ to $\mathbb{R}$. $u$ is not a map from $\mathbb{R}$ to $\mathbb{R}$
anymore and $u_t$ does not denote the derivative of $u$ anymore but turned into the name of a coordinate.
Having taken this step, we can now analyze the \ti{algebraic} solution surface
in $\mathbb{R}^3$ imposed by the algebraic equation $u_t=u$.
The surface is visualised in yellow on the left hand side of figure (\ref{fig:ut1u}).\\
Now that we have this surface, we would like to find a way to recover
the usual notion of a (differential) solution of our differential equation in a geometric way.
To this end, note that if we define the bundle
\begin{equation}
  \pi:\mathbb{R}^3\ra\mathbb{R},\qquad (t,u,u_t)\mapsto t
  \label{eq:ut1uBundle}
\end{equation}
then a section $s:\mathbb{R}\ra\mathbb{R}^3,~t\mapsto (t,s_1(t),s_2(t))$ of this bundle
only lies in the yellow solution surface if $s_1(t)=s_2(t)$. Because
$u,u_t$ are coordinate functions now, we can use them to write $u(t):=u(s(t))=s_1(t)$ and $u_t(t):=u_t(s(t))=s_2(t)$.
In particular, if one chooses $u(t)=A\exp(t)=u_t(t)$ for some $A$,
then we obtain a section that traces out a line that corresponds to our usual solution.
Furthermore, observe on the right hand side of figure (\ref{fig:ut1u})
where many such sections are plotted for different values of $A\in\mathbb{R}$
that they \ti{foliate} our solution surface, i.e. they do not intersect and their union is the whole plane $u_t=u$.
However, they are not the only sections whose image lies in our solution surface.
As remarked above, all sections for which $s_1=s_2$ lie inside, for example
$s(t):=(t,\sin(t),\sin(t))$ is also a section lying in the yellow surface
but it is not a solution of our differential equation because
for this section, it is not true that $ds_1(t)/dt = s_2(t)$. Therefore, we need to
introduce another geometric object that \ti{singles out} those sections in our yellow surface
that are solutions of the PDE. This geometric construction is what is called a
\ti{Cartan distribution}.\\
To obtain it, observe that any vector in a tangent space at a point in $\mathbb{R}^3$ is of the form
\begin{equation}
  v = a \frac{\partial}{\partial t} + b\frac{\partial}{\partial u} + c\frac{\partial}{\partial u_t}
  \label{eq:vecfield}
\end{equation}
where $\partial_t,\partial_u,\partial_{u_t}$ are taken as basis vectors of the tangent space
and $a,b,c$ are any coefficients in $\mathbb{R}$. But
if we take the derivative of any section $s$ at $t$, it has the form
\begin{equation}
  \frac{ds}{dt} = \frac{d}{dt}(t,s_1(t),s_2(t))
  = 1\frac{\partial}{\partial t}
  + s_1'(t) \frac{\partial}{\partial u}
  + s_2'(t) \frac{\partial}{\partial u_t}
  \label{eq:derSection}
\end{equation}
Now if we additionally require that we only want to have sections for which $s_1'(t)=s_2(t)$,
then \ti{their} vector fields must in general be of the form
\begin{equation}
  v_C = \br{\frac{\partial}{\partial t} + u_t \frac{\partial}{\partial u}} + b \frac{\partial}{\partial u_t}
  \label{eq:CartanDisExample}
\end{equation}
for any $b\in\mathbb{R}$. \ti{Conversely}, if we define a sub-bundle of the
tangent bundle by
\begin{equation}
  \mathcal{C} := \te{span}
  \bbr{\br{\frac{\partial}{\partial t} + u_t \frac{\partial}{\partial u}},~ \frac{\partial}{\partial u_t}}
  ~\subset~ TM =
  \te{span}\bbr{ \frac{\partial}{\partial t}, \frac{\partial}{\partial u}, \frac{\partial}{\partial u_t}}
  \label{eq:cartanDisutu}
\end{equation}
then all curves (images of sections $s$) that are tangent to this sub-bundle \ti{do} fulfill the condition
that $ds_1(t)/dt=s_2(t)$ for all $t\in\mathbb{R}$.
Because $\mathcal{C}$ is a sub-bundle of $TM$, it has a coordinate invariant meaning
and is exactly the geometric object we were looking for. It is the Cartan distribution.\\
In figure (\ref{fig:cartanDisutu}),
the Cartan distribution is visualised
on the left-hand side
while the right-hand side displays how the exponential function is the only one that simultaneously lies
in the solution surface and is tangent to the Cartan distribution.
The black curve on the right-hand side is a section of the form $s(t)=(t,\sin(t),\sin(t))$
and is not tangent to the Cartan distribution even though it lies in the algebraic solution surface.
Curves that are tangent to the Cartan distribution and lie in the solution
surface of a PDE are called \ti{integral submanifolds}. They correspond to
the (usual, differential) solutions of our PDE.
The algebraic solution surface endowed with the Cartan distribution is therefore all that
is needed to geometrise a differential equation.\\
Finally note that the Cartan distribution only depends on the number of independent
and dependent variables and the order of a PDE. For example, if we would
geometrise the PDE $\exp(u_t)=\sin(t)u^2-\sinh(u)$, which is still
a first order PDE with an algebraic solution surface in $\mathbb{R}^3$,
the Cartan distribution would be exactly the same.

\subsubsection{General theory of Jet Spaces}

\begin{figure}
  \centering
  \includegraphics[width=.5\textwidth]{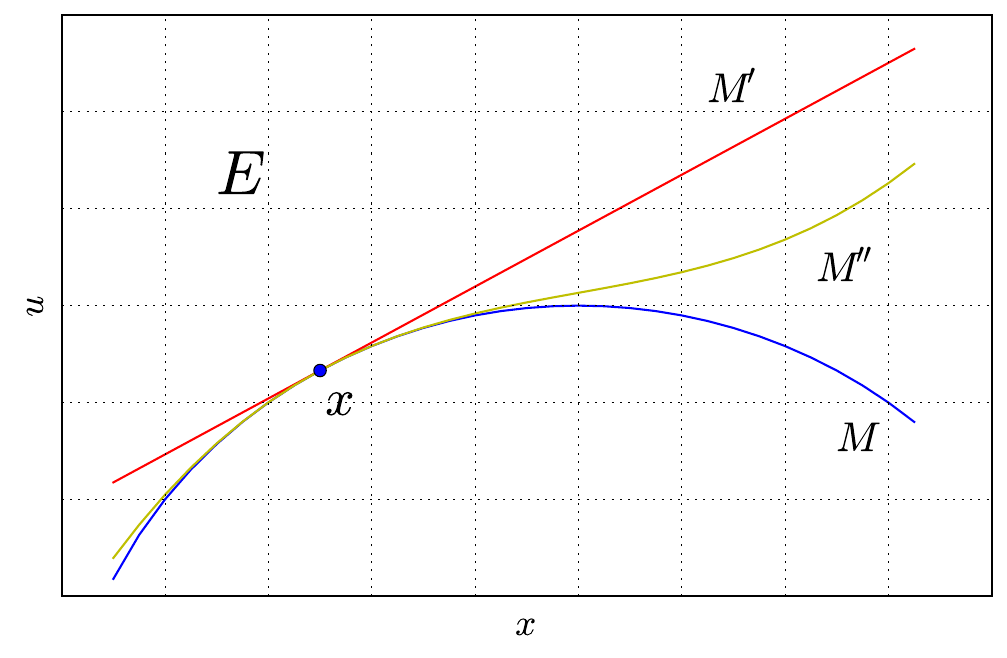}
  \caption{$M$ and $M'$ have the same 1-jet while $M$ and $M''$ have the same 3-jet.}
  \label{fig:jets}
\end{figure}

Now that the basic ideas were exemplified, let us enter the beautiful realm of the general theory.
In doing so, we follow \cite{vinogradov2001} and \cite{Krasilshchik1999}
and take into account \cite{luca2010}. We will also use some proofs of \cite{saunders1989}.
All objects and morphisms considered below are considered in the category of smooth manifolds.\\
As a first step, we need a coordinate-invariant definition of the idea
to promote higher derivatives to new coordinates.
This can be initiated with the definition of a Jet.
\gd{def:jet}{
  Let $E$ be an $m+e$-dimensional smooth manifold.
  Two $m$-dimensional submanifolds $M,M'$ of $E$ are said to have the same $k$-th order
  \ti{Jet} $[M]_p^k$ at
  $p\in M \cap M' \subset E$ if they are tangent
  up to order $k$.
}
To be ``tangent up to order $k$'' means that if one locally
describes the submanifolds as images of sections, then the derivatives
of those sections agree up to order $k$.
This can be made precise as follows.
Choose a point $p\in E$. By definition of a submanifold, around $p$ there
is always a neighbourhood $U$ that is small enough such
that one can choose coordinates that are \ti{adapted} to
$M$.
This means that one can write them in the form $(x^i,u^j)$ where $x^i,u^j\in\mathbb{R}$,
$i\in\bbr{1,\cdots,m}$, $j\in\bbr{1,\cdots,e}$
such that $U_M:=U\cap M = \bbr{ (x^i,u^j)~|~u^j = f^j(x^1,\cdots,x^m) }$
where $f^j$ are (smooth) functions.
\yr{rem:coord}{
  The convention is used that tuples like $(x^i,u^j)$ stand for tuples like $(x^1,\cdots,x^m,u^1,\cdots,u^e)$.
}
Note that this means that if one defines a smooth manifold
$O:=\bbr{(x^i)}$
that consists
only of points described by the $x^i$-coordinates,
then $s:O\ra U_M,~x^i\mapsto s(x^i):=(x^i,f^j(x^i))$ defines a diffeomorphism onto $U_M$.
Therefore, if we denote by the same letter the map $s:O\ra U$, then the image of this
section is the submanifold that locally corresponds to $M$ (as said above).
Note that this means that after a choice of adapted coordinates, we locally have a fibered smooth manifold $\pi:U\ra O$.\\
If $M$ and $M'$ intersect such that $p\in M\cap M'$, then one
can find an adapted coordinate chart such that
$U_M$ is as above and $U_{M'} := U\cap M' = \bbr{ (x^i,u^j)~|~u^j = (f')^j(x^1,\cdots,x^m) }$.\footnote{
If one would like to consider $M$ alone, then one could set $f^\alpha=0$ but if
there are two submanifolds $M$, $M'$ which intersect ($p\in M\cap M'$),
then one can usually not find coodinates such that $f^\alpha=0=(f')^\alpha$.}
As a result, if $O$ is defined as above, then $s':O\ra U,~x^i\mapsto (x^i,(f')^j(x^i))$
is another section of $\pi:U\ra O$ whose image is the submanifold $U_{M'}$.\\
Now that we have clearly defined the two sections whose images are the submanifolds,
it remains to say that their derivatives agree up to order $k$. To say this in
a convenient way, one usually introduces the so-called \ti{multi-index notation}.
This is done as follows.
$\alpha = \alpha_1 \cdots \alpha_n$ denotes a multi-index. It is a tuple of $n\in\mathbb{N}_0$
    numbers $\alpha_i\in \bbr{0,1,\cdots,m=\te{dim}(M)}$
    for which one defines the \ti{length} $|\alpha|=n$.
    One defines a multiplication for multi-indices as follows:
    \begin{equation}
        \alpha\sigma := \alpha_1 \cdots \alpha_n \sigma_1 \cdots \sigma_l
        \qquad\Ra\qquad |\alpha\sigma|=n+l.
      \label{eq:multiplicationOfMultiIndicesIntro}
    \end{equation}
    If $s$ is a section of our fibered smooth manifold $\pi:U\ra O$ as above
    and $i\in\bbr{0,\cdots,m}$ an index and $\alpha=\alpha_1 \cdots \alpha_n$ a multi-index, then define
    \begin{equation}
      s^j_i := \frac{\partial s^j}{\partial x^i},\qquad
      s^j_\alpha := \frac{\partial^{n} s^j}{\partial x^{\alpha_1} \cdots \partial x^{\alpha_n}}
      \qquad
      \Ra\qquad
      s^j_{\alpha i} = \frac{\partial^{n+1} s^j}{\partial x^{\alpha_1} \cdots\partial x^{\alpha_n} \partial x^i}.
      \label{eq:multiIndexDifferentiationIntro}
    \end{equation}
    Now with all those notions, we can finally say that
  $M$ is \ti{tanget to $M'$ up to order $k$} at $p\in M\cap M'\subset E$ if
\begin{equation}
  s^j_\sigma(\pi(p))
  =(s')^j_\sigma(\pi(p)),~|\sigma|\le k.
  \label{eq:kthOrderTangency}
\end{equation}
It really just means what was said at the beginning: All derivatives of the
local sections that locally describe the submanifolds agree up to order $k$.
A visualisation of this idea is shown in figure (\ref{fig:jets}). At the point $x$, the manifold $M$ is tangent to $M'$ up to
first order and therefore they have the same 1-jet and $M$ is tangent to $M''$ up to
third order and therefore they have the same 3-jet.
\\
Importantly, one can show that if eq. (\ref{eq:kthOrderTangency}) holds for one choice of coordinates,
then it holds for all possible choices, see e.g. \cite{saunders1989}. Therefore, jets
are geometric, coordinate invariant objects.\\
To have the same jet up to some order is an equivalence relation because
it is defined via the equation (\ref{eq:kthOrderTangency}). Thus if this equation holds
for sections $s_1$ and $s_2$ and for sections $s_2$ and $s_3$, then it holds for $s_1$ and $s_3$.
In other words, a Jet $[M]_p^k$ is an equivalence class.
Using Jets, we can define Jet Spaces.
\gd{def:jetSpace}{
  The \ti{Jet Space} $J^k(E,m)$
  is defined as the set of all
  jets of order $k$ of $m$-dimensional submanifolds in $E$ at all points of $E$, i.e.
  \begin{equation}
    J^k(E,m) := \bbr{ [M]^k_p~|~M \ni p,~ \te{dim}(M)=m,~p\in E}
    \label{eq:kJetSpaceAtPoint}
  \end{equation}
}
One can show that Jet Spaces are naturally endowed with the structure of a smooth manifold
(see for instance \cite{saunders1989} again). It is thus justified to call them
smooth manifolds of jets if desired. In particular, they are objects in the category of smooth manifolds.\\
Now in order to understand that this is really the mathematical structure we were looking for,
it is important to note that a jet $[M]^k_p$ is completely determined by the derivatives
of the local section that describes $M$ around $p$. In particular, if $U_M=s(O)$ and $\pi(p)=x=(x^1,\cdots,x^m)$
is fixed and one specifies
the tuple $(x^i,s^j(x),(\partial^{|\sigma|}s^j/\partial x^\sigma)(x))$, then
the Jet $[M]^k_p$ is completely determined by this tuple. Now, if one chooses other values for
any derivative of $s$ at $x^i$, then
one will get another section that describes another submanifold of $E$ at the same point.\footnote{
  In fact, any value of $\mathbb{R}$ can be chosen for the value of the derivatives at a point which means
  that the fibered smooth manifold $\pi^{k+1}_{k}:J^{k+1}(E,m)\ra J^k(E,m)$ that will be defined below is affine
  for $k\ge 1$ (meaning that the fibers of the projection are vector spaces).
}
Thus, if one considers the union of all of those tuples at all points in the neighbourhood $U$,
one will get any tuple of the form $(x^i,u^j,u^j_\sigma)$, $1\le |\sigma|\le k$ where
$x^i,u^j,u^j_\sigma$ are now \ti{coordinate functions} that locally describe the Jet Space $J^k(E,m)$.\footnote{
  In particular, it is emphasized again that exactly as in the motivating example,
  $u^j_\sigma$ now is a name of a coordinate
  and does not denote
  the derivative of $u^j$ like in the case of a section $s^j_\sigma(x)$
  which does represent a derivative of $s^j$ of order $|\sigma|$ at $x$
  as defined in eq. (\ref{eq:multiIndexDifferentiationIntro}).
}
(For example, note that $J^0(E,m)=E$ and $J^1(E,m)$ is the space of all $m$-dimensional
subspaces of the tangent spaces $T_pE$ at all points $p\in E$).
As a consequence, one locally indeed recovers exactly what one was looking for: A
space with as many coordinates as there are derivatives up to order $k$. But at the same
time, globally a Jet Space is a union of Jets which are coordinate invariant objects.
Therefore a Jet Space is the right notion with which one can geometrise a PDE.
\\~\\
To obtain the dimension of a Jet Space,
we need to count the number of possible derivatives of $e$ dependent variables with respect to $m$ independent
variables. For the $u^j$ there are as many possibilities
to take derivatives of order $k$ as there are possibilities to put $k$ balls in between
$m-1$ sticks. Therefore
\begin{equation}
  \begin{split}
  &\te{dim}(J^k(E,m))-\te{dim}(J^{k-1}(E,m))=
  e
  \begin{pmatrix}
    m-1+k\\
    k
  \end{pmatrix}\\
  &\te{dim}(J^k(E,m)) = m+ e\sum_{i=0}^k
  \begin{pmatrix}
    m-1+i\\
    i
  \end{pmatrix}
  = m+e
  \begin{pmatrix}
    m+k\\
    k
  \end{pmatrix}\\
  \end{split}
  \label{eq:dimJetSpace}
\end{equation}
Next define projections and prolongations
\gd{def:projAndProl}{
  A \ti{projection} between Jet Spaces is defined by
  \begin{equation}
    \pi^{k}_l:J^{k}(E,m)\ra J^{l}(E,m),\qquad [M]_p^{k}\mapsto [M]_p^l,\qquad 0\le l\le k
    \label{eq:projectionOfJetSpaces}
  \end{equation}
  A \ti{$k$-Jet prolongation} is defined as a map from submanifolds of $E$ to $J^k(E,m)$ by
  \begin{equation}
    j^k(M):M\subset E\ra J^k(E,m),\qquad p\mapsto [M]_p^k
    \label{eq:jetMap}
  \end{equation}
  Furthermore, say that $M^{(k)}:= \te{im}(j^k(M))$ is a \ti{prolongation} of the submanifold $M$
  and note that $\pi^k_l\circ j^k = j^l$.
}
Recall that above, we locally chose adapted coordinates
$(x^i,u^j)$ around every point in $E$ and then defined a smooth manifold $O=\bbr{(x^i)}$
in order to obtain a fibered smooth manifold $\pi:U\ra O$. If $s(O)=U_M$, $p=(x,u)$,
then in local coordinates, one can write
\begin{equation}
  j^k(s)(x):= j^k(s(O))(s(x)) = j^k(U_M)(p) =
  j^k(M)(p) = \br{x^i,s^j_\sigma(x)}
  \label{eq:jetMapLocalCoord}
\end{equation}
In the special case where we can fix a surjective submersion $\pi:E\ra M$,
that means where we can consider a fibered smooth manifold globally, we can
do everything in the same way as above but have to observe
that $\pi$ is now globally fixed. In this case the image of every
(possibly local) section $s:O\subset M\ra E$ is again
a smooth $m$-dimensional submanifold of $E$. But \ti{not} every $m$-dimensional submanifold
of $E$ can be written as the image of such a local section,
namely those which are \ti{not horizontal} to the projection $\pi$ (which is globally fixed) can not.\\
However, the space of the Jets of the images of those sections form a dense subset
in $J^k(E,m)$ which is denoted by $J^k(\pi)$ or $J^k(E)$ and is called a \ti{Jet Bundle}.
It is thus a less general construction than a Jet Space. However,
for many purposes this construction will suffice and in
the literature it is often the only case treated.
\\~\\
Finally, the last important notion that we must introduce in this subsubsection is
that of repeated Jets. Suppose we already have a Jet Space $J^k(E,m)$.
Then we can regard $J^k(E,m)$ itself as a usual smooth manifold $E':=J^k(E,m)$
and look at \ti{its} Jet Space $J^l(E',m)$.
Locally, $J^k(E,m)$ might be described by the coordinates $(x^i,u^j_\sigma)$.
When considering it as a usual smooth manifold $E'$, then the coordinates
of $J^l(E',m)$ are $(x^i, (u^j_\sigma)_\alpha)$ where $|\sigma|\le k$ and $|\alpha|\le l$.
(Note that this is not the same as $u^j_{\sigma\alpha}$ because one ``double-counts''
those coordinates that arise from Jets of sections whose derivatives would usually commute.)\\
One would like to identify the subset of $J^l(J^k(E,m),m)$
that consists of repeated Jets.
To do so, one defines an embedding
\begin{equation}
  i_{k,l}:J^{k+l}(E,m)\ra J^l(J^k(E,m),m),~[M]^{k+l}_p\mapsto [ [ M ]^k ]^l_p
  \label{eq:repeatedJetA}
\end{equation}
In local coordinates, this embedding is
$(x^i,u^j_{\sigma\alpha}=s^j_{\sigma\alpha}(x))\mapsto (x^i,(u^j_\sigma)_\alpha=s^j_{\sigma\alpha}(x))$.
One can show that it is well defined (see \cite{saunders1989} again).\\
This embedding is important for the following reason. If one has a submanifold $J\subset J^k(E,m)$,
then one can prolong it to $J^l(E',m)$ by using $j^l(J)$ in the usual way.
To understand how to prolong it into $J^{k+l}(E,m)$,
one can first take the intersection $\te{im}(j^l(J))\cap i_{k,l}(J^{k+l}(E,m))$. In this intersection
are only points of the form $[[M]^k]^l_p$ and therefore the projection
$p:\te{im}(j^l(J))\cap i_{k,l}(J^{k+l}(E,m))\ra J^{k+l}(E,m),~[ [ M ]^k]^l_p\mapsto [M]_p^{k+l}$ is well-defined.
Thus, one can make the following definition.
\gd{def:prolSubmfd}{
  Let $J$ be a submanifold of $J^k(E,m)$. Its $l$-th \ti{prolongation (into $J^{k+l}(E)$)}
  is defined by
  \begin{equation}
    P^l(J) := p(\te{im}(j^l(J))\cap i_{k,l}J^{k+l}(E,m))
    \label{eq:prolSubmfdJetSpaceA}
  \end{equation}
}
\yr{rem:prolMightEmpty}{
  Such a prolongation might \ti{not be smooth or might not always exist}
  because the intersection $J^l(J)\cap i_{k,l}(J^{k+l}(E,m))$
  might not be smooth or empty.}
\ye{example:emptyProlong}{
  Let us consider the submanifold $M=\mathbb{R}$ of $E=\mathbb{R}\times\mathbb{R}$.
  Locally choose a projection $\pi:U\ra O$ projecting
  a subset of $\mathbb{R}\times\mathbb{R}$
  onto the first factor.
  If we choose the local coordinate $t$ for $O$ and $(t,u)$ for $U$,
  then the Jet Space $J^1(E,1)$ has local coordinates $(t,u,u_t)$
  and $J^2(E,1)$ has local coordinates $(t,u,u_t,u_{tt})$.\\
  If we consider the Jet Space $J^1(J^1(E,1),1)$, it has local coordinates
  $(t,u,u_t,(u)_t,(u_t)_t)$. To avoid confusion, it is emphasized that $u_t\ne (u)_t$
  because they are by definition different coordinates.\\
  Consider now the subspace $J$ of $J^1(E,1)$
  that is locally described by $(t,\sin(t),\sin(t))$. Its prolongation $\te{im}(j^1(J))$
  is the subset of $J^1(J^1(E,1),1)$ that
  has local coordinates $(t,\sin(t),\sin(t),\cos(t),\cos(t))$. Note that the
  third and the fourth component of this tuple are different.
  Meanwhile $i_{1,1}(J^2(E,1))$ is the subspace with local coordinates $(t,u,u_t,u_t,u_{tt})$.
  The third and the fourth component of this tuple are equal. Therefore, the prolongation
  $P^1(J)=p(J^1(J)\cap i_{1,1}(J^2(E,1)))$ consists only of those discrete points in $J^2(E,1)$
  where $\sin(t)=\cos(t)$.
}
\yr{rem:borelLemma}{
  The Borel lemma says that for every point $\theta\in J^k(E,m)$, one can
  locally find a section $s:O\ra U$ such that $\theta=j^k(s)(\pi\circ\pi^k(\theta))$.\\
  The above example shows that this does not mean that given a submanifold $J$ of $J^k(E,m)$
  with dimension $m$, one can always find a section such that $j^k(s)(O)=J$. In general this is not possible.\\
  In fact sections of the form $j^k(s)$ are exactly those that are singled
  out by the Cartan distribution as explained in the previous subsubsection and
  as will be explained in subsubsection
  [\ref{sec:CartanDis}]
  about Cartan Distributions.
}

\subsubsection{Differential Operators}

\begin{figure}
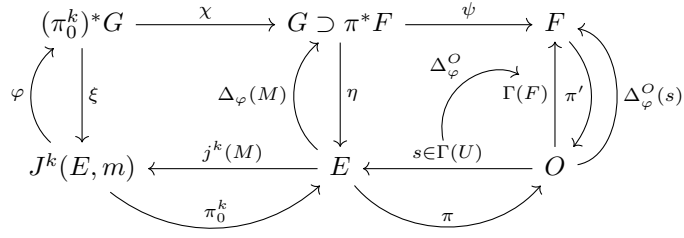

  \centering
  \bt[huge]
  (\pi^k_0)^*G \ar{r}{\chi}\ar{d}{\xi} \& G \ar{d}{\eta} \supset \pi^*F
  \ar{r}{\psi}\& F \ar[""{name=piEta},bend left=40,']{d}{\pi'}\\
  J^k(E,m) \ar[bend right=40]{r}{\pi^k_0} \ar[bend left=50]{u}{\varphi} \&
  E \ar[""{name=pi},bend right=50]{r}{\pi} \ar[']{l}{j^k(M)} \ar[bend left=50]{u}{\Delta_\varphi(M)} \&
  O \ar[',""{name=secU}]{l}{s\in\Gamma(U)}
  \ar[bend right=80,']{u}{\Delta_\varphi^O(s)} \ar[""{name=secFO}]{u}{\Gamma(F)}
  \ar[from=secU, to=secFO, shorten <=6pt,shorten >=10pt, controls={+(-0.2,1) and +(-0.7,0.4)}]{r}{\Delta_\varphi^O}
  \et
  \caption{Relationships of a differential operator acting on submanifolds $\Gamma^m(E)$ or local sections
  $\Gamma(U)$.}
  \label{fig:diffOpGeom}
\end{figure}

We will need the definition of a pullback bundle.
\gd{def:pullbackBundle}{
  If $Y$ is a smooth manifold and $f:Y\ra X$ is a smooth map and $\pi:N\ra X$ is a fiber bundle
  with fibers denoted by $N_x,~x\in X$,
  then $f^*N$ denotes the pullback bundle over $Y$. It is defined as follows:
  \begin{equation}
    f^*N := \bbr{(y,n)\in Y\times N~|~f(y)=\pi(n)} = \bbr{ N_{f(y)}~|~y\in Y}.
    \label{eq:pullbackBundleIntr}
  \end{equation}
  This means, to each point $y\in Y$, we attach the fiber $N_{f(y)}$ that would usually be attached
  to the point $x=f(y)\in X$.\footnote{
    For more details on pullback bundles and a proof that the pullback bundle
    is also the pullback in a category theoretical sense
    see \cite{tu2017differential}, section 20.4.
  }
  With the maps $\xi:f^*N\ra Y,~(y,n)\mapsto y$ and $\chi:f^*N\ra N,~(y,n)\mapsto n$,
  the following diagram commutes
  \begin{center}
  \bt
  f^*N \ar{r}{\chi} \ar{d}{\xi}\& N \ar{d}{\pi}\\
  Y \ar{r}{f}\& X
  \et
  \end{center}
}
Let $\eta:G \ra E$ be a fiber bundle. Now we use the definition above and set $N=G$, $\pi=\eta$, $X=E$,
$Y=J^k(E,m)$ and $f=\pi^k_0$ to pull back $G$ to $(\pi^k_0)^*G$.
So to each point $[M]^k_p$ of $J^k(E,m)$ we attach a fiber diffeomorphic to $G_{p}$
(where $p=\pi^k_0([M]^k_p)$).
Now we can define a differential operator as follows.
\gd{def:diffOp}{
  A \ti{differential operator} (of order $\le k$ acting on submanifolds of $E$) is a (local) section
  \begin{equation}
    \varphi: J\subset J^k(E,m) \ra (\pi^k_0)^*G
    \label{eq:diffOp}
  \end{equation}
}
(Here, $G$ serves to determine the codomain of the operator and in particular the dimensions of the fibers of $\eta:G\to E$ determine the number of differential equations that locally describe the kernel of such an operator, as we will see below in prop. \ref{prop:diffEqIsKerDiffOp}.) To understand why this definition makes sense, let us
denote by $\Gamma^m(E)$ the space of all submanifolds of dimension $m$ of $E$,
and denote by $\Gamma(G|_M)$ the space of sections with domain $M$ of $\eta:G\ra E$.
Then we can
define the operator
\begin{equation}
  \Delta_\varphi:\Gamma^m(E)\ra \Gamma(G|_M),~M\mapsto \chi\circ \varphi\circ j^k(M):
  M\ra G,~p\mapsto \Delta_\varphi(M)(p)
  \label{eq:usualDiffOp}
\end{equation}
To obtain an overview about the various maps involved, see figure (\ref{fig:diffOpGeom}).
\\
Now let us express the action of $\varphi$ and $\Delta_\varphi$ in local coordinates.
Depending on which smooth manifold $M\subset E$ we locally want to look at,
we locally single out an adapted chart $(x^i,u^j)$ which
gives rise to the local coordinates $(x^i,u^j_\sigma)$ on the Jet Space $J^k(E,m)$
and to a fibered smooth manifold structure $\pi:U\subset E \ra O$ in such a way
that there exists a local section $s:O\ra E$ such that $s(O)=U_M$.
Thus, first of all, we locally obtain for this choice of coordinates
\begin{equation}
  \varphi(x^i,u^j_\sigma)=(x^i,u^j_\sigma,\varphi^n(x^i,u^j_\sigma))
  \label{eq:diffOpLocalCoord}
\end{equation}
where $n\in\bbr{1,\cdots,N}$ and $N=\te{dim}(G)_g-\te{dim}(G)_{\eta(g)}$
is the dimension of the fiber of $\eta:G\ra E$ at $g$ (which is assumed to be locally constant).
Now recall that
$j^k(M)(p)=j^k(s(O))(s(x))=(x^i,s^j_\sigma(x))=:j^k(s)(x)$,
so that we obtain $\varphi(j^k(s)(x))=(j^k(s)(x),\varphi^n(j^k(s)(x)))$.
As a consequence, $\chi(\varphi(j^k(s)(x)))=(j^0(s)(x),\varphi^n(j^k(s)(x)))$ (where $j^0(s)(x)=s(x)$), i.e.
\begin{equation}
  \Delta_\varphi(M)(s(x))=(x^i,s^j(x),\varphi^n(x^i,s^j_\sigma(x)))
  \label{eq:diffOpLocalCoordUsual}
\end{equation}
But recall that $s^j_\sigma(x)=\partial^{|\sigma|} s^j/\partial x^\sigma(x)$ denotes the derivatives of
$s$ up to order $k$ at $x$! Therefore locally
$\Delta_\varphi(M)$
maps sections (that locally describe submanifolds $M$ of $E$)
to any function of derivatives up to order $k$ of those sections.\\
To recover the usual definition completely,
let us use again the fibered structure
$\pi:U\ra O$.\footnote{
  Again, this does not mean that all submanifolds in $U$ are local sections of $\pi:U\ra O$,
  it just means that no matter at which submanifold $M$ in $E$ we are looking,
  we can locally choose a suitable $\pi:U\ra O$ such that locally in $U$, $M$ is the image of one of its sections.
}
Thus we can define the fiber bundle over $O$ that consists of
the fibers $G_{s(x)}$ of $G$ over $U_M$, i.e. $\pi':F:=\bbr{ F_{s(x)} ~|~x\in O}\ra O$.
This means that $G$ restricted to the points $g\in G$ for which $\eta(g)=p\in U_M$ is the pullback bundle of $F$.
More concretely, $\pi^*F=\bbr{ F_{\pi(p)}~|~p\in U_M} = \bbr{ g\in G~|~\eta(g) \in U_M}$
and therefore $\pi^*F$ can be identified with a subset of $G$. Because $G$ is a fiber bundle over $E$,
every point can locally be written $g=(p,g_p)$ and there is a natural projection
given by $\psi:\pi^*F\ra F,~\psi(g)=\psi(p,g_p)=(\pi(p),g_p)$.\\
Now define another operator that locally acts on the sections of $\pi:U\ra O$ by
\begin{equation}
  \Delta_\varphi^O:\Gamma(U)\ra \Gamma(F),~s\mapsto \psi\circ\chi\circ\varphi\circ j^k(s(O))\circ s:
  O\ra F,~x\mapsto \Delta_\varphi^O(s)(x)
  \label{eq:veryUsualDiffOp}
\end{equation}
Having defined all this, we use (\ref{eq:diffOpLocalCoordUsual}) to write down
$\Delta_\varphi^O(s)(x)=\psi(\Delta_\varphi(M)(s(x)))$
in local coordinates:
\begin{equation}
  \Delta_\varphi^O(s)(x) = (x^i,\varphi^n(x^i,s^j_\sigma(x)))
  \label{eq:diffOpUsual}
\end{equation}
This is exactly a differential operator imposed on sections of a fiber bundle in the usual sense.
In particular, if we only look at those submanifolds
that are
all $\pi$-horizontal sections
of some globally fixed projection of fibered smooth manifolds $\pi:E\ra M$,
then $\pi^*F = G$ and we can define $G$ in terms of $F$.
Thus, if one defines $\pi^k:=\pi\circ \pi^k_0$, then in this case $(\pi^k_0)^*G=(\pi^k)^*F$.
Furthermore, the Jet space
of this fibered smooth manifold is $J^k(\pi)=:J^k(E)$.
Therefore, if we define $\Phi:=\psi\circ\chi\circ\varphi$, then we obtain the diagram
\begin{equation}
  \bt
  J \subset J^k(E) \ar{drr}{\Phi} \ar{d}{\pi^k_0}
  \ar{rr}{\varphi}
  \& \& (\pi^k)^*F \ar{d}{\psi\circ\chi} \\
  E \& \& F \\
   \&\& \\
   \& M
   \ar[""{name=E}]{uul}{\Gamma(E)}
   \ar[',""{name=F}]{uur}{\Gamma(F)}
   \ar[from=E, to=F, bend left=50, shorten <=6pt,shorten >=6pt]{r}{\Delta_\varphi^M}
\et
  \label{eq:diffOpAnalyticViewRecovered}
\end{equation}
The reader is recommended to contrast this with the diagram (\ref{eq:diffOpAnalyticView}) of the analytic
view. In particular, note the important fact that
$\varphi$ and $\Phi$ are \ti{morphisms in the category of smooth manifolds}
while $\Delta_\varphi$ and $\Delta_\varphi^M$ are not (they are operators on infinite-dimensional spaces of sections).
\yr{rem:difOp}{
  If $\pi:E\ra M$ is a fibered smooth manifold,
  then, for every section $\varphi:J^k(E)\ra (\pi^k)^*F$ one obtains
  a differential operator $\Delta_\varphi^M$. Conversely, given a usual differential operator
  $\Delta:\Gamma(E)\ra\Gamma(F),~s\mapsto\Delta(s)$ of order $k$, one can
  define $\varphi:J^k(E)\ra (\pi^k)^*F$
  by $\varphi(j^k(s)(x)):=(j^k(s)(x),\Delta(s)(x))$. Then $\Delta=\Delta_\varphi^M$.
Therefore, there is a 1-to-1 correspondence between usual differential operators and
the geometric operators $\Delta_\varphi^M$.\\
However the geometric definition [\ref{def:diffOp}]
is more fundamental because one can impose the operator $\Delta_\varphi$ on any submanifolds
of $E$ and not only those that are $\pi$-horizontal to a projection
$\pi:E\ra M$.
}
Because the case of a fibered smooth manifold is especially important in practice, we take the freedom
to cause some confusion by
calling $\Phi$ a differential operator as well. By definition, any morphism of fibered smooth manifolds\footnote{
  A morphism of fibered smooth manifolds with the same base is a map that preserves the base space.
  If $\pi:E\ra M$ and $\pi':E'\ra M'$ are fibered smooth manifolds,
  then a map $\Phi:E\ra E'$ is called a morphism of fibered smooth manifolds
  if there exists a map $\phi:M\ra M'$ such that $\pi'\circ\Phi=\phi\circ\pi$.
  In the case where $M=M'$, one can choose $\phi=\te{id}$, i.e.
  then a map is a morphism of fibered smooth manifolds if $\pi'\circ\Phi=\pi$.\\
  In the present case $\Phi:J\ra F$ is indeed a morphism of fibered smooth manifolds
  because $\Phi(x^i,j^k(s)(x))=(x^i,\varphi^n(j^k(s)(x)))$, i.e. the
  $x^i$-components are preserved.
}
is in correspondence with some $\Phi=\psi\circ\chi\circ\varphi$, and therefore the following
definition makes sense.
\gd{def:diffOpFiber}{
  If $\pi:E\ra M$ and $\pi':F\ra M$ are fibered smooth manifolds,
  any morphism of fibered smooth manifolds
  \begin{equation}
    \Phi: J\subset J^k(E)\ra F
    \label{eq:diffOpFiber}
  \end{equation}
  is also called a \ti{differential operator} (of fibered smooth manifolds of order $\le k$).
}
A very important concept related to differential operators is that of a prolongation
because it will be used to determine whether systems of differential equations
that we will encounter later are consistent (formally integrable).
\gd{def:prolongation}{
  The $l$-th \ti{prolongation} of a differential operator $\Phi:J\ra F$ of fibered smooth manifolds is the map
  \begin{equation}
    \begin{split}
      &p^l(\Phi): P^l(J) \ra J^l(F),\qquad
      j^{k+l}(s)(x)\mapsto j^l(\Phi(j^k(s)(x)))
    \end{split}
    \label{eq:prolongDiff}
  \end{equation}
}
Note that by remark [\ref{rem:prolMightEmpty}], the prolongation of $\Phi$ might not exist
if $P^l(J)$ is empty.\\
To express the prolongation in local coordinates, we will need the
total differential operator.
\gd{def:totalDiffOp}{
  The \ti{total differential operators} $D_i,~i\in\bbr{1,\cdots,m=\te{dim}(M)}$ are locally defined by
  \begin{equation}
    D_i := \frac{\partial}{\partial x^i} +
    \sum_{j=1}^e \sum_{|\sigma|\ge 0} u^j_{\sigma i} \frac{\partial}{\partial u^j_\sigma}
  \label{eq:totalDiffOp}
  \end{equation}
  Write $D_i^2 = D_i\circ D_i$ and
  if $\alpha=\alpha_1 \cdots \alpha_n$ is a multi-index, write
  $D_\alpha := D_{\alpha_1} \circ \cdots \circ D_{\alpha_n}$.
}
(If one defines $J^\infty(E)$ as the inverse limit of the sequence
\begin{equation}
  \bt
  J^0(E) \& \ar[']{l}{\pi^1_0} J^1(E) \& \ar[']{l}{\pi^2_1} J^2(E) \& \ar[']{l}{\pi^3_2} \cdots
  \et
  \label{diag:jInftySeq}
\end{equation}
then $D_i$ can be understood as vector fields on $J^\infty(E)$.)
Now we express the prolongation in local coordinates.
If, locally, $\Phi$ is written
$\Phi(x^i,u^j_\sigma)
=(x^i,\Phi^h(x^i,u^j_\sigma))$, where $|\sigma|\le k$, $h\in\bbr{1,\cdots,H}$ and $H$ is the dimension
of the fiber of $\pi':F\ra M$, then we locally have
\begin{equation}
p^l(\Phi)(x^i,u^j_{\sigma \alpha}) = (x^i,D_\alpha \Phi^h(x^i,u^j_{\sigma})),
  ~
  0 \le |\sigma| \le k,~
  0 \le |\alpha| \le l.
  \label{eq:prolonDiffOpExplicit}
\end{equation}
To define a differential equation in the next subsubsection, we will need the notion of kernel.
\gd{def:kernelOfDifOp}{
  If $s$ is a local section of $\eta:G\ra E$,
  the \ti{kernel} $\ker_s(\varphi)$ of a differential operator $\varphi:J\subset J^k(E,m)\ra (\pi^k_0)^*G$
  is locally defined by
  \begin{equation}
    \ker_s(\varphi):= \bbr{ \theta \in J~|~\chi(\varphi(\theta))=s(\pi^k_0(\theta))}
    \label{eq:kerDiffOp}
  \end{equation}
}
In particular, if we have a fibered smooth manifold $\pi:E\ra M$, and an operator $\Phi:J\ra F$
such that $\Phi=\psi\circ\chi\circ\varphi$, then for a section $s:M\ra F$, we have
\begin{equation}
  \ker_s(\Phi) = \bbr{ \theta \in J~|~\Phi(\theta)=s(\pi^k(\theta))}
  \label{eq:kerDiffOpFibered}
\end{equation}
In local coordinates, we have
\begin{equation}
  \ker_s(\varphi) = \bbr{ (x^i,u^j_\sigma) \in J~|~\varphi^n(x^i,u^j_\sigma)=s^n(x^i)}
  \label{eq:kerLocal}
\end{equation}
where $n\in\bbr{1,\cdots,N}$ and $N$ is the dimension of the fiber $\eta:G\ra E$
which is the same as the dimension of the fiber $\pi':F\ra M$ in the fibered case.
If we have a vector bundle $\pi:E\ra M$, then there is a zero section $0:M\ra E$
given by $0(x^i):=(x^i,0)$ and we define $\ker_0(\varphi)=:\ker(\varphi)$.
In the general case, such a section can only be defined locally after a choice of coordinates.\\

\subsubsection{Differential equations}

\gd{def:diffEq}{
A \ti{differential equation} (of order $\le k$) is a submanifold $\mathcal{E}\subset J^k(E,m)$.
}
In particular, note that $E=J^0(E,m)$, so any submanifold and therefore also any smooth manifold
is a differential equation of order $0$.\\
Locally one can always describe a smooth submanifold by a set of equations
imposed on the coordinates of the embedding smooth manifold. Therefore, we can locally describe
a smooth smooth manifold $\mathcal{E}$ by a set of equations imposed on the coordinates of $J^k(E,m)$.
Such a set of equations can always be expressed as the kernel of a map $\varphi:J^k(E,m)\ra (\pi^k_0)^*G$
for an appropriately chosen fibered smooth manifold $\eta:G\ra E$. Therefore, the following proposition holds
(see also \cite{goldschmidt1967})
\bp{prop:diffEqIsKerDiffOp}{
  If $\varphi:J\subset J^k(E,m)\ra (\pi^k_0)^*G$ is a differential operator
  and $s:U\ra G$ is a local section of $\eta:G\ra E$ such that
  \begin{equation}
    s(U)\subset \chi(\varphi(J))\te{ and }\te{rank}(\varphi)\te{ is locally constant}
    \label{eq:bedDiffOp}
  \end{equation}
  or if $\Phi:J\subset J^k(E)\ra F$ is a differential operator
  and $s:U\ra F$ is a local section of $\pi':F\ra M$ such that
  \begin{equation}
    s(U)\subset \Phi(J)\te{ and }\te{rank}(\Phi)\te{ is locally constant}
    \label{eq:bedDiffOpFibered}
  \end{equation}
  then $\ker_s(\varphi)$ or $\ker_s(\Phi)$ is a differential equation (of order $\le k$).
}
Note that $\ker_s(\Phi)$ corresponds to the definition of a differential equation
in the usual sense while $\ker_s(\varphi)$ is more general.
Another very important concept needed later is that of the prolongation of an equation.
\gd{def:prolEq}{
  The $l$-th \ti{prolongation $P^l(\mathcal{E})$ of a differential equation}
  $\mathcal{E}\subset J^k(E,m)$ is defined by $P^l(\mathcal{E}):=\mathcal{E}^{(l)}$.\footnote{
    See definition [\ref{def:prolSubmfd}] for the definition of $P^l(\mathcal{E})$.
  }
}
By remark [\ref{rem:prolMightEmpty}], such a prolongation might not exist. If $\mathcal{E}=\ker_s(\varphi)$,
then by definition $P^l(\mathcal{E}) = \ker_{p^l(s)}(p^l(\varphi))$.
Then we can locally write
\begin{equation}
  P^l(\mathcal{E}) = \ker_{p^l(s)}(p^l(\varphi)) =
  \bbr{ \theta \in J^{(l)}~|~D_\alpha\varphi^n(\theta) = D_\alpha s^n(\pi^{k+l}_0(\theta)),~|\alpha|\le l}
  \label{eq:prolongKerDiffEqLocalCoords}
\end{equation}
In order to define solutions of a PDE, we go on to define the Cartan distribution in the next subsubsection.

\subsubsection{Cartan Distribution}
\label{sec:CartanDis}

As already explained in the detailed motivating example in section
\ref{sec:chap3utu}, one desires to have a geometric object called Cartan Distribution that
defines in an intrinsic way what solutions of a differential equation are.
In the motivating example, we declared a sub-bundle of the tangent bundle
as the Cartan Distribution.
In particular, we defined it as the span of lines tangent to prolonged sections.
Below we will do the same but for arbitrary dimensions and
Jet Spaces. Instead of lines tangent to prolonged sections, we will define planes
tangent to prolonged submanifolds.\\
Recall that if $\theta\in J^k(E,m)$ and $M$ is a submanifold of $E$, then its prolongation is
denoted (cf. [\ref{def:projAndProl}]) by $M^{(k)}=\te{im}(j^k(M))$. Use this for the following
\gd{def:Rplane}{
  An \ti{$R$-plane} at a point
  $\theta \in J^k(E,m)$
  is defined to be a subspace of the tangent space $T_\theta(J^k(E,m))$
  of the form $T_\theta(M^{(k)})$
  for any submanifold $M$ of $E$ (whose prolongation contains the point $\theta$).
}
\gd{def:cartanDis}{
  The span of all $R$-planes at a point $\theta\in J^k(E,m)$ is denoted by $\de{C}_\theta$.
  The map
  \begin{equation}
    \de{C}:J^k(E,m)\ra TJ^k(E,m),\qquad \theta\mapsto \de{C}_\theta \in T_\theta(J^k(E,m))
    \label{eq:CartanDistribution}
  \end{equation}
  is called \ti{Cartan Distribution} (on $J^k(E,m)$).
}
In particular, as $J^k(E)$ is dense in $J^k(E,m)$, its Cartan distribution is obtained by
restricting the Cartan Distribution of $J^k(E,m)$, i.e. $\de{C}_\theta$ is replaced by
$\de{C}_\theta\cap T_\theta(J^k(E))$ and at every point it is the span of $R$-planes
of prolongations of sections of the fibered smooth manifold $\pi:E\ra M$ whose Jet Bundle is $J^k(E)$.\\
Next, recall that in the motivating example, solutions were simply defined as those
sections of $\pi^1:J^1(E)\ra M$ whose image curves were tangent to the Cartan Distribution.
Here we will define solutions in a more general way by replacing curves with $m$-dimensional submanifolds.
\gd{def:integralSubmfd}{
  A submanifold $W\subset J^k(E,m)$ is said to be \ti{integral (for the Cartan
  Distribution)} if $T_\theta W\subset \de{C}_\theta$ for all $\theta\in W$.
  \\
  An integral submanifold is said to be \ti{locally maximal} if no open subset
  of $W$ can be embedded into an integral submanifold of greater dimension.
}
By definition, the prolongations $M^{(k)}$ of submanifolds $M$ of $E$ are integral submanifolds
of the Cartan distribution. 
This shows that the Cartan distribution is the geometric structure we were looking for.
Similarly to expression (\ref{eq:cartanDisutu}),
one can express the Cartan Distribution $\de{C}$ of $J^k(E,m)$ in the general case in local
coordinates. If one locally chooses a surjective submersion $\pi:U\subset E\to O\subset M$, and coordinates that are adapted to this projection, $(x^i,u^j_\sigma)$ with $|\sigma|\le k$ for $J^k(E,m)$,
then one has
\begin{equation}
  \de{C} = \te{span}\br{ D_i^T, \frac{\partial}{\partial u^j_\gamma} },\qquad|\gamma|=k,\qquad
  D_i^T := \frac{\partial}{\partial x^i} + \sum_{j=1}^e \sum_{|\sigma| < k}u^j_{\sigma i}
  \frac{\partial}{\partial u^j_\sigma}
  \label{eq:CartanDisLoc}
\end{equation}
where the $D_i^T$ are called \ti{truncated total derivative} operators.
Note that in the case where $k$ is big, there are in general many derivatives of the form
$\partial/\partial u^j_\gamma$ with $|\gamma|=k$. However, when passing to the limit
$k\ra \infty$, then locally $\de{C}=\te{span}(D_i)$ and thus $\de{C}$ becomes $m$-dimensional.
\\
One can also look at the Cartan Distribution of a submanifold of $J^k(E,m)$ without
the need to consider it inside $J^k(E,m)$. To do so, one defines the restriction
of the Distribution to a submanifold of $J^k(E,m)$ as follows.
\gd{def:cartanDisSubmfd}{
  If $J\subset J^k(E,m)$, then its Cartan Distribution is defined by
  \begin{equation}
    \de{C}(J):=\bbr{ \de{C}_\theta\cap T_\theta(J)~|~\theta\in J}
    \label{eq:cartandissubmfd}
  \end{equation}
}
In this way, a PDE can be thought of as a manifold $J$, equipped with a certain distribution $C(J)\subset TJ$ and can be studied without considering $J^k(E,m)$.

\subsubsection{Solutions of a PDE}
\label{sec:solutionsOfPde}

\gd{def:solutionPDE}{
  A \ti{solution} of a differential equation $\mathcal{E}\subset J^k(E,m)$ is a submanifold $S \subset \mathcal{E}$ which is integral for the Cartan distribution and locally maximal.\footnote{One could also more briefly write: 'A \ti{solution} of a PDE $\mathcal{E}$ is a locally maximal integral submanifold of $\mathcal{E}$.'}
}
As mentioned in the previous section, every prolongation $M^{(k)}$ is a locally maximal integral submanifold of the Cartan distribution. As a consquence, whenever one can find a submanifold $M\subset E$
 such that $\te{im}(j^k(M))\subset \mathcal{E}$, then one has found a solution of $\mathcal{E}$.
\yr{rem:singSolutions}{
 Please note that the above definition \ref{def:solutionPDE}, in contrast to the non-geometric definition, is 
general enough to accomodate the treatment of certain singular solutions as well, like
for example shock wave solutions, while staying in the category of smooth manifolds. This is due to the fact that such singularities vanish in higher order jet spaces where the solutions become smooth submanifolds. The best way to understand this is to consider an example. The reader is thus highly encouraged to take a close look at example \ref{ex:singularSolutions} in the main text.
}
Finally, let us express the above defined notion of a solution in local coordinates to see that it is actually corresponding to the usual analytic notion of a solution that we would expect, when considering the appropriate special case. To this end, choose a local neighbourhood $U$ of $E$ and a surjective submersion $\pi:U\ra O$, which is such that the image of the section $s:O\to U$ locally describes the manifold $M$ and additionally assume that the solution of $\mathcal{E}\subset J^k(E,m)$ is given by $M^{(k)}$ for this $M\subset E$. Under those conditions, the prolongations of $s$ locally describe the prolongations of $M$ and the solution is thus the section $s:O\ra U$
such that $j^k(s)(O)\subset \mathcal{E}$.
This notion agrees with the usual notion of smooth solution in analysis if $\mathcal{E}=\ker_{s'}(\varphi)$
(for a section $s':O\ra F$)
because then equation (\ref{eq:kerLocal}) holds locally and
thus a solution is a section that fulfills $\varphi^n(x^i,s^j_\sigma(x))=(s')^n(x^i)$
which is precisely the solution of a system of $N$ differential equations of order $\le k$ in the usual sense.
\ye{ex:diffEq}{
  Consider the fibered smooth manifold (in fact vector bundle) $\pi:E:=\mathbb{R}\times\mathbb{R}\ra M:=\mathbb{R}$
  again. $J^1(E)$ has local coordinates $(t,u,u_t)$.
  Define the differential equation $\de{E}$ as the subset
  $\de{E}:=\bbr{(\rho,\lambda,\lambda)~|~\rho,\lambda\in\mathbb{R}}$.\\
  It can be written as the kernel of the differential operator
  $\varphi:J=J^1(E)\ra F:=\mathbb{R}\times\mathbb{R}=E$ if we define
  $\varphi(t,u,u_t):=(t,u_t-u)$. Then $\ker_0(\varphi)=\bbr{\theta\in J~|~(t,u_t-u)=(t,0)}=\de{E}$.\\
  The prolongation $J^{(1)}=(J^1(E))^{(1)}=p(J^1(J^1(E))\cap i_{1,1}(J^2(E))=J^2(E)$
  with local coordinates $(t,u,u_t,u_{tt})$. Thus
  the prolongation of the operator $\varphi$ is a map $p^1(\varphi):J^{(1)}\ra J^1(F)$
  with action $p^1(\varphi)(t,u,u_t,u_{tt})=(t,u_t-u,u_{tt}-u_t)$.
  The kernel of the prolongation is $\de{E}^1=\bbr{(\rho,\lambda,\lambda,\lambda)~|~\rho,\lambda\in\mathbb{R}}$.\\
  The solutions of the equation are $s(t):=(t,A\exp(t))$
  where $A\in\mathbb{R}$. Those are sections $s:M\ra E$ such that
  $j^1(s)(t) = (t,A\exp(t),A\exp(t)) \subset \de{E}$ for all $t$, i.e. $j^1(s)(M)\subset \de{E}$.
  There are no other sections that fulfill this requirement.
  Their images are 1-dimensional submanifolds of the 2-dimensional smooth manifold $\de{E}$
  which is itself a submanifold in the 3-dimensional Jet Space $J^1(E)$.\\
  Furthermore, note that a point $\theta\in \de{E}$ is a solution of the differential equation
  up to first order. For example, if one chooses the point $\theta=(2,4,4)$, then
  we can define $t_0:=t(\theta)=2$, $u_0:=u(\theta)=4$ and $(u_t)_0:=u_t(\theta)=4$.
  Using them, we can define the taylor expansion $T(t):=u_0+(u_t)_0(t-t_0)$.
  Prolonging $T(t)$ once results in $T_t(t) = (u_t)_0 = u_0$. Thus up to first order,
  they both equal $u_0$.
  This can be carried through for a prolongation of any order. At second order,
  we could use a point $\theta\in \de{E}^1$ to define a taylor expansion that agrees up to order two.
  If we go up all the way until infinity, then a single ``point'' of $\de{E}^\infty$
  gives a taylor expansion that solves the equation exactly around the projection of that point,
  in this case the taylor expansion would result in the
  exponential function with some fixed coefficient $A\in\mathbb{R}$.\\
  Therefore, intuitively, the space $\de{E}^\infty$ can be thought of as the
  (formal) ``space of solutions'' of $\de{E}$. 
  Note that if a differential equation has no formal solutions,
  then the limit $\de{E}^\infty$ does not exist.
}

Finally, it is remarked that having geometrised our differential equations, one can also define them
in a category-theoretical way. As they are the kernels of morphisms in the category of smooth manifolds,
one can use the category-theoretical definition of a kernel.
To do this, we need to introduce the notion of an equalizer. Following \cite{leinster2014}, we give the following
definition.
\gd{def:equalizer}{
  A \ti{fork} in some category consists of objects and maps
  \begin{equation}
    \bt
    A \ar{r}{f} \& X \ar[bend left=30]{r}{s}\ar[bend right=30,']{r}{t} \& Y
    \et
    \label{diag:fork}
  \end{equation}
  such that $s\circ f=t\circ f$.\\
  An \ti{equalizer} of $s$ and $t$ is an object $E$ together with a map $i$
  such that
  \begin{equation}
    \bt
    E \ar{r}{f} \& X \ar[bend left=30]{r}{s}\ar[bend right=30,']{r}{t} \& Y
    \et
    \label{diag:equalizer}
  \end{equation}
  is a fork, and with the property that for any other fork (\ref{diag:fork}) there exists a unique
  map $\bar f: A\ra E$ such that the following diagram commutes.
  \begin{equation*}
    \bt
    A \ar{d}{\bar f} \ar{dr}{f}\&\\
    E \ar{r}{i} \& X
    \et
  \end{equation*}
}
If two maps $s$ and $t$ are transversal,
their equalizer in the category of smooth manifolds exists and
is isomorphic to the object $E:=\bbr{ x \in X~|~s(x)=t(x)}$ with the inclusion map $i:E\ra X$
Then, by (\ref{eq:kerDiffOp}), we can recover the submanifold $\de{E}$
corresponding to the kernel of a differential operator as the equalizer of the diagram
\begin{equation}
  \bt[large]
  \de{E}\ar[hookrightarrow]{r}\& J^k(E,m) \ar[bend left=30]{r}{\chi\circ\varphi}
  \ar[bend right=20,']{r}{s\circ\pi^k_0} \& G
  \et
  \label{eq:equalizersmooth manifolds2}
\end{equation}
Note that such an equalizer \ti{does not necessarily exist}. It does only exist
if the condition (\ref{eq:bedDiffOp}) or (\ref{eq:bedDiffOpFibered}) of proposition
[\ref{prop:diffEqIsKerDiffOp}] holds. 

\subsubsection{Diffiety and Vinogradov sequence}
\label{sec:diffieties}

As we already saw in example [\ref{ex:diffEq}], the infinite prolongation of
a differential equation $\de{E}$ plays a very important role because
it corresponds to the space of formal solutions of the differential equation.
This infinite prolongation is the object that takes into account
all differential consequences, that means all equations that arise
from the system $\de{E}$ by taking any number of total derivatives or prolongations.
In algebraic geometry so-called \ti{varieties} are the central object of study.
They are algebraic ideals
that take into account all algebraic consequences
obtained by algebraically manipulating an equation, for instance by using multiplication or addition.
Therefore, and because of the fact that all differential consequences are
related to the solution space of a differential equation,
one expects that the differential ideal that takes into account those consequences
must also be a central object in the theory of differential equations
(at least those that are known to have formal solutions).
Therefore, one would like to define something like a \ti{diff}erential var\ti{iety},
or in short \ti{diffiety}. Such an object was indeed invented, namely by A. M. Vinogradov, see \cite{vinogradov1984diffiety}.
We follow him here to define at least so-called elementary diffieties using the notion of prolongation already introduced above.
\gd{def:diffiety}{
  If $\de{E}$ is a $k$-th order differential equation,
  its \ti{elementary diffiety} is the pair $(\de{E}^\infty,\de{C}(\de{E}^\infty))$.
}
A diffiety is an object that ``locally is'' an elementary diffiety. What exactly
this means is however more difficult to define and will not be done here.
Furthermore, by remark [\ref{rem:prolMightEmpty}] and the example below, $\de{E}^\infty$ might not exist.
To show that it is well-defined also requires a careful introduction
of the concept of $J^\infty(E)$ because it is infinite dimensional
(such an introduction is given e.g. in \cite{Krasilshchik1999}, chapter 4).
$J^\infty(E)$ can actually be shown to be a so-called \ti{profinite dimensional smooth manifold}, see
\cite{gueneysuPflaum2017}. All this will not be discussed further here.\\
Given a diffiety, one can use it to analyze the properties of the formal solution space of
a differential equation. In particular, one can obtain the symmetries and conservation
laws of a differential equation. Noether's theorem relates continuous symmetry transformations
to conservation laws but in order to obtain a conservation law in this way,
one must know which symmetry transformations exist. This can be done
studying a diffiety. In particular, in many cases, one can
determine \ti{the complete set} of symmetries and conservation laws.\\
In order to extract information from a diffiety, Vinogradov invented the
so-called $\de{C}$-spectral sequence. It already came up in the 70's and 80's but
the interested reader is referred to \cite{vinogradov2001}, \cite{Krasilshchik1999}
and \cite{Krasilshchik1998homMeth}.
In the following, the $\de{C}$-spectral sequence is briefly sketched and referred to simply as the \ti{Vinogradov sequence}.\\
As a first step, let $\de{O}:=\de{E}^\infty$ be the formal solution space of a diffiety
$(\de{E}^\infty,\de{C}(\de{E}^\infty))$.
      Now define $\Lambda(\mathcal{O}):=\sum_{i\ge 0} \Lambda^i(\mathcal{O})$ to
      be the algebra of differential forms over $\mathcal{O}$.
      Consider the corresponding de Rham complex:
      \begin{equation}
      \bt
        C^\infty(\mathcal{O}) \ar{r}{\text{d}_0}
        \& \Lambda^1(\mathcal{O}) \ar{r}{\text{d}_1}
        \& \Lambda^2(\mathcal{O}) \ar{r}{\text{d}_2}
        \& \cdots
      \et
        \label{eq:deRhamComplexDiffiety}
      \end{equation}
      Its cohomology groups $H^i(\mathcal{O}):=\text{ker}(\text{d}_i)/\text{im}(\text{d}_{i-1})$
      contain topological (structural) information about the PDE.\\
      Even more information can be extracted when taking the Cartan distribution into account.
      This is what the Vinogradov sequence will facilitate.
      To this end, let $\mathcal{C}\Lambda(\mathcal{O})=\sum_{i\ge 0}\de{C}\Lambda^i(\de{O})$
      be the submodule of
      differential forms $\Lambda(\mathcal{O})$ over $\de{O}$
    whose restriction to the distribution vanishes.
    This means
    \begin{equation}
      \mathcal{C}\Lambda^p(\de{O}) \ni w\text{ iff }w(X_1,\cdots,X_p)=0~\forall~X_1,\cdots,X_p\in \mathcal{C}(\de{O})
      \label{eq:filtrationOfSpecSeq}
    \end{equation}
    Let $\mathcal{C}^k\Lambda(\de{O})$ be its $k$-th power, i.e.
    the linear subspace of $\mathcal{C}\Lambda$ generated
    by $w_1 \wedge \cdots \wedge w_k,~w_i\in \mathcal{C}\Lambda$.\\
        Now one obtains a filtration
        \begin{equation}
          \Lambda(\mathcal{O}) \supset
          \mathcal{C}\Lambda(\mathcal{O}) \supset
          \mathcal{C}^2\Lambda(\mathcal{O}) \supset
          \cdots
        \end{equation}
        and all ideals $\de{C}^k\Lambda$ are stable because
        $\text{d}(\mathcal{C}^k\Lambda^i(\mathcal{O}))\subset \mathcal{C}^k\Lambda^{i+1}(\mathcal{O})$.
        Therefore, the filtration of modules completely determines a spectral sequence.
        (For more information on how this works, see some book on homological algebra, for example
          \cite{rotman1979introduction}.
        Spectral sequences in general simplify the calculation of cohomology classes.)
        We denote this sequence by
        \begin{equation}
          \mathcal{C}E(\mathcal{O})=\bbr{E^{p,q}_r,\text{d}_r^{p,q}}\qquad\te{where}\qquad
          E^{p,q}_0 := \frac{\de{C}^p\Lambda^{p+q}(\de{O})}{\de{C}^{p+1}\Lambda^{p+q}(\de{O})},\qquad\te{and}\qquad
          E_{r+1}^{p,q} := H(E_{r}^{p,q},d_r^{p,q})
          \label{eq:specSeq}
        \end{equation}
        The filtration above is finite in each degree, that means
        \begin{equation}
          \Lambda^k(\de{O}) \supset \de{C}^1\Lambda^k(\de{O}) \supset \cdots \supset \de{C}^{k+1}\Lambda^k(\de{O})
          = 0
          \label{eq:specSeqFiniteDegree}
        \end{equation}
        If the filtration is finite in this sense, then the spectral sequence converges
        (see also \cite{rotman1979introduction}, chapter 10.3 for instance)
        to the de Rham cohomology $H(\de{O})$ (of the diffiety).
        Therefore, one can now analyze the terms of the spectral sequence order by order.
        This is done for example in chapter 5 of \cite{Krasilshchik1999}. Here, it is only summarized which
        information is contained in the Vinogradov sequence.
    \begin{enumerate}
      \item
        $E_1^{0,n}$ corresponds to action functionals constrained by the PDE $\mathcal{E}$
        and for $\mathcal{L}\in E_1^{0,n}$, the corresponding Euler-Lagrange equation
        is $\text{d}_1^{0,n} \mathcal{L}=0$.
      \item
        $E_1^{0,n-1}$ corresponds to conservation laws for solutions of $\mathcal{E}$.
      \item
        $E_2$ is interpreted as characteristic classes of bordisms of solutions of $\mathcal{E}$.
      \item
        There are still many terms awaiting an interpretation.
    \end{enumerate}
  In this article, the Vinogradov sequence is not explicitly required. It is however
  of conceptual importance because it can be used to investigate any differential equation
  with a well-defined formal solution space.
  Therefore, as soon as an intersection of theories turns out to be consistent,
  we know that it can be turned into a diffiety and that those strong
  homological methods described above can be applied to investigate it.
  \\~\\
  Back to section \ref{sec:notationPreliminaries}.

  \newpage
  \section{Program to calculate rank of symbols of tensorial systems}
  \label{app:formalIntSolver}
  
  As systems of linear tensorial equations arise in the process of determining formal integrability,
  A program was written that generates the matrix of the linear system
  and calculates its rank.\\
  The commutation relations of derivatives require to make use of permutation functions
  and are what makes the generation of the matrix a bit complex.\\
  The implementation is sympy-based to facilitate symbolic calculation.
  This makes it possible to calculate the rank of the symbol of non-linear equations
  because (tensorial) coefficients can be taken into account. One can also set them to an arbitrary value
  after the generation of the matrix to see how the rank changes.
  The sympy implementation is of course not as efficient as a numpy (or C$^{++}$, Fortran, etc. based)
  implementation but as the
  matrices that arose above have at most a few thousand entries, this is perhaps okay.
  (Another implementation would perhaps have made support for symbolic calculations more difficult.)\\
  Below are also given examples of how to execute the code.
  
  \subsection{Program code}

  \pythonexternal{formalIntegrabilityPythonCode/formalIntegrabilitySolver.py}
  ~
  \subsection{Examples of application of program}
  
  \subsubsection{Calculation of symbol $g^{2,3}$ of Maxwell-equations}
  \pythonexternal{formalIntegrabilityPythonCode/formalIntegrabilitySolverExecutionMWg23.py}
  \includegraphics[width=.45\textwidth]{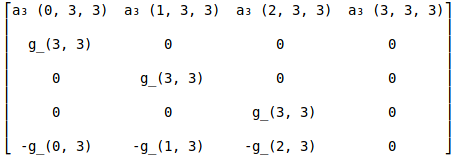}\\
  ('4 x 4 Matrix.', 'Rank: 3')
  \subsubsection{Calculation of symbol $g^3$ of the intersection of magneto-statics and hydro-dynamics}
  \pythonexternal{formalIntegrabilityPythonCode/formalIntegrabilitySolverExecutionHydrog3.py}
  ('64 x 60 Matrix.', 'Rank: 44')
  
\newpage
\section{Axiomatic derivation and empirical limits of Maxwell's equations}
\label{app:axForm}

\subsection{Axiomatic derivation}
\label{sec:axDeri}
The present subsection shall provide a more in-depth understanding of the laws of
Electrodynamics by providing an axiomatic derivation of Maxwell's equations.

\bigskip

One beautiful axiomatic approach to electrodynamics is given by
\cite{Zirnbauer1998}.
He explains in detail why differential forms are the natural candidates for observables in electrodynamics
and provides prescriptions on how to measure the formal quantities in principle.
Below we follow him and summarize his more detailed account.
\begin{enumerate}
  \item axiom:
    There exists a quality that we call (electric) charge
    and its amount per volume can be quantified in terms of a space- and time-dependent density.
  \item axiom:
    The spacetime
    we live in can
    be modeled as 4-dimensional Lorentzian manifold
    equipped with a metric $g$ to measure lengths and angles.
    Here it will be necessary to refer to both, a 3-dimensional space $M_3$ equipped
      with a euclidean metric $g_3$ (with signature (1,1,1)) to discuss electro-/magnetostatics
      and a
      4-dimensional manifold $M_4$ equipped with a Lorentzian metric $g_4$ (with signature (-1,1,1,1))
      to discuss electrodynamics.\\
      In spacetime (either $M_3$ or $M_4$),
    charge in an infinitesimal volume can fulfill the first axiom if it is described
    as a covariant 3-form (a form is an alternating cotensor field) 
    $\rho=\rho_{123}dx_1\wedge dx_2\wedge dx_3$.
    In the following, we write
    ``$\te{d}$'' 
    for an exterior derivative on $M_3$ and
    ``$\te{d}_t$'' 
    for the exterior derivative on $M_4$.\\
    Similarly, $\star_t$ shall denote the Hodge star operator on $M_4$ and $\star$ the operator on $M_3$.
  \item axiom: Charge is conserved, that means, whenever charge changes inside a given volume over time, then
    this change must be equal to the flow of charge $j$ out of the surface of the volume.\\
    In the above defined terms, charge conservation reads
    \begin{equation}
      \dot \rho+\te{d}j=0.
      \label{eq:chargeConservation}
    \end{equation}
    This axiom results, together with the Poincare Lemma, in the inhomogeneous Maxwell equations.
    The Lemma states that on any contractible domain, a form is exact iff it is closed.\footnote{
      On a manifold that is not contractible, de Rham cohomology has to be considered
    and the equations then only hold locally.}
    As $\rho$ is a 3-form, $\te{d}\rho=0$ which implies, by the Poincare-Lemma
    that there is a 2-form $D$ (in the spatial 3D subspace) which satisfies $\te{d}D=\rho$. Thus,
    \begin{equation*}
      \rho=\te{d}D~\overset{(\ref{eq:chargeConservation})}{\Ra}~\te{d}(\dot D+j)=0~
      \overset{\te{(Poincare Lemma)}}{\Ra}~\dot D+j=\te{d}H.
    \end{equation*}
    The forms $D$ and $H$ are called \ti{electric excitation} and \ti{magnetic excitation} respectively.\footnote{
    Following the precise understanding of the great physicist A. Sommerfeld.}
  \item axiom: Existence of $E$ field and Coulomb-force.\\
    Observations show 
    that a charged body in a space with other charged bodies is subject to a force
    $K^{(e)}$.
    The cause for this force is attributed to the existence of
    a space filling quality called the \ti{electric field}
    $E$ (which in turn is related to the charge distribution by the constitutive equations below).\\
    Now if one moves a charged test body in an electric field, one has to spend work/energy $-W_e$ depending
    on the way $\gamma$ in which the charge is moved and proportional to the
    amount of charge $q$. Therefore, $-W_e(\gamma)=q \int_\gamma E$.
    As $W_e$ is a scalar function, this shows that $E$ must be described as a 1-form, $E=E_1 dx_1+E_2dx_2+E_3dx_3$.\\
    Using the usual definition of work, relating it to force, $W_e(\gamma)=-\int_\gamma K^{(e)}$
    (so force is also a 1-form),
    one may take an infinitesimal $\gamma$ to write the force law as $K^{(e)}_p=qE_p$ at a point $p$.
    It is also called \ti{Coulomb-force}.
    Experiments reveal that under electrostatic conditions (i.e. no time dependence of
    the field and the charge density), $E_p$ can be expressed explicitly in terms of the
    density:
     If one denotes the distance of some point $p'$ (of our affine space) to a point $p$ by
      $r_{p'}(p):=|p'-p|$, 
      then the unit covector that points
      from $p'$ to $p$ is just $dr_{p'}(p)$. For a given charge density
      $\rho'=\rho(p')dx_1'\wedge dx_2'\wedge dx_3'$, the expression is then given by
      \begin{equation}
        K_p^{(e)}=qE_p,\qquad E = \frac{Q}{4\pi \varepsilon_0}\frac{dr_{p'}}{r_{p'}^2}
         = \frac{\sum_i q_i}{4\pi \varepsilon_0}\frac{dr_{p'}}{r_{p'}^2}
         \ra \frac{\int_V \rho'}{4\pi \varepsilon_0}\frac{dr_{p'}}{r_{p'}^2}
        \qquad
        \te{(no integration over $dr$)}.
        \label{eq:Coulomb}
      \end{equation}
  \item axiom: Existence of $B$ field and magnetic Lorentz-force.\\
    Observations also show that moving charges (currents) near magnets are subject to a force. The cause
    is thus attributed to the existence of a \ti{magnetic field} $B$.\footnote{
      It is in fact possible to derive the form of Maxwell's equations for non-accelerated
      source-charges entirely from Coulomb's law and the
      axioms (and formalism) of Special Relativity. This is done in great detail in
      the interesting treatment
      by \cite{Haskell2003}. However, I would not subscribe to his conclusions about accelerated charges
      because he did not seem to have taken covariance into account.
    }
    \\
    If one moves a current (e.g. a current carrying wire)
    in this field, one has to spend work $-W_m$ that depends on
    the surface $S$ through which the current (carrying wire) is moved and which is proportional
    to the amount of current $I$. Therefore, $-W_m(S)=I\int_S B$. The expression $\int_SB$ is also
    called \ti{magnetic flux} through $S$.
    This shows that $B$ must be described as 2-form, $B=B_1dx_2\wedge dx_3+B_2dx_3\wedge dx_1+B_3dx_1\wedge dx_2$.\\
    Now, if one considers an infinitesimal part of the current (carrying wire)
    described by the vector $\varepsilon u$ and an infinitesimal movement in the magnetic field
    by another vector $\varepsilon u_2$, then $\varepsilon u$ and $\varepsilon u_2$ span the
    infinitesimal surface $S$ through which the current moves infinitesimally.\\
    As a consequence,
    $I\int_S B = IB_p(\varepsilon u_2,\varepsilon u) + \mathcal{O}(\varepsilon^3)$.
    Therefore, the force that acts on the infinitesimal part of the current is $IB_p(\cdot,\varepsilon u)$.
    Thus if the wire shape can be parameterised by $\gamma=a(t)\partial_1+b(t)\partial_2+c(t)\partial_3$, the force
    in a $B$ field is given by
    $K_\gamma^{(m)}=I \int_\gamma B_\gamma(\cdot,\partial\gamma/\partial t)$\footnote{
      Note that the result of the integral is a 1-form because one integrates along $\gamma$ which
      is 1-dimensional. If one specifies $B$ to be the field caused by another current-carrying wire, then
      one obtains \ti{Ampére's law}.}
      and if we assume that at a certain point of the
    wire, the whole current is given by
    $qv=I\varepsilon u$, then the force on that point is given by $K_p^{(m)}=qB(\cdot,v)=-qv\lrcorner B$.
    This is also called the \ti{magnetic Lorentz-force}.
  \item axiom: Superposition and (total) Lorentz-force.\\
    The forces of the electric and magnetic field add up linearly and the (total) \ti{Lorentz-force}
    is given by
    \begin{equation*}
      K_p=q(E_p-v\lrcorner B_p).
    \end{equation*}
    (Because parts of $E$ are transformed into parts of $B$ and vice versa upon Lorentz transformations,
    such a linear superposition is also necessary to make the physics the same for all observers.)
  \item axiom: Faraday's law of induction (resulting in the homogeneous Maxwell equations).\\
    Faraday's law says that magnetic flux is conserved in the sense that
    whenever flux changes inside a given surface over time, then
    this induces a change in field tension $E$ at the border of this surface
    that acts against\footnote{The negative sign is crucial to obtain Lenz's law
      in combination with $\te{d}H=j+\dot D$.}
    the change of the flux: $\int_S \dot B = - \int_{\partial S} E$, or, in differential
    form:
    \begin{equation*}
      \dot B = -\te{d}E~\Ra~\te{d}B=\te{const}.
    \end{equation*}
  \item Magnetic fields are source-free.\\
    In the regime of classical electrodynamics, it is not possible to find magnetic monopoles. Instead one
    finds
    \begin{equation*}
      \te{d}B=0.
    \end{equation*}
    Alternatively, one can combine the relativity principle with the preceding axiom
    to obtain the same consequence as follows.
    The principle states ``All laws of nature are the same
    in all inertial frames.''\footnote{The justification for this much stronger assumption
    is that laws are observed to be invariant in experiments.
    Note however that the conclusion, $\te{d}B=0$, follows from applying the relativity principle
    to the last axiom that says $-\te{d}E=\dot B$.
    If, for some reason, 
    nature would provide a way for the law to become
    $-\te{d}E=\dot B+j_m$, analogous to $\te{d}H=\dot D+j$, then these more symmetric Maxwell equations
    would be consistent. 
    In that case, one would have magnetic monopoles. In other words, Relativity does not
    forbid magnetic monopoles per se, it only does so in combination with $-\te{d}E=\dot B$.
}
    \\
    If this is true, then $\te{d}B=\te{const}$ implies $\rho_m:=\te{const}=0$ (we can call
    $\rho_m$ \ti{magnetic density} because it is a 3-form)
    because otherwise an observer in a moving frame would see a time varying magnetic density $\rho_m=\rho_m(t)$
    contradicting the principle $\Ra~\te{d}B=0$.
  \item axiom: Constitutive equations (relating $E$ with $D$ and $B$ with $H$).\\
    $E$ and $D$ are likely to be related because if $E$ is the cause for forces on charged particles
    and these only experience forces in presence of other charged particles, then the
    excitation $D$ that they cause should in turn be connected to the cause of their forces $E$.
    An analogous argument makes the connection of $H$ and $B$ with respect to charge currents plausible.
    In any case, empirically the following equation is found to hold:
    \begin{equation*}
      D=\varepsilon_0\star E,~H=\mu_0^{-1}\star B
    \end{equation*}
    where $\varepsilon_0$ is the \ti{dielectric constant} and $\mu_0$ is the \ti{magnetic (vacuum) permeability}.
    One can show that this is the only way to relate these quantities if one demands
    their relation to be linear, local and invariant under the action of the Poincare group.\footnote{The
      relation is only true when considering \ti{all}
    charges and currents in space. For cases in which one can only
    describe the distribution effectively, for example in materials, one finds the relations
    \begin{equation*}
      D=\varepsilon_0\star E+P[E],~H=\mu_0^{-1}\star B-M[B].
    \end{equation*}
  However, fields inside materials will not be considered in this treatment.}
\end{enumerate}
    This concludes our axiomatic approach. Combining the equations, we are left with Maxwell's equations
    \begin{equation}
      \begin{split}
        \te{d}D&=\rho,~\te{d}H=\dot D+j\\
        \te{d}B&=0,~\te{d}E=-\dot B\\
        D = \varepsilon_0&\star E,~H=\mu_0^{-1}\star B\\
        K&=q(E-v\lrcorner B)
        \label{eq:maxwellsEq}
      \end{split}
    \end{equation}
    A reformulation using $\te{d}_t$ is very helpful to identify the symmetry groups whose action leaves
    Maxwell's equations invariant.
    To this end, one can define the Faraday form $F:=E\wedge dt+B$,
    the Maxwell form $G:=D-H\wedge dt$,
    the 4-current $J:=\rho - j\wedge dt$
    (which is conserved: $\te{d}_tJ=(\dot \rho+\te{d}j)\wedge dt=0$),
    the constant $k:=-\sqrt{\mu_0/\varepsilon_0}$
    and the
    relativistic four velocity, $u=dx/ds$ (where $ds$ is the infinitestimal distance
    between spacetime-points)
    such that Maxwell's equations
    take the form
    \begin{equation}
      \te{d}_tF=0,~\te{d}_tG=J,~ \star_t F = kG\qquad\te{ or }\qquad
      \te{d}_tF=0,~\te{d}_t\star_t F=kJ~\te{ with force }~
      K_t=q~
      u\lrcorner F.
      \label{eq:ED}
    \end{equation}
    In practice, it is notationally convenient
    to set $\varepsilon_0$ and $\mu_0$ to one. Furthermore, if one only refers to forms on $M_4$,
    one will drop the $t$ in $\te{d}_t$. As $\te{d}F=0$, we can invoke Poincaré lemma again
    to deduce, on a star shaped region, the existence of a vector potential $A$ such that $\te{d}A=F$.
    Using it, we obtain the equations
    \begin{equation}
      \te{d}\star\te{d}A=J\qquad\Ra\qquad\te{d}\star F=J,~\te{d}F=0
      \label{eq:MWbothForm}
    \end{equation}
    They are called Maxwell's equations in terms of vector potentials and Faraday tensors respectively.

\bigskip

Back to subsection \ref{chap:applications}.

\subsection{History and empirical limits of the laws of Electrodynamics}
  \label{sec:phenomenaED}
  To conclude the analysis of Electrodynamics, a short overview is given about some aspects of its history and about
  how its laws were experimentally verified.
  No pretension is made that the overview is complete in any way. However, clarifying the validity bounds
  of a theory is important to understand the transition from one theory to another. The discussion below is mainly provided in order to show how difficult a precise determination of such bounds is.\\
  Here the validity bounds of a physical phenomenon in a classical field theory are understood to be
  the empirical bounds of its corresponding law in terms of differential equations. 
  Importantly, every law that is formulated in terms of differential equations
  already underwent a process of extrapolation. To see this clearly,
  it is important to realise that experiments are always only conducted under specific circumstances.
  What allows the extrapolation to a law is the demand for \ti{consistency} with a set
  of many but finitely many experiments.
  In Electrodynamics one could validate a law for a specific charge distribution, say on a cylinder or helicoid.
  When conducting an experiment for this specification, one could validate
  Maxwell's equation for this particular charge distribution.
  However, this would not validate Maxwell's equation in general because one would have to provide
  empirical bounds for all the other possible geometries in which charge can be arranged
  (and these are uncountable). \\
  Or one could construct a small motor by exploiting Faraday's law of induction and thus
  validate the law for a specific magnet and a specific coil carrying a specific current.
  But this would not be an experimental validation of Faraday's law in general.
  Facing these issues, it seems in principle impossible
  to validate a theory as a whole, or to reduce phenomena to a relevant set.
  The reason is not only that every measurement has an associated measurement error
  but that the range of parameters for which systematic measurement data
  is available is much smaller than one might think at first.
  \\
  However, if we would agree to be satisfied with an approximation of the empirical bounds of the most general laws, then one could take these laws
  to be our relevant phenomena (which provide at least an approximation for the validity
  of all phenomena. Thus, we'd obtain something weaker than in the ideal proposal but still useful).
  Especially Electrodynamics is experimentally well approachable because
      the interpolation of the results of a range of experiments was
      taken as starting point for the very construction of the theory.
  That
  empirical considerations
      are at the heart
      of the formulation,
  is apparent in the above axiomatic approach
  where the conservation of charge, the Coulomb-law, the Lorentz-force law 
  and Faraday's law are directly referred to as starting points for subsequent definitions and constructions.\\
  Take, for example, Coulomb's law (\ref{eq:Coulomb}).
      It says that the force due to any charge distribution is the sum of the distribution of
      the individual charges.
      This means that an estimate of the validity of Coulomb's law is also an estimate
      for the validity of the principle of superposition for electric fields.
      This principle
      breaks down in non-linear optics and for very strong fields
      but it is known to be quite accurate in the weak field limit outside of
      matter. Furthermore, electrostatics rests on Coulomb's law, and one can therefore easily show that
      it is equivalent to the differential equations in electrostatics. 
      And this in turn means that
      it suffices to test Coulomb's law with distributions over a sufficiently wide range to
      estimate the bounds of electrostatics.
      \\
      Something else that makes
      Electrodynamics especially tractable is the fact that
      in free space it has no ``non-fixed parameters''
      which must be determined by experiment
      except for $\varepsilon_0$ and $\mu_0$.
      That only they are appearing in the equations
      is rather special if one compares this situation to, for example,
      electrodynamics in matter, where every material has its own special properties and therefore can
      only be described effectively by introducing additional fitting parameters, like the
      resistance of a conductor (which in turn is only approximately constant and depends on other
      parameters like temperature). Of course, if one assumes that all materials are themselves made of
      atoms that are made of protons and electrons that interact electromagnetically, then one could
      develop the idea that in the end, if one takes Maxwell's equation in free space and
      puts in the charge distribution that is constituted by all particles in a given material,
      the effective parameters should come out as predictions. But on one hand, this idea does not take into account quantum effects (quantum electrodynamics and other QFTs) and on the other hand, it would in any case require the
      knowledge about the states of billions of particles, which is  (in practice) not possible and hence
      one can often only validate
      the description of effective macroscopic phenomena.
      Moreover, for practical purposes the effective models
      with their fitting parameters remain the most efficient approach.\footnote{
        One can also employ models that act as ``mediators'' between parts of Maxwell's equations and
        equations in matter, e.g. the Drude model for ``deriving'' Ohm's law. There one assumes
        the force $\dot p = qE-p/\tau$, and looks at the equilibrium state in which $\dot p=0$
        such that $qE=p/\tau$ which results, together with a current $j=nqv$, $p=m_ev$, $E=U/d$
        and $j=I/A$ in a relation $U\propto I$. However, $\tau$ is clearly a new fitting parameter,
        thus making the model also an approximation. For more information see \cite{mcelroy}.}
        
        \bigskip
        
        For another example, note that
      the standard model of particle physics also requires a lot of free parameters that are
      determined via fits after a complex renormalisation procedure.
      For such a model, it is far more difficult to provide actual bounds for its validity because
      perturbation theory is involved to obtain (most) results that can be verified by experiment.
      In other words, the correspondence between measurable observables (usually
      related to correlation functions) and
      the most general equations (e.g. the Schwinger-Dyson equation or the path integral formalism) is much less direct as in electrodynamics.
      \\~\\
      One could certainly discuss more complexities that arise when determining empirical bounds
      but here, the remainder of the subsection shall content itself with providing references to some experimental evidence
      for Maxwell's equations which shall act as the underlying laws of our relevant phenomena.
      \begin{enumerate}
        \item
          \textbf{Charge conservation:} Was already formulated in 1747 by Benjamin Franklin:
          ``It is now discovered and demonstrated, both here and in Europe,
          that the Electrical Fire is a real Element,
          or Species of Matter, \ti{not created} by the Friction, but \ti{collected} only.'' \cite{Franklin}.
          Faraday provided first solid experimental evidence for it in 1843 (see \cite{Heilbron1979}).\\
          And until today, there is no experiment that states any evidence for non-conservation of
          charge, even in regimes where electrodynamics is not used for a description of nature
          down to scales of particle physics. There, charge can be destroyed and created but
          it is always created and destroyed in pairs, i.e. for every positive charge that
          is created or destroyed, a negative charge is created or destroyed,
          see e.g. the particle physics review by \cite{0954-3899-37-7A-075021}.\\
          Concrete empirical bounds for charge non-conservation are provided in
          \cite{Belli}. For decays of electrons with a decay time of $\tau_e>2.4\cdot 10^{24}$ years
          (with a confidence interval of 90\%), they predict that the parameter $\varepsilon_W^2$, which
          is the square of the ratio of the coupling that does not conserve charge to the conserving Fermi coupling,
          has magnitude $\varepsilon_W^2 <2.2\cdot 10^{-26}$.\\
          This puts charge conservation in the macroscopic realm of electrodynamics on a very solid footing.
          Thus, if one accepts axiom 2 of subsection \ref{sec:axDeri} (stating that spacetime can be modeled as affine space or
           manifold in which the Poincaré lemma is applicable and
          charge as a 3-form), then the inhomogeneous Maxwell equations $\te{d}D=\rho$ and $\te{d}H=j+\dot D$
          (note that the equations are not the same as $\te{d}\star E=\rho/\varepsilon_0$ and
          $\mu_0^{-1}\te{d}\star B=j+\varepsilon_0\star \dot E$ which are also sometimes referred to as
          inhomogeneous Maxwell equations because $\star$ requires uncertainty bounds on the metric
          and the relation between $D,H$ and $E,B$ requires uncertainty bounds on the constitutive equations)
          can be viewed as having the same empirical bound as charge conservation (because they are
          directly derived from it).\\
          As a consequence, a natural follow-up question is whether axiom 2 is well verified.
          That charge can be modeled as 3-form is merely a consequence of the idea that space
          can be modeled as affine space or on a suitable
          manifold and the existence and form of charge (axiom 1). Axiom 1 is
          evident by the very existence of the
          multitude of electromagnetic phenomena. Even if charge is just an emergent
          quality of some deeper ramifications of nature that are yet to be discovered, we can define
          it as the emergent quality that shows all the effective behaviours that we do observe
          when manipulating it according to an understanding of electromagnetism and in this sense
          it surely does exist as is manifest by the countless applications of electromagnetism in our
          daily life. \\
          That space can be modeled as affine space or a suitable
          manifold is less obvious. Indeed,
          according to Einstein's general relativity,
          it might be the case that the topology of spacetime is such that the Poincaré lemma is not
          applicable globally. Perhaps however, the equations would still always hold locally.
          Furthermore, one could justifiably ask why it is a good idea to
          model spacetime locally as a Euclidean affine or vector space at all
          (as is done by definition when using a manifold).\\
          This question is probably not easy to answer and goes back to the philosophers and mathematicians
          who created this idea. Perhaps it is already predated by the ideas of the ancient greek Euclid himself (who formulated geometry in a synthetic manner), followed by Descartes who invented the cartesian coordinates, and scientists like Newton and Leibniz who
          used these ideas to formulate their theories. And today all areas of science and engineering use
          these concepts to produce well working methods. Thus, the idea that space is locally Euclidean (and globally a manifold) is macroscopically very well verified
          and at the same time, it is hard to quantify any error of this assumption. 
          Even general relativity is locally euclidean and one would thus
          need to find some failure of the assumed topology of a manifold in 
          observed data. 
          In any case, no relevant impact on the scales of interest
          for electrodynamics has been detected for otherwise one would have noted this behaviour in applications
          or there would be proposals for experiments that test these ideas.\\
          Another
          reason why it is hard to quantify a possible error in space
          being locally euclidean is that space itself can only be envisioned by studying the qualities
          therein and thus can not directly be put to a test.\\
          Hence, even without a quantitative factor, one might take the working applications
          that were developed with the model of local Euclidean space as evident enough to trust in axiom 2 (at least down to quite small scales) almost
          as much as in charge conservation and therefore
          decide to take as overall empirical bound on the inhomogeneous equations a factor close to $\varepsilon_W$.
        \item
          \textbf{Coulomb-law:} Was first systematically studied by Cavendish in the 1770s. However, he
          was an extraordinarily shy man and thus did not publish his writings.
          Coulomb studied the same effect and published his findings in 1784. Maxwell published
          Cavendish's really great writings in 1879, see the new reprinted edition \cite{Maxwell2016}.
          A more recent (though still nearly 50 year old) account is provided by \cite{bartlett1970}.
          Surprisingly, their experimental design was still quite similar (though of course
          more precise) to Cavendish's. They were also testing the validity of the square law
          by using concentric spheres and estimated that the force $F$ was proportional to $r^{-2+q}$
          where $|q|\le 1.3\cdot 10^{-13}$. This is a very tight bound on the inverse square in the law.
          However, note that they were not interested (and it is experimentally very
          difficult) to test this proportionality for a wide range of varying parameters of the charge $Q$
          and the radius $r$. Their 5 concentric spheres, just like the 2 concentric spheres of Cavendish,
          were placed near to each other. This, as noted above, tests the law for a rather confined set of
          radii and charges. Would the same factor $q$ be found if the spheres had radii of several kilometres
          in extend? It is tempting to answer with ``yes'' because it is not obvious
          what should change if the problem is just
          scaled up. But if we were to apply a very rigorous standard, everything that
          the experiment really shows is the estimate of the parameter $q$ for their particular spheres.\\
          This is the problem that was outlined at the beginning of the section and the reason why
          one can only obtain an estimate for the bounds of all phenomena. It is reasonable that
          experiments should at least be interpolated slightly around verified parameters
          (i.e. other values of $Q$ and $r$ that do not deviate too much) because
          if something would suddenly change at some smaller perturbation, then this would probably quickly have
          become evident to a careful experimentalist or in later applications. \\
          But for larger parameter changes such interpolations should be enjoyed with more care.
          For, example, increasing the charge very much is known to result in non-linear effects
          (and was of course
          also known to \cite{bartlett1970}) or when making the distance between charges very small,
          then quantum effects become important.
          To make the estimates on the bounds of Coulomb's law somehow more substantial,
          references to at least two other sources are therefore provided below.\\
          If we acknowledge quantum electrodynamics as the more general theory from which
          electrodynamics can be derived, then the experiments that verify quantum electrodynamics (QED)
          give us another source for verifying Maxwell's equations.
          In particular, it is believed to be possible to estimate the validity of QED by measuring
          the fine structure constant $\alpha$.\footnote{
      In QED, there is some subtlety in this verification,
      because theoretical predictions require as input an extremely precise value of $\alpha$,
      which can only be obtained from another precision QED experiment.
      Because of this, the comparisons between theory and experiment are usually quoted as
      independent determinations of $\alpha$.
      QED is then confirmed to the extent that these measurements of $\alpha$
    from different physical sources agree with each other.}
          These measurements are said to be among the most accurate in the world.
          \cite{alpha} for instance have determined a very precise value
          that has an error of only 0.7 parts per billion.\footnote{Interestingly,
            it has been proposed and investigated if the fine structure ``constant''
            allows some variation over time and at least has been found to
            do so by \cite{alphaOverTime}.}
            Of course, to really take
          this value as evidence for the part of QED from which electrodynamics can be derived
          would require further discussion but
          this will not be done here.
          For now, let us assume that the verification allows us to draw some further conclusions
          about the validity of Coulomb's law.
            In quantum field theory, one possibility for deriving
            the coulomb potential for two point charges is to
            add a mass parameter $m$ for the photon to the usual Lagrangian of electrodynamics,
            $\mathcal{L}\ra -\frac{1}{4}F \wedge G+\frac{1}{2}mA\wedge \star A+A\wedge J$
            to avoid singularity issues with the propagator resulting from
            the treatment with the path integral. The resulting tree level potential
            then looks like a Yukawa-potential
            \begin{equation*}
              V = \frac{e^{-mr}}{4\pi r} \qquad \te{(natural units employed)}
            \end{equation*}
            which goes to the usual em-potential $V \ra (4\pi r)^{-1}$ in the limit of $m\ra 0$,
            see for example \cite{Zee2010}, section I.5. Therefore, if we take this derivation to be correct,
            then an indirect upper bound for large distances (not for
            short distances because it only takes into account the tree level order of vacuum polarisation)
            of the Coulomb-law is provided by the photon mass - if the photon mass is experimentally
            verified to vanish, then the Coulomb law does not decrease in any way at large distances according
            to quantum field theory (though this evidence is of course of another quality than
            a direct measurement of the Coulomb law at large distances, e.g. using geophysical methods).
            And there are upper bounds for the photon mass in the literature, e.g.
            \cite{Accioly2010} estimated it to be $m<3.5\cdot 10^{-11}$MeV.
            $\approx 3.5\cdot 10^{-11}\cdot(1.79\cdot 10^{-30})$kg.\footnote{Furthermore, on p. 6,
            they cite other references with even smaller limits.}\\
            At small distances, vacuum polarisation becomes important.
        The general formula for the interaction potential between two point electric charges which
        contains the lowest order corrections to the vacuum polarization is derived in
        e.g. \cite{Frolov2011}. They use a sum of the
        Uehling and Wichmann-Kroll potentials $U(r)+W_K(r)$ to express these corrections.
        It is apparent that
        effects become quite strong close to the compton length of the electron, $\lambda_c \propto 10^{-12}$m.
        Actually it is rather surprising that even at these distances, the Coulomb law is still taken as
        an important term in the description.\\
        However, both the above sources were mainly cited to show that neither
        for large nor for small distances, quantum field theory imposes
        strong conceptual restrictions
            to an inverse square law within the bounds in which it can be directly measured.\\
            As a direct measurement still provides the most reliable source for empirical bounds,
            it might be a bit more secure to finally settle down on the values
            provided in \cite{Jackson1999} who notes after an interesting
            discussion in section I.2 (with further references) that
            ``The laboratory and geophysical tests show that on length scales of order
            $10^{-2}$ to $10^{7}$m,
            the inverse square law holds with extreme precision.
            At smaller distances we must turn to less direct evidence often
            involving additional assumptions.'' He also does consider these other assumptions
            reasonable and concludes by saying that
            ``The inverse square law is known to hold over at least 25 orders of magnitude in the length scale!''
            but for now, it might be more secure to stick
            with the bounds $\varepsilon_C=([10^{-2},10^7]$m,$|q|$). Assuming, for this range, similar bounds on $q$ as those estimated by \cite{bartlett1970}, this would at least imply  that in the intervall from $10^{-2}$ to $10^7$m, it seems reasonable to trust Coulomb's law
            to be valid within an error not too far from $|q|\le 1.3\cdot 10^{-13}$ for the inverse square as described above.\\
          The Coulomb law is not only important for verifying the electro-static force relation $K_p=qE_p$ but
          as the static equation $\te{d}\star E=\rho/\varepsilon_0$ follows from Coulomb's law,
          one could argue that it is not completely unsensible to take its
          bound to be approximately the same.
          Furthermore, $\rho=\te{d}D$, thus this would also establish
          a bound for the constitutive relation $\varepsilon_0\te{d}\star E=\te{d}D$ in electro-statics.
        \item \textbf{Lorentz-force law, Faraday's law, constitutive equations
          and Ampère's circuital law:}\\
          The magnetic part of the Lorentz-force law was first formulated in its present form by
            \cite{Heaviside1889} who invented modern vector notation to express Maxwell's equations.
            In 1895, \cite{Lorentz1895} formulated the law including the electric forces and also
            showed that Maxwell's equations are invariant under Lorentz transformations.\footnote{
              Interestingly, according to \cite{Huang1993}, it is not completely clear
              whether the \ti{relativistic} Lorentz-force law has been experimentally well tested or not.}\\
          Faraday's law of induction is described in detail in the ninth series of his \ti{Researches}, see
          \cite{Faraday1834}.\footnote{
            Faraday was a full-blood experimentalist. Among his investigations for induction, he
            described the following test:
            ``On placing the
            tongue between two plates of silver connected by wires with the parts which the
          hands had heretofore touched (1064.),
        there was a powerful shock on breaking contact, but none on making contact.''}
        However, his description is not formulated using a mathematical formalism.
        (Maxwell later uses this series to collect evidence for the fact that fields
        can carry momentum, see \cite{Maxwell1873}, Chapter V.)\\
        The meaning of the constitutive equations is usually understood to lie in the fact that
        they may vary for Maxwell's equations in matter. Thus, to account for different constitutive equations
        was a long lasting endeavor involving the development of electrodynamics in several subareas
        of solid state physics, in magnetohydrodynamics, plasma physics and others.
        To handle the equations there, it was necessary to develop linear response theory and other tools.
        Therefore, their development can not be dated back to a single contributor.\footnote{
          Of course all contributions are to a certain extend the product of the scientific community as
        a whole and the combined efforts of individuals.} Additional information about deviations that
        may be taken into account are given in e.g. \cite{Mackay2010}.
        Ampère's circuital law, $\mu_0^{-1}\te{d}B=\varepsilon_0 \star \dot E+j$, (not to be
        confused with Ampère's (force) law), was actually invented by Maxwell as well
        (using methods of hydrodynamics), \cite{maxwellAmpereCircuitalLaw}.
        \\
        All the above named laws were introduced together for the reason that it was hard in all
        cases to find more recent accounts for their experimental validity.
        It was rather surprising that it was difficult to find
        publications that test Coulomb's law
        and that it was not possible to find any systematic experimental test that refers to e.g.
        Faraday's law. There are millions and millions of copies
        of Maxwell's equations in books, scripts, notes and websites on electrodynamics but very rarely,
        almost never,
        the exact empirical bounds are discussed. There are even thousands of manuals on how students should
        construct experiments that test Maxwell's equations but these are all descriptions of
        experiments that can take place in a small lab and are similar in construction.
        The author has not found reports on recent experiments that test Faraday's law or Ampère's law on a large scale.
        One might try to study the experimental and technical designs of early inventors (like Nikola Tesla or Guglielmo Marconi) or  one could look into the literature that deals with  present applications, e.g. in the
        telecommunication or astrophysical sector.
        But the interpretation of these applications would be different in quality and it would use up a lot of time. 
      \end{enumerate}
        Mostly, books and websites will quote that at the quantum level, quantum electrodynamics is
        the more precise description. But the transition between electrodynamics and
        quantum electrodynamics is not at all that clear.
        The above considerations regarding the
        Coulomb potential and its transition to the Uehling and Wichmann-Kroll potentials
        already show how complicated approximations become
        and it gets much more complicated for multiparticle systems. One must
        always make additional (mostly statistical) assumptions to derive the simple laws of electrodynamics
        in the macroscopic limit and a direct grasp of what ``really happens'' at the microscopic level
        is not naturally obtained.\\
        When discussing with other scientists, they confirmed that
        a central experimental database should ideally exist
        but probably does not exist because of the work that would be needed to curate the data. Furthermore, one would perhaps have to go deep into the history of science
        and this would again take a lot of time.\\
          What is interesting about this is that the scientific community does not seem to be particularly
          bothered by the fact that laws are not presented together with their bounds.
          Instead of providing definite scale intervals of validity,
          one (sometimes) finds lists of effects that one theory does describe and another does not.
        Maybe the necessity to mention the bounds is not seen until there is some counter-evidence, some effect
          that requires the development of another theory. But actually that is not good practice
          because it makes the transition between theories blurrier.\\
        Due to the possibility to find some evidence
        on the bounds of the Coulomb law but not in the same way on the other laws, it seems to me
        that it is usually assumed that the laws of electrodynamics must break down together
        once certain scales are reached.
        This might not even be a false assumption because, as already mentioned in an earlier footnote,
        \cite{Haskell2003} shows how to derive the form of Maxwell's equations from Coulomb's law and the
        formalism of Special Relativity (SRT). Thus, if one is willing to trust in SRT
        for the evidence that has been found regarding the non-additivity of the speed of light,
        the length contraction and time dilation phenomena, one could 
        at least argue that
        the bounds of Coulomb's law
        can represent the bounds of electrodynamics.
        
        \bigskip
        
Presumably, the above discussion already gives the reader an understanding of the
complexities that arise if one tries to specify the validity bounds of a law.
The experimental discoveries, the communication of this understanding, the extrapolation
of the laws involved and the theoretical advances are all subject to convoluted historical developments.
There is no centrally organised database that curates data for all kinds of theories and effects.
At least in particle physics, there is the particle database but for classical field theories
the history is simply too long and the scientific fields have developed so many subfields that
it is hard to summarise all this.\\
Therefore, if the transition of theories is studied, one can either work on the subject like a historian,
going through all those developments or one can stay on a more formal
level and consider similarities in the mathematical formalism
in the hope of finding deeper consistency conditions that can help to
clarify which possible models should be considered as those that have a tight relationship
with reality.
 
 \bigskip
 
 Back to subsection \ref{chap:applications}.

\parskip 7pt
\bibliography{bib}{}
\bibliographystyle{apalike}

\end{document}

%% file: programmingCodeDisplay.tex
\DeclareFixedFont{\ttb}{T1}{txtt}{bx}{n}{8} 
\DeclareFixedFont{\ttm}{T1}{txtt}{m}{n}{8}  

\usepackage{color}
\definecolor{deepblue}{rgb}{0,0,0.5}
\definecolor{deepred}{rgb}{0.6,0,0}
\definecolor{deepgreen}{rgb}{0,0.5,0}
\definecolor{backcolour}{rgb}{0.95,0.95,0.92}

\definecolor{codegray}{rgb}{0.5,0.5,0.5}
\definecolor{codepurple}{rgb}{0.58,0,0.82}
\usepackage{listings}

\newcommand\pythonstyle{\lstset{
  language=Python,
  commentstyle=\color{deepgreen},
  numberstyle=\tiny\color{codegray},
  stringstyle=\color{codepurple},
  backgroundcolor=\color{backcolour},
  numbers=left,
  basicstyle=\ttm,
  otherkeywords={self},             
  keywordstyle=\ttb\color{deepblue},
  emph={MyClass,__init__},          
  emphstyle=\ttb\color{deepred},    
  stringstyle=\color{deepgreen},
  showstringspaces=false            %
}}

\lstnewenvironment{python}[1][]
{
\pythonstyle
\lstset{#1}
}
{}

\newcommand\pythonexternal[2][]{{
\pythonstyle
\lstinputlisting[#1]{#2}}}

\newcommand\pythoninline[1]{{\pythonstyle\lstinline!#1!}}